\documentclass[journal]{elsarticle}
\journal{Journal of Systems and Software}
\usepackage{float}
\usepackage{multirow}
\usepackage{savesym}
\usepackage{amsmath}
\savesymbol{iint}
\savesymbol{iiint}
\usepackage{txfonts}
\restoresymbol{TXF}{iint}
\restoresymbol{TXF} {iiint}
\usepackage{graphicx}
\usepackage{changepage}
\usepackage{url}
\usepackage{amssymb}
\usepackage{ulem}
\usepackage[table]{xcolor}
\usepackage{tabularx}
\usepackage{array}
\newcolumntype{P}[1]{>{\centering\arraybackslash}p{#1}}
\usepackage{booktabs}
\usepackage{wasysym}
\usepackage{listings}
\usepackage{fontawesome}
\usepackage{multirow}
\usepackage{adjustbox}
\usepackage{pdflscape}
\usepackage{rotating}
\usepackage[utf8]{inputenc}
\usepackage{tikz,lipsum,lmodern}
\usepackage[most]{tcolorbox}
\usepackage{longtable}
\usepackage{pgfplots}
\usepackage{pgf-pie}
\usepackage[toc,page]{appendix}
\usepackage{textcomp}
\usepackage{epstopdf}
\usepackage{csquotes}
\usepackage{dirtytalk}
\usepackage[document]{ragged2e}
\usepackage{parskip}
\usepackage{siunitx}
\usepackage{amssymb}
\usepackage{tikz}
\usepackage{hyperref}
\hypersetup
{
  colorlinks   = true, 
  urlcolor     = red, 
  linkcolor    = blue, 
  citecolor    = blue 
}

\usepackage[colorinlistoftodos,prependcaption,textsize=tiny]{todonotes}
\usepackage[utf8]{inputenc}
\usepackage[english]{babel}
\usepackage{geometry}
\pgfplotsset{compat=1.9}

\usepackage{lineno}

\usepackage{mdframed}
\usepackage{lipsum}
\newmdenv[topline=false, leftline=true, rightline=true, bottomline=false,%
  linewidth=2pt, innerleftmargin=12pt, rightmargin=-4pt,%
  innerrightmargin=12pt, skipabove=8pt, skipbelow=8pt]{leftrightbar}%

\begin{document}

\begin{frontmatter}

\title{Design, Monitoring, and Testing of Microservices Systems:\\ The Practitioners' Perspective}

\author[mymainaddress]{Muhammad Waseem}
\ead{m.waseem@whu.edu.cn}
\address[mymainaddress]{School of Computer Science, Wuhan University, 430072 Wuhan, China}

\author[mymainaddress]{Peng Liang\corref{mycorrespondingauthor}}
\cortext[mycorrespondingauthor]{Corresponding author at: School of Computer Science, Wuhan University, China. Tel.: +86 27 68776137; fax: +86 27 68776027.}
\ead{liangp@whu.edu.cn}

\author[mysecondaryaddress]{Mojtaba Shahin}
\ead{mojtaba.shahin@monash.edu}
\address[mysecondaryaddress]{Department of Software Systems and Cybersecurity, Faculty of Information Technology, Monash University, 3800 Melbourne, Australia}

\author[mythirdaddress]{Amleto Di Salle}
\ead{amleto.disalle@univaq.it}
\address[mythirdaddress]{Department of Information Engineering, Computer Science and Mathematics, University of L’Aquila, I-67100 L'Aquila, Italy}

\author[myfourthaddress]{Gast\'{o}n M\'{a}rquez}
\ead{gaston.marquez@usm.cl}
\address[myfourthaddress]{Department of Electronics and Informatics, Federico Santa Mar\'{i}a Technical University, 4030000 Concepci\'{o}n, Chile}

\begin{abstract}
\justifying
\textbf{Context}: Microservices Architecture (MSA) has received significant attention in the software industry. However, little empirical evidence exists on design, monitoring, and testing of microservices systems.

\noindent\textbf{Objective}: This research aims to gain a deep understanding of how microservices systems are designed, monitored, and tested in the industry.

\noindent\textbf{Method}: A mixed-methods study was conducted with 106 survey responses and 6 interviews from microservices practitioners.

\noindent\textbf{Results}: The main findings are: (1) a combination of domain-driven design and business capability is the most used strategy to decompose an application into microservices, (2) over half of the participants used architecture evaluation and architecture implementation when designing microservices systems, (3) API gateway and Backend for frontend patterns are the most used MSA patterns, (4) resource usage and load balancing as monitoring metrics, log management and exception tracking as monitoring practices are widely used, (5) unit and end-to-end testing are the most used testing strategies, and (6) the complexity of microservices systems poses challenges for their design, monitoring, and testing, for which there are no dedicated solutions.

\noindent\textbf{Conclusions}: Our findings reveal that more research is needed to (1) deal with microservices complexity at the design level, (2) handle security in microservices systems, and (3) address the monitoring and testing challenges through dedicated solutions.
\end{abstract}

\begin{keyword}
\texttt Microservices Architecture \sep Design \sep Monitoring\sep Testing\sep Industrial Survey
\end{keyword}

\end{frontmatter}

\justifying
\section{Introduction}
\label{sec:introduction}

Microservices Architecture (MSA)\footnote{All the abbreviations used in this paper are provided in Table \ref{tab:Abbreviations}.} inspired by Service-Oriented Architecture (SOA) has gained immense popularity in the past few years \cite{dragoni2017microservices}. The International Data Corporation (IDC) predicted that by the end of 2021, 80\% of cloud software will be developed using the MSA style \cite{balalaie2016microservices}. With the MSA style, an application is designed as a set of business-driven microservices that can be developed, deployed, and tested independently \cite{taibi2019monolithic}. MSA follows a “share-as-little-as-possible” architecture approach and communicates with each other through an API layer \cite{SamarpitTulie, richards2015microservices}. In contrast, SOA adopts a “share-as-much-as-possible” architecture approach and typically uses the Enterprise Service Bus for communication purpose \cite{SamarpitTulie, richards2015microservices}. Another key difference between MSA and SOA is that MSA advocates one data storage per microservice, while SOA uses one data storage for the whole system \cite{SamarpitTulie}. Moreover, the systems that adopt the MSA style (i.e., microservices systems) have a better fault-tolerance than SOA-based systems \cite{rajput2018hands}. The advantages of the MSA style are enormous, and the key drivers of adopting MSA are faster delivery, improved scalability, and greater autonomy compared to other architectural styles (e.g., SOA) \cite{jamshidi2018microservices}.

There is a growing body of literature investigating different aspects of MSA \cite{waseemMSAdevops, waseemtestingMSA, di2019architecting, alshuqayran2016systematic, soldani2018pains, pahl2016microservices, LI2020106449}. One notable research area is how to migrate a monolithic system to MSA (e.g., \cite{balalaie2016microservices,fan2017migrating, taibi2017processes, di2018migrating}). Others propose solutions for securing microservices systems (e.g., \cite{dragoni2017microservices, Pereira2019SecMec}). Researchers have recently shown an increasing interest in understanding how microservices systems are developed and migrated in the industry (e.g., \cite{pahl2018architectural, kang2016container, taibi2018continuous}). Knoche and Hasselbringa conducted a survey study to investigate the drivers, barriers, and goals of adopting MSA \cite{knoche2019drivers}. Ghofrani and Lübke reported the services boundaries identification techniques (e.g., Domain-Driven Design (DDD)) and notations used to design and develop microservices systems \cite{ghofrani2018challenges}. The survey study conducted by Viggiato et al. reports the popular programming languages, technologies, advantages, and challenges in microservices systems \cite{viggiato2018microservices}. Haselbock et al. surveyed MSA experts to identify essential design areas (e.g., infrastructure design) and design challenges of microservices systems \cite{haselbock2018expert}. 

In another study, Baškarada et al. empirically explored the practitioners' perspective on the advantages of MSA over monolithic architectures and the challenges faced by MSA practitioners when developing microservices systems \cite{bavskarada2018architecting}. Zhang et al. examined the characteristics, benefits, and costs of MSA in practice \cite{zhang2019microservice}. Zhou et al. investigated the fault analysis and debugging of microservices through an industrial survey \cite{zhou2018fault}. In the same line, Wang et al. conducted a mixed-method study (i.e., interview and survey) with practitioners to collect the best practices, challenges, and solutions for the successful development of microservices systems \cite{Wang2020}. Taibi et al. conducted a mixed-method study (i.e., interview and survey) to investigate the motivations (e.g., maintainability, scalability), issues (e.g., monolith decoupling, database migration, and data splitting), and benefits (e.g., maintainability improvement, scalability improvement) of microservices architecture \cite{taibi2017processes}. They also developed a migration process framework which is composed of three parts (i.e., system structure analysis, definition of the system architecture, and prioritization of feature development) \cite{taibi2017processes}. The mixed-method study by Di Francesco et al. focused on the activities (e.g., domain decomposition, services identification) and challenges (e.g., high coupling among parts of the legacy systems, identification of the boundary of each service) that practitioners face when migrating towards MSA \cite{di2018migrating}. 

Despite these efforts, it is little empirically known how the software industry designs, monitors, and tests microservices systems. These three aspects play a significant role in succeeding with microservices systems \cite{rajput2018hands, newman2015building}. The design of a microservices system (i.e., MSA design) provides a basis for communicating architectural decisions among stakeholders, exposing system structure, realizing use cases, and addressing both functional and quality requirements \cite{richardson2016microservices}. However, to maintain the benefits of microservices systems (e.g., flexibility and separation of concerns), it is essential to reduce complexity at the design level. Monitoring and testing help to achieve a better understanding of the system and improve quality throughout the entire development process. However, microservices systems are increasingly complex and need more intensive monitoring and testing than other types of systems (e.g., SOA, monolith) \cite{rajput2018hands}. This stems from the unique characteristics of microservices systems: greater complexity, polyglot programming languages, heterogeneous and independent microservices, communication behavior at runtime, deployment infrastructure model (e.g., application server vs. container), frequent changes, and several others \cite{PeterArijs,DaveSwersky2018monitoring}. 
Hence, software development organizations need to develop new monitoring and testing strategies or evolve traditional ones to match the unique characteristics of microservices systems. 

This study aims to bridge the gap mentioned above by providing a comprehensive understanding of how microservices systems are designed, monitored, and tested in practice. To this end, we conducted a mixed-methods study, including a relatively large-scale survey with 106 responses and 6 interviews from microservices practitioners. Overall, our results suggest that a combination of DDD and business capability is the most effective strategy to decompose an application into microservices and addressing security and availability is a significant concern for MSA practitioners. We found that ``API gateway'' and ``Backend for frontend'' patterns are the most used MSA patterns for providing the entry point to clients’ requests of microservices systems. We observed that MSA practitioners consider ``clearly defining the boundaries of microservices'' as the most critical challenge when designing microservices systems. Concerning monitoring microservices systems, most of the respondents collect metrics that support monitoring quality aspects, such as performance (e.g., resource usage), availability (e.g., ratio of the system available time to the total working time), scalability (e.g., load balancing), and reliability (e.g., service level agreement). About testing microservices systems, we found that End-to-End (E2E) testing and contract related (e.g., consumer-driven) testing are gaining space in the industry. Our findings also indicate that the complexity of microservices systems poses challenges for their design, monitoring, and testing.

This study makes the following three major contributions: 

\begin{enumerate}

\item It brings quantitative and qualitative evidence of how practitioners design, monitor, and test microservices systems.
\item It identifies the challenges that practitioners face and the solutions employed by them when designing, monitoring, and testing microservices systems.
\item It provides an in-depth discussion on the findings from the microservices practitioners' perspective and implications for researchers and practitioners.
\end{enumerate}

The rest of this paper is structured as follows: Section \ref{sec:background} presents the background of the survey. Section \ref{sec:ResearchMethod} describes the research method in detail. Section \ref{sec:SurveyResults} report the survey results. Section \ref{sec:discussion} and Section \ref{sec:threats} discuss the findings and threats to the validity of the survey respectively. Section \ref{sec:relatedWork} summarizes related work. Section \ref{sec:conclusions} concludes this survey with future research directions.

\section{Background}
\label{sec:background}
Microservices is an architecture style where an application is built as a set of independent services. The main characteristics of the MSA style are (i) \textit{small}, i.e., a service accomplishes a single business capability; and (ii) \textit{autonomous}, i.e., a single deployable unit contains all needed data to be independently executed~\cite{newman2015building}. In contrast, a monolithic system implements features as a single unit to be deployed as a whole~\cite{newman2020monolith}. In recent years, microservices have become popular, and the software industry has moved from a monolithic architecture to MSA. A substantial body of literature (e.g., \cite{jamshidi2018microservices, ponce2019migrating, dragoni2017, esposito2016challenges, Yarygina, cerny2018contextual}) has concluded that MSA is significantly different from monolithic architecture and SOA in terms of design, implementation, test, and deployment. MSA contains a suite of fine-grained services working together as a single application that can be executed independently, while in a monolithic architecture, all functionalities are encapsulated into one single application that cannot be executed independently \cite{ponce2019migrating}. MSA is more fault-tolerant, highly scalable, and loosely coupled than monolithic architecture \cite{ponce2019migrating}. MSA and SOA are also different in several aspects. For instance, MSA-based systems can use shared or single database dedicated to a specific microservice, whereas SOA-based systems can only use a shared database \cite{cerny2018contextual}. In MSA, each service can run its own processes and communicate via lightweight mechanisms (e.g., HTTP, REST), whereas in SOA, each service shares a common communication mechanism (e.g., Enterprise Service Bus) \cite{richards2015microservices}.

This section presents different aspects related to design, monitoring, and testing of microservices systems that are investigated in this study.

\subsection{Design of microservices systems}
According to Bass et al. \cite{bass2003software} ``\textit{the software architecture of a system is the set of structures needed to reason about the system, which comprises software elements, relations among them, and properties of both}''. The design elements of microservices systems are (micro)services, views (i.e., solutions), processes, tools, culture, and organization structure \cite{nadareishvili2016microservice}. This survey investigates several aspects of the MSA design, such as application decomposition and implementation strategies, architecting activities, description methods, design patterns, Quality Attributes (QAs), skills required to design and implement MSA, and MSA design challenges and their solutions. We briefly describe some of these aspects below.

\textbf{Architecting activities}: The architecture of a software-intensive system is the result of various architecting activities that are performed by related stakeholders. Hofmeister et al. defined a general model of architecture design, which is composed of a set of architecting activities, including architectural analysis, synthesis, and evaluation \cite{hofmeister2007general}. Architecting activities are widely explored in the context of monolithic architecture, but the activities are rarely investigated in the context of MSA. We identified only one Systematic Mapping Study (SMS), i.e., \cite{di2019architecting}, that reports architecting activities in the context of MSA. In this study, we investigated the level of agreement and disagreement of the practitioners about the general architecting activities employed in the context of MSA. According to the literature (e.g., \cite{hofmeister2007general, TANG2010352, LI2013777}), the general architecting activities we considered for MSA include Architectural Analysis (AA), Architectural Synthesis (AS), Architectural Evaluation (AE), Architectural Implementation (AI), and Architectural Maintenance and Evolution (AME).

\textbf{MSA description methods}: Architecture description methods consist of a set of practices that are used to express the architecture of a system-of-interest \cite{iso2011ieee}. To describe the architecture of software systems, architects may use various notations \cite{malavolta2012industry}, such as boxes and lines, Unified Modeling Language (UML), Architectural Description Languages (ADLs), and Domain-Specific Languages (DSLs). In this study, we investigated the description methods and diagrams used by microservices practitioners to describe and represent the architecture of microservices systems. 

\textbf{MSA design patterns}: Patterns are reusable solutions to occurring problems in specific contexts \cite{gamma1995design}. Since the emergence of MSA, dozens of patterns have been documented for implementing different aspects (e.g., data management, deployment) of microservices systems \cite{alshuqayran2016systematic, richardson2018microservices, marquez2018actual, marquez2018review,taibi2018architectural}. We investigated how often microservices practitioners use various patterns to address the problems related to design of MSA-based systems, such as application decomposition, data management, deployment, services communication, server-side service discovery, and security in the context of microservices systems.

\textbf{Quality attributes}: A QA is a “\textit{measurable or testable property of a system that is used to indicate how well the system satisfies the needs of its stakeholders}" \cite{bass2003software}. Architecture design of software systems is the first phase of software development, in which QAs can be satisfied. Some empirical studies (e.g., \cite{di2019architecting, alshuqayran2016systematic, pahl2016microservices, LI2020106449}) on MSA have taken QAs into consideration, but the results reported are not from the practitioners' perspective. The ISO/IEC 25010 standard regulates a comprehensive set of QAs of software systems \cite{international2016systems}, but these QAs are considered with different priorities in the design of microservices systems \cite{waseemMSAdevops}. Thus, we decided to investigate the importance of those QAs that are reported in the existing studies (e.g., \cite{marquez2018actual, o2007quality, LI2020106449}) in the context of MSA.

Besides the aspects described above, we also investigated several other aspects, such as application decomposition strategies (e.g., DDD), development processes (e.g., DevOps, Scrum, XP), architectural components for microservices systems, and skills required to design MSA. Also, we identified the MSA design challenges, solutions, as well as the impact of the MSA challenges on monitoring and testing of microservices systems.

\subsection{Monitoring of microservices systems}
A microservices system is primarily based on isolated microservices that can be developed, deployed, and operated independently \cite{fowler2014microservices}. Microservices interact with each other at runtime to exchange information. The runtime architecture of microservices systems is continuously evolved. Thus, information on the system architecture and service interaction and behavior needs to be generated and updated automatically \cite{mayer2017dashboard}. The dynamic nature of microservices systems needs monitoring infrastructure to diagnose and report errors, fault, failure, and performance issues. In this study, we investigated the monitoring metrics, monitoring practices, tools, challenges, and solutions for monitoring microservices systems. 

\textbf{Monitoring metrics} are numerical representations of data measured over intervals of time that generally come from resource usage and system behavior \cite{sridharan2018distributed, DaveSwersky2018monitoring}. Monitoring metrics can be used to predict the system's future behavior. We identified 12 monitoring metrics from grey literature, e.g.,~\cite{DaveSwersky2018monitoring, Susan17, Apurva16} (see Figure~\ref{fig:SQ29}) that are used for infrastructure-level monitoring and microservices monitoring. Specifically, metrics such as resource usage (e.g., CPU, memory), threads, the status of dependencies, and database connections are used for infrastructure-level monitoring. In contrast, language-specific metrics (e.g., garbage collector behaviour, database latency), availability, Service-Level Agreement (SLA), endpoint success, and errors and exceptions are used to monitor microservices~\cite{Susan17}.

\textbf{Monitoring practices} concern with how to monitor microservices systems. We identified six monitoring practices from grey literature (e.g.,~\cite{rajput2018hands, DaveSwersky2018monitoring, Carstensen_2019}), i.e., log management, exception tracking, health check API, log deployment, audit logging, and distributed tracking. Each of these practices consists of several activities. For instance, log management is related to generating, collecting, centralizing, parsing, transmitting, archiving, and disposing log data (e.g., errors, failure) produced by applications servers, network devices, and other components~\cite{Carstensen_2019}. Similarly, exception tracking is a monitoring practice with which exceptions are identified, understood, and resolved with monitoring tools. Health check API is a service implemented within microservices systems to get the operational status and dependencies of microservices. It also returns the performance information (e.g., connection time) of the microservices~\cite{richardson2018microservices}.

\textbf{Monitoring tools} can be classified into two categories: libraries and platforms. Monitoring libraries are used during the development of microservices and permit collecting the application data. In contrast, monitoring platforms allow gathering and analyzing data from different sources, such as the hosts, infrastructure, and microservices~\cite{DaveSwersky2018monitoring}. We identified 13 candidate monitoring tools (see Table \ref{tab:MonitoringofMSA}), which are listed in our survey question, from the literature (e.g., ~\cite{rajput2018hands, waseemMSAdevops, DaveSwersky2018monitoring, Carstensen_2019}).



\subsection{Testing of microservices systems}
Software testing is intended to identify problems and to fix them for improving the quality of software systems \cite{saha2008understanding}. However, the testing of microservices systems requires extra effort due to, for example, the polyglot code base in multiple repositories, feature branches, and database per service \cite{JakeLumetta2018MSATesting}. Some other aspects that bring difficulties in microservices testing are inter-process communication, inter-service dependencies, cloud infrastructure, and third-party services components \cite{waseemtestingMSA, sotomayor2019comparison}. We identified several studies (e.g., \cite{rajput2018hands, JakeLumetta2018MSATesting, rahman2015reusable, heorhiadi2016gremlin, meinke2015learning, clemson2014, DawidAdrian2020}) in which microservices testing was discussed. However, we did not find any study that presents the empirical evidence on testing of microservices systems from the practitioners' perspective. Our previous work also indicates that there is a lack of knowing the industrial reality concerning testing for MSA-based applications \cite{waseemtestingMSA}. Therefore, this industrial survey explores different aspects of testing microservices systems, such as testing strategies and skills required to test microservices systems.

\textbf{Testing strategies}: A testing process for microservices systems includes a set of testing strategies \cite{clemson2014}. These strategies contain functional (e.g., unit testing) and non-functional testing strategies (e.g., performance testing) \cite{DawidAdrian2020}. The aim of using multiple testing strategies in the testing process is to make sure a developed application successfully operates across various environments and platforms. We identified 12 testing strategies (see Table \ref{tab:TestingofMSA}), which are listed in our survey question, from the literature (e.g., \cite{waseemtestingMSA, rajput2018hands, JakeLumetta2018MSATesting, rahman2015reusable, heorhiadi2016gremlin, meinke2015learning, clemson2014, DawidAdrian2020}).

\textbf{Testing tools}: Microservices practitioners may use various testing tools to detect bugs and evaluate applications' behavior, depending on the chosen testing strategy, for instance, tools to perform unit, performance, load, and contract testing. We identified 20 candidate testing tools (see Table \ref{tab:TestingofMSA}), which are listed in our survey question, from the literature (e.g., \cite{waseemtestingMSA, rajput2018hands, JakeLumetta2018MSATesting, rahman2015reusable, heorhiadi2016gremlin, meinke2015learning, clemson2014, DawidAdrian2020}).

\section{Research method} \label{sec:ResearchMethod}
Our research aims to provide a comprehensive insight into how microservices systems are designed, monitored, and tested in practice. Given this objective and inspired by the guidelines for selecting empirical methods for software engineering research \cite{Easterbrook2002Selecting}, we decided to use a mixed-methods study, which collected data from a survey and interviews. We first conducted a survey to reach out to a wide range of practitioners to gain a broad overview of how microservices systems are designed, tested, and monitored. We then carried out in-depth interviews to understand the reasons behind the key findings of the survey. 
We also followed the guideline proposed by Kitchenham and Pfleeger \cite{Kitchenham2008Personal} and used the survey template recommended by evidence-based software engineering \cite{SurveyTemplate}.

\subsection{Research questions} \label{sec:RQs}
We formulated the Research Questions (RQs) in Table \ref{tab:ResearchQuestions} based on the study aim presented in Section \ref{sec:introduction}. The findings from this set of RQs are expected to contribute to the growing body of evidential knowledge about the state of the practice of design, monitoring, and testing of microservices systems.

{\renewcommand{\arraystretch}{1}
\centering
\footnotesize
    \begin{longtable}{|p{5cm}|p{8cm}|}
     \caption{Research questions and their rationale}
    \label{tab:ResearchQuestions}
         \\ \hline
         \textbf{Research Questions} & \textbf{Rationale}\\
         \hline
          \textbf{RQ1}: How microservices systems are designed in the industry? & The answer to this RQ helps to understand various aspects of designing microservices systems from the practitioners' perspective. This \textbf{RQ aims} to investigate the team structure, application decomposition strategies, MSA implementation strategies, architecting activities, description methods, design components, QAs, MSA design patterns, skills required to design and implement MSA, design challenges and their solutions, as well as the reasons for the key findings related to the design of microservices systems.\\\hline

         \textbf{RQ2}: How microservices systems are monitored in the industry? & Monitoring microservices systems is challenging because a large number of independent services are communicating with other independent services in many ways. This \textbf{RQ aims} to investigate microservices monitoring metrics, practices, tools, monitoring challenges and their solutions, as well as the reasons for the key findings related to the monitoring of microservices systems.\\\hline
         
         \textbf{RQ3:} How microservices systems are tested in the industry? & Microservices systems are challenging to test because of several factors, such as polyglot code base, feature branches, database per service. This \textbf{RQ aims} to investigate microservices testing strategies, tools, skills, testing challenges and their solutions, as well as the reasons for the key findings related to the testing of microservices systems.\\\hline

\end{longtable}}

\subsection{Survey}
This section briefly discusses the design of the survey.

\subsubsection{Survey type}
We conducted this survey as a descriptive survey with a cross-sectional design. Cross-sectional design surveys are appropriate for gathering information at a set point of time \cite{Kitchenham2008Personal}. Surveys can be administered in many ways, including online questionnaires, phone surveys, or face-to-face interviews \cite{lethbridge2005studying}. We decided to conduct the survey study through an online questionnaire because (1) this type of survey is time and cost-effective, (2) our potential participants were from different countries, and (3) online questionnaires are convenient for collecting the responses from a large number of participants \cite{lethbridge2005studying}.

\subsubsection{Sample and population}
\label{Samplepopulation}

We did not limit our target population to a specific region or country (e.g., China). We invited the potential practitioners working in the software industry, and whose organizations had adopted or planned to adopt MSA. We asked several questions about the background of the participants (see Table \ref{tab:Demographics}), such as their main responsibilities, number of years working in the IT industry, number of years working with microservices systems, and professional training related to MSA. We used the following approaches to reach the target population.
\begin{enumerate}
    \item \textbf{Personal contacts and authors of industrial track research papers and web blogs}: We sent an invitation email to our personal contacts who work on microservices in the IT industry and the authors of the industrial track research papers (e.g., conference proceedings) and web blogs related to microservices. We also requested the respondents to disseminate our survey to their colleagues who are eligible to answer the questions or share the contacts of relevant practitioners with us.
    \item \textbf{Microservices practitioners on GitHub}: We executed a search string (i.e., “microservice” OR “micro service” OR “micro-service”) in GitHub. We sorted the search results as “best match” to our search string and collected 9,000 publicly available email addresses of MSA-based project contributors. Then we sent an invitation email with a survey link to all practitioners whose email addresses were collected. 
    \item \textbf{Social and professional platforms}: The authors are active members of several microservices groups on LinkedIn, Facebook, and Google, where microservices practitioners from different countries share their issues, experiences, and knowledge. We posted our survey to all the groups in Table \ref{tab:platforms}.
\end{enumerate}

{\renewcommand{\arraystretch}{1}
\centering
\footnotesize
    \begin{longtable}{|l|p{6cm}|p{4cm}|l|}
\caption{Microservices practitioners social and professional platforms used to post our survey}
\label{tab:platforms}\\
\hline
\textbf{\#} & \textbf{Group Name}                                   & \textbf{URL}                             & \textbf{Platform}         \\ \hline
\endfirsthead
\endhead
1           & Microservice   architecture                           & https://tinyurl.com/42pazdhj             & \multirow{5}{*}{LinkedIn} \\ \cline{1-3}
2 & Microservices   meet DevOps in a Docker Container and Kubernetes world & https://tinyurl.com/akwt44jj &  \\ \cline{1-3}
3 & Cloud,   Kubernetes, Docker, microservices, Gitops discussions         & https://tinyurl.com/pfatv6s4 &  \\ \cline{1-3}
4           & .Net   people                                         & https://tinyurl.com/3an82ktz             &                           \\ \cline{1-3}
5           & Software   architects and enterprise architects Group & https://tinyurl.com/2r7y4cnw             &                           \\ \hline
6           & Microservice   architecture                           & https://tinyurl.com/fh2evwwc             & \multirow{5}{*}{Facebook} \\ \cline{1-3}
7           & Microservice   Java Spring                            & https://tinyurl.com/35c4xsf2             &                           \\ \cline{1-3}
8           & Developing   microservices                            & https://tinyurl.com/8nbs9p2f             &                           \\ \cline{1-3}
9           & Spring   boot and microservices developers            & https://tinyurl.com/yuc658tv             &                           \\ \cline{1-3}
10          & Microservices   developer                             & https://tinyurl.com/3tzbmz85             &                           \\ \hline
11          & Microservices   community                             & microservices-community@googlegroups.com & Google   groups           \\ \hline
\end{longtable}
}

\subsubsection{Obtaining a valid sample}
\label{obtainingsample}
To identify valid answers from the collected responses, we applied the inclusion and exclusion criteria presented in Table \ref{tab:inlu_exclu}. We used the criterion I1 to include the participants' responses and evaluated this criterion by the answers of survey questions DQ2 (we asked the participants their major responsibilities in the companies), DQ3 (we asked the participants their usage of MSA for developing applications), and DQ5 (we asked the participants their experience (in years) working with microservices systems). On the other hand, we used the criterion E1 to exclude those responses that were inconsistent, randomly filled, or meaningless. We also used the criterion E2 to exclude the entire response from a respondent if more than 5 SQs have inconsistent, randomly filled, or meaningless answers. This is because practitioners had started our survey but he/she might not have sufficient knowledge or intention to answer all the survey questions. Inconsistent responses mean that the answers to two or more survey questions are conflicting. For example, a participant chose ``\textit{we use MSA style for building all applications in our organization/team}'' to survey question ``\textit{DQ3: Regarding the usage of MSA for developing applications.}'', and he or she also answered “\textit{I have no idea}" to survey question “\textit{SQ12: How do you decompose an application into microservices?}". We also found several answers that were randomly selected. For example, a respondent answered the usage of all the MSA design patterns as “\textit{Very often}", answered all the quality attributes as “\textit{Very important}", and rated the severity of all the challenges as “\textit{Catastrophic}”. Moreover, we considered logically senseless answers to some or all the open-ended survey questions as meaningless. For example, a participant wrote the answer ``\textit{It is easy to get}'' to survey question ``\textit{SQ26: What solutions (e.g., practices, tools, team structures) can be employed to address or mitigate the challenges related to architecting for MSA-based systems?}''.

{\renewcommand{\arraystretch}{1}
\centering
    \footnotesize
    \begin{longtable}{|l|p{7.6cm}|}
    \caption{Inclusion and exclusion criteria for selecting valid responses}
\label{tab:inlu_exclu}
        \\ \hline
         \textbf{Inclusion} & \textbf{I1}: The respondent must have responsibilities associated with MSA-based system development and operation. 
        \\ \hline
         \textbf{Exclusion} & \textbf{E1}: An inconsistent, randomly filled, or meaningless response is excluded.
         
         \textbf{E2}: If more than 5 survey questions (SQs) have inconsistent, randomly filled, or meaningless answers, the whole response is excluded.
        \\ \hline
\end{longtable}
}

\subsubsection{Sampling method}
The non-probabilistic sampling method was used to obtain a valid sample for this survey. We used the non-probabilistic sampling because the target population was specific to microservices practitioners, which were hard to identify through a random selection. As a type of non-probabilistic sampling, we also used convenience and snowball sampling. Convenience sampling ``\textit{refers to acquiring responses from available people who are willing to participate}'' \cite{Kitchenham2008Personal}. The reason for using the convenience sampling method for this survey is that it provides easy access to the target population through personal contacts and social and professional platforms (e.g., GitHub, LinkedIn, Facebook). To maximize the chances of getting more responses, we also used the snowball sampling technique, which asks the participants in a survey to invite other people who have relevant expertise and agree to participate in the survey \cite{Kitchenham2008Personal}.

\subsubsection{Survey instrument preparation}
In this section, we describe the survey instrument and questionnaire format. Through reviewing the grey literature (e.g., \cite{fowler2014microservices, DaveSwersky2018monitoring, clemson2014, JakeLumetta2018MSATesting}), peer-reviewed literature (e.g., \cite{di2019architecting, alshuqayran2016systematic, taibi2018continuous, soldani2018pains, pahl2016microservices}), and conducting SMSs (i.e., \cite{waseemMSAdevops, waseemtestingMSA}) on microservices systems, we developed a questionnaire consisting of 39 survey questions (see Appendix B). The survey questions are organized into the following four groups.
\begin{itemize}
    \item \textit{Demographic information}: We used ten demographics questions (DQ1-DQ10, see Table \ref{tab:Demographics}) to collect comprehensive demographic information about practitioners (e.g., country, responsibilities, experience) and their respective companies (e.g., size, work domain, development methods). We asked for demographic information to determine whether the respondents represent microservices practitioners for results generalization purpose. 
    \item \textit{Design of microservices systems}: We formulated 17 survey questions with the purpose of exploring the practitioners’ perspective on the design of microservices systems (SQ11-SQ27, see Table \ref{tab:DesignofMSA}). The questions are about individuals or teams who are responsible for designing the architecture of microservices systems (SQ11), application decomposition strategies (SQ12), MSA design implementation strategies (SQ13), MSA architecting activities (SQ14-SQ18), MSA description methods (SQ19, SQ20), MSA design components (SQ21), quality attributes (SQ22), and MSA design patterns (SQ23). Moreover, we asked three questions to explore the MSA design challenges (SQ24), the impact of MSA design challenges on monitoring and testing of microservices systems (SQ25), and solutions for the MSA design challenges (SQ26). Finally, the skills required to design and implement MSA are investigated through SQ27.
    \item \textit{Monitoring of microservices systems}: The third part of our survey consists of six questions that aim to explore practitioner’s perspectives on monitoring microservices systems (SQ28-SQ33, see Table \ref{tab:MonitoringofMSA}). We explored the monitoring metrics (SQ28), monitoring practices (SQ29), monitoring tools (SQ30), monitoring challenges (SQ31), the severity of monitoring challenges (SQ32), and the solutions for monitoring challenges (SQ33).
    \item \textit{Testing of microservices systems}: The survey questions (SQ34-SQ39, see Table \ref{tab:TestingofMSA}) are used to solicit the participants' perspectives on the testing strategies (SQ34), tools (SQ35), and challenges (SQ36), the severity of testing challenges (SQ37), skills required to test microservices systems (SQ38), and the solutions for testing challenges (SQ39). 
\end{itemize}

\subsubsection{Questionnaire format}
The survey starts from the Welcome page to explain the purpose and design of the survey and provides our contact information (see Appendix B). Our survey questionnaire has both closed-ended (e.g., SQ11, SQ12, SQ13) and open-ended questions (e.g., SQ26, SQ33, SQ39). The closed-ended questions of our survey are composed of multiple-choice, rating scale, Likert scale, matrix, and drop-down questions. The open-ended questions in our survey allow the participants to provide their solutions for design, monitoring, and testing challenges. We used Google Forms for the implementation of the survey questionnaire.

\subsubsection{Pilot survey} We performed a pilot survey to ensure that (i) the length of the survey is appropriate, (ii) the terms used in the survey are clear and understandable, and (iii) the answers to survey questions are meaningful. We recruited 10 practitioners for the pilot survey through our personal contacts and sending emails to publicly available email addresses of microservices practitioners on GitHub. We got only a few suggestions to refine some adopted terms, and we incorporated them in the final version of the survey instrument.

\subsection{Interviews} 
To get deeper insights into the survey results, we conducted follow-up interviews with microservices practitioners. We present the detailed procedure of the interview process below.

\subsubsection{Participant selection}

We recruited microservices practitioners from IT companies in five countries: Pakistan, Chile, Netherlands, Sweden, and Colombia. Interviewees were recruited by sending an email to some of the survey's original respondents and personal contacts in each country. Since the interviews were expected to last about 40 to 55 minutes in order to ensure a sufficient time allocation to each question, we decided to focus on gaining a deep insight into the key findings obtained from the survey. We informed the participants that this interview is entirely voluntary with no compensation. We finally recruited 6 interviewees and asked them to provide their demographics information (see Table \ref{tab:IQDMinfo}), and the information about the interviewees is presented in Table \ref{tb:Interviewees}. 

{\renewcommand{\arraystretch}{1}
\footnotesize
\begin{longtable}{|l|l|l|c|c|l|}
\caption{Interviewees and their information}
\label{tb:Interviewees}\\
\hline
\multirow{2}{*}{\textbf{\#}} &
  \multirow{2}{*}{\textbf{Responsibilities}} &
  \multirow{2}{*}{\textbf{Domain}} &
  \multicolumn{2}{c|}{\textbf{Experience in}} &
  \multirow{2}{*}{\textbf{Country}} \\ \cline{4-5}
   &                               &           & \textbf{Microservices} & \textbf{IT Industry} &             \\ \hline
\endfirsthead
\endhead
P1 & Design and development        & Telecom   & 6 years             & 17 years                & Pakistan    \\ \hline
P2 & Design and development        & E-commerce & 5 years             & 15 years                & Sweden      \\ \hline
P3 & Design, development, and maintenance & Banking & 4 years              & 4 years                & Netherlands \\ \hline
P4 & Design and development        & Telecom   & 3 years              & 9 years                & Pakistan    \\ \hline
P5 & Development                   & Telecom      & 2 years              & 5 Years                 & Colombia       \\ \hline
P6 & Design and development        & Telecom      & 1 year              & 20 Years                 & Chile       \\ \hline
\end{longtable}
}

The roles of the interviewees are mainly related to the design and development of microservices systems. However, they also actively participated in other microservices system development stages, such as testing, monitoring, and deployment. Their average experience in the IT industry is 11.7 years (minimum 4 years and maximum 20 years). The interviewees' average experience on microservices is 3.5 years (minimum 1 year and maximum 6 years). We refer to the interviewees as P1 to P6.

\subsubsection{Interviewing process} The first, fourth, and fifth authors of the study conducted the interviews. The average time for all interviews is 53 minutes. The interviews were conducted in Urdu, Italian, and Spanish, and later were translated into English. The interviews are semi-structured and divided into four sections.
\begin{itemize}
    \item \textit{Demographic information}: We asked five questions regarding interviewees' demographics, such as experience in the IT industry and microservices, responsibilities, and organization domains.
    \item \textit{Design}: We asked 13 open-ended questions to get a deeper understanding of seven key findings concerning microservices system design. The questions include: (i) confirmation of results about the combination of DDD and business capability, (ii) reasons for using a combination of DDD and business capability strategy, (iii) decomposition strategies other than DDD and business capability, iv) reasons for creating the only high-level design for microservices systems, (v) possible risks if practitioners do not create a design or detailed design for microservices systems, (vi) reasons for the frequent use of Architecture Evaluation (AE) activity, (vii) reasons for paying more attention to architecting activities close to the solution space instead of the problem space, (viii) reasons for not using ADLs and DSLs, (ix) reasons for using informal and semi-formal languages, (x) reasons for considering security, availability, performance, and scalability as most important, (xi) reasons for using API gateway and Backend for frontend, Database per service, and Access token patterns as most often, (xii) reasons for the most prominent challenges of designing microservices systems, and (xiii) solutions for the most prominent design challenges.
    \item \textit{Monitoring}: We asked 5 open-ended questions to get a deeper understanding of four key findings from the survey regarding microservices system monitoring. The questions include: (i) reasons for frequent use of the monitoring metrics, like resource usage, load balancing, and availability, (ii) reasons for frequent use of the monitoring practices, like log management, exception tracking, and health check APIs, (iii) opinions about using Jira for monitoring purpose, (iv) reasons for the most prominent challenges of monitoring microservices systems, and (v) solutions for the most prominent monitoring challenges.
    \item\textit{Testing}: The interviewees were asked four questions to elaborate on the key findings obtained from the survey regarding microservices system testing. The questions include: (i) reasons for frequent use of the testing strategies, like unit testing, integration testing, and E2E testing, (ii) reasons for frequent use of the testing tools, like Junit, JMeter, and Mocha to test microservices systems, (iii) reasons for the most prominent challenges of testing microservices systems, and (v) solutions for the most prominent testing challenges.
\end{itemize}

\subsection{Data analysis}

\subsubsection{Survey data analysis}
\label{SurveyDataanAlysis}
We used two techniques for analyzing the obtained data from survey participants. First, a descriptive statistic technique was used to analyze the closed-ended questions' responses (e.g., SQ11, SQ12, SQ13). Second, we used open coding and constant comparison techniques from Grounded Theory (GT) \cite{glaser2017discovery, hoda2017becoming} to analyze the qualitative response to the open-ended questions (e.g., SQ26, SQ33, SQ39). The relationship between survey questions, Interview Questions (IQs), data analysis methods, and RQs are shown in Table \ref{tab:Questions_AnalysisMethods}.

\subsubsection{Interview data analysis}

The main goal of this step is to identify the reasons obtained from the interviews for the key findings of the survey results. The interviews were translated and transcribed by the first, fourth, and fifth authors from audio recordings to text transcripts. Before applying the open coding and constant comparison techniques, the second and third authors read the transcribed interviews and highlighted the irrelevant information. We dropped several sentences that are not related to “design, monitoring, and testing of microservices systems”. After removing the extraneous information from the transcribed interviews, the first author read and coded the interview transcripts’ contents to get the answers to the interview questions. The second and third authors of the study verified the coding results and provided suggestions to improve the quality of coding results. We used the MAXQDA tool\footnote{\url{https://www.maxqda.com/}} for analyzing and coding the qualitative data.

{\renewcommand{\arraystretch}{1}
\centering
\footnotesize
    \begin{longtable}{|l|l|l|}
     \caption{Relationship between questions, data analysis methods, and research questions}
    \label{tab:Questions_AnalysisMethods}
         \\ \hline
         \textbf{RQs/Section} & \textbf{Survey and interview questions}&\textbf{Data Analysis Method} \\
         \hline
          Demographics & DQ1-DQ10 & Descriptive Statistics \\
         \hline
         RQ1 & SQ11-SQ25 & Descriptive Statistics \\
         \hline
         RQ2& SQ28-SQ32 & Descriptive Statistics \\
         \hline
         RQ3 &SQ34-SQ38 & Descriptive Statistics \\
         \hline
         RQ1 & SQ27 & Open Coding and Constant Comparison \\ \hline
         RQ2 &SQ33 & Open Coding and Constant Comparison \\
         \hline
         RQ3 &SQ39 & Open Coding and Constant Comparison \\
         \hline
         Demographics& IQ1.1-IQ1.5 & Descriptive Statistics \\\hline
         RQ1& IQ2.1-IQ2.7 & Open Coding and Constant Comparison \\\hline
         RQ2& IQ3.1-IQ3.4 & Open Coding and Constant Comparison \\\hline
         RQ3& IQ4.1-IQ4.3 & Open Coding and Constant Comparison \\\hline
\end{longtable}}

\subsubsection{Result comparison} To better understand practitioners' perspective on design, monitoring, and testing of microservices systems, we divided the responses about Likert scale survey questions (i.e., SQ14-SQ18, SQ19, SQ22, SQ23, SQ29, SQ32, SQ34, SQ37) into three major demographic groups (i.e., Experienced-based comparison, Organization size-based comparison, MSA style-based comparison).

\begin{itemize} 
    \item Participants who have $\le$ 2 years of experience with microservices systems (63 responses)
    \item Participants who have \textgreater{} 2 years of experience with microservices systems (43 responses)
    \item Participants whose organization or team does not use the MSA style at all (15 responses)
    \item Participants whose organization or team uses the MSA style for building some or all applications (91 responses)
    \item Participants who are working in a relatively small and medium organization ($\le$ 100 employees) (51 responses)
    \item Participants who are working in a relatively large organization (\textgreater{} 100 employees) (55 responses)
\end{itemize}

We computed the statistically significant difference (i.e., P-value) for the three groups. We summarized the survey responses to Likert scale survey question statements, SQ\#, mean values, P-values, and effect size in Table \ref{tab:SigDiffArchDscrip}, Table \ref{tab:SigDiffQAs}, Table \ref{tab:SigDifPatterns}, Table \ref{tab:SigDifMonitorPractices}, Table \ref{tab:SigDifMonitorChallenges}, Table \ref{tab:SigDiffTestingStrategies},  and Table \ref{tab:SigDiffTestingchallenges}. The detail about the gaps between the different demographic groups is reported in Section \ref{GapsMSAdesign}, Section \ref{GapMSAmonitoring}, and Section \ref{GapsMSAtesting}.
In this survey, we used five points Likert scale for obtaining levels of Agreement (e.g., strongly agree, agree), Frequency (e.g., very often, often), Importance (e.g., very important, important), and Severity (e.g., catastrophic, major) from the practitioners about several aspects of microservices systems (see Likert scale survey questions in Table \ref{tab:DesignofMSA}, Table \ref{tab:MonitoringofMSA}, and Table \ref{tab:TestingofMSA}). We analyzed each Likert scale level to corresponding survey question option to identify the statistically significant differences between the two chosen groups (e.g., experience $\le$ 2 years vs. experience \textgreater{} 2 years) at 95\% confidence level. We first use the independent sample T-test by assuming equal variances to test the statistically significance difference between two group means (e.g., experience $\le$ 2 years vs. experience \textgreater{} 2 years), and then we applied Bonferonni correction for the multiple comparisons on the means of survey question statements (e.g., see Table \ref{tab:SigDiffArchDscrip}). We also calculated the effect size \cite{nakagawa2007effect} to quantify the difference between two groups. We used different colors to indicate participants from which groups have more responses about options to survey questions. The responses to survey questions are sorted according to the number of responses. The dark grey color shows that the first group is more likely to agree with survey question options, and the light grey color shows that the second group is more likely to agree with survey question options. We use the following symbols to show the statistically significant difference of the chosen groups.

\begin{itemize}
    \item \faWrench{} symbol indicates a significant difference between participant groups who have \textbf{experience $\le$ 2 years} vs. \textbf{experience \textgreater{} 2 years}.
    \item \faGears{} symbol indicates a significant difference between participant groups who use the \textbf{MSA style} vs. \textbf{No MSA style}.
    \item \faGroup{} symbol indicates a significant difference between participant groups who are from an organization that has \textbf{Employees size $\le$ 100} vs. \textbf{Employees size \textgreater{} 100}.
\end{itemize}

\section{Results} 
\label{sec:SurveyResults}
In this section, we provide the our mixed-methods study results according to the collected responses and answer the RQs defined in Section \ref{sec:RQs}.

\subsection{Demographics}\label{sub_sec:demographics}

\begin{figure*}[hbt!]
  \centering
  \includegraphics[scale=0.55]{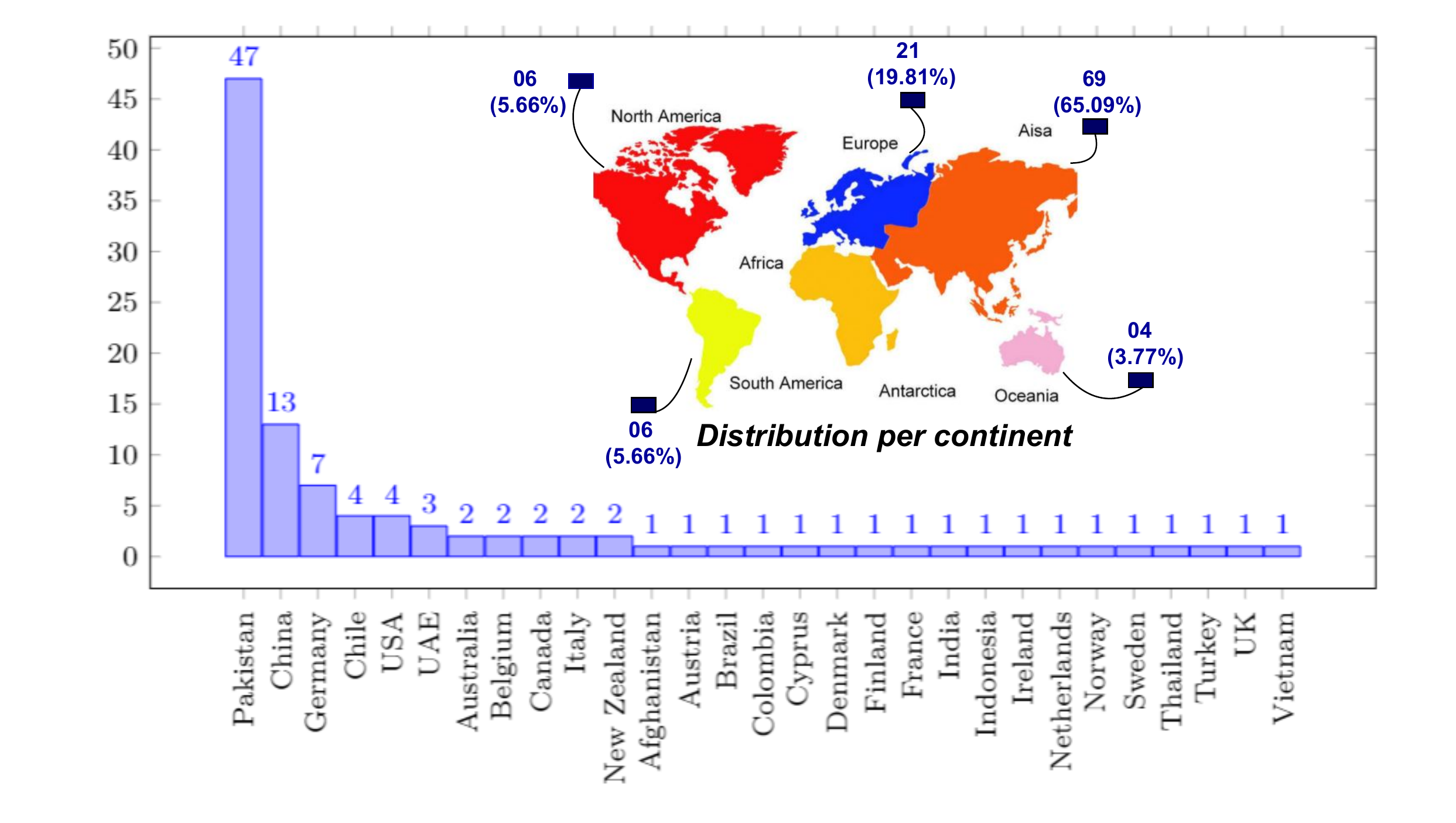}
  \caption{The number of respondents from each country}
  \label{fig:countries}
\end{figure*}

\textbf{Countries}: We received 106 responses to the survey (indicated by R1 to R106), interviewed 6 microservices practitioners (indicated by P1 to P6) from 6 organizations in 6 countries (see Table \ref{tb:Interviewees}). Respondents came from 29 countries of 5 continents (see Figure \ref{fig:countries}) working in diverse teams and roles to develop microservices systems. The majority of them are from Pakistan (47 out of 106, 63.2\%), China (13 out of 106, 12.2\%), and Germany (7 out of 106, 6.6\%), and a close look of the results shows that the respondents from other countries tend to answer similarly to the respondents from the top three countries (i.e., Pakistan, China, Germany).

\begin{figure}[hbt!]
\begin{centering}
\begin{tikzpicture}  
\begin{axis}[
footnotesize,
	xbar, 
	width=11.5cm, height=6.5cm, 
	enlarge y limits=0.02,
	enlargelimits=0.07,  
	symbolic y coords={Others, Business analyst, Operations staff, Database administrator, C-level executive, Software quality engineer, Consultant, System analyst, Database developer, DevOps engineer, Architect, Application developer},
  	ytick=data,
	nodes near coords, nodes near coords align={horizontal},
	every node near coord/.append style={font=\footnotesize},
    xlabel={\textbf{Num of participants}},
]
	\addplot coordinates {(69,Application developer) (45,Architect) (21,DevOps engineer) (21,Database developer) (17,System analyst) (11,Consultant) (11,Software quality engineer) (9,C-level executive) (6,Database administrator) (4,Business analyst) (4,Operations staff) (4,Others)};
\end{axis}
\end{tikzpicture}  
\caption{Survey participants' responsibilities in their companies}
\label{fig:roles}
\end{centering}
\end{figure}
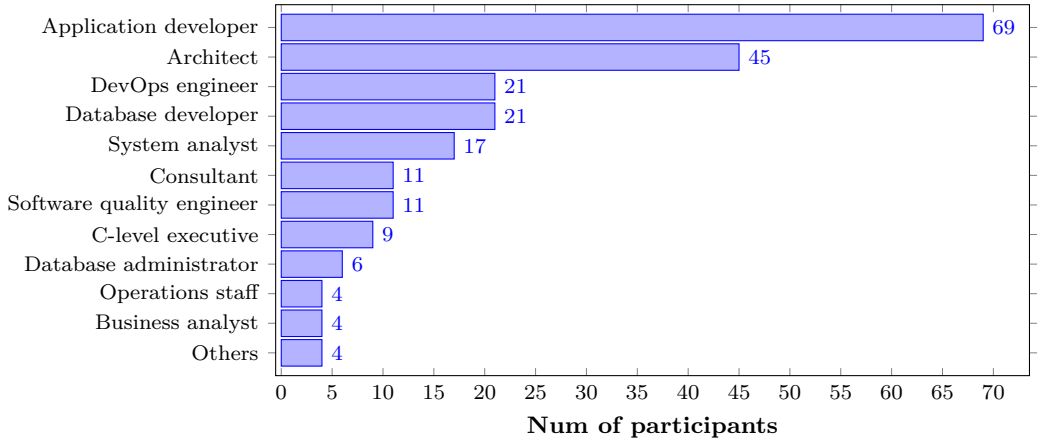

\textbf{Responsibilities}: Figure \ref{fig:roles} shows that the majority of participants were application developer (69 out of 106, 65.1\%), architect (45 out of 106, 42.4\%), and DevOps engineer (21 out of 106, 19.8\%). Note that one participant may have multiple major responsibilities in the company (DQ2), and consequently the sum of the percentages exceeds 100\%. 

\begin{figure*}[hbt!]
  \centering
  \includegraphics[scale=0.55]{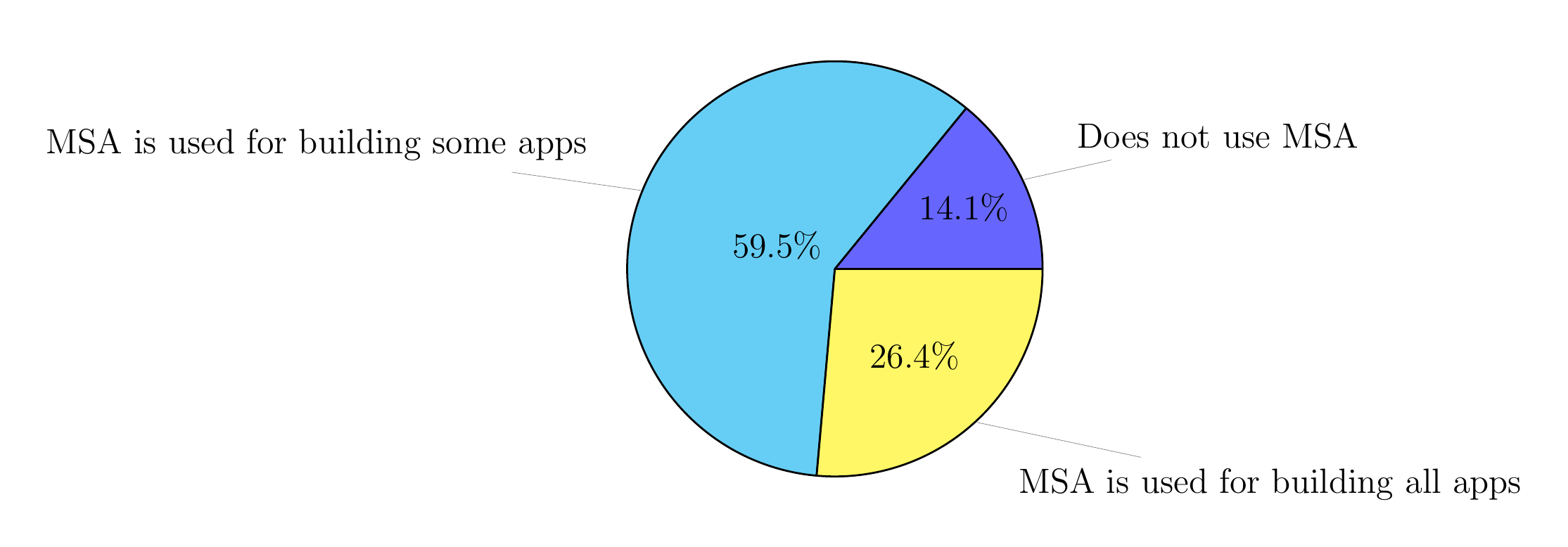}
  \caption{Status of MSA adoption across organization}
  \label{fig:SQ3}
\end{figure*}

\textbf{MSA adoption across organization}: As shown in Figure \ref{fig:SQ3}, 59.5\% (63 out of 106) of the respondents said that they only used MSA to develop specific applications in their respective organizations. 26.4\% (28 out of 106) of the respondents answered that they used MSA to develop all the applications in their organizations. Overall, almost 86\% of the respondents are from organizations that have adopted MSA to develop applications. Note that although 14.1\% (15 out of 106) of the respondents answered that their organizations did not use MSA to develop applications, these respondents still have experience and expertise with MSA-based systems, which is one of the criteria to answer this survey (see the Welcome page of the survey in Appendix B).

\begin{figure*}[hbt!]
  \centering
  \includegraphics[scale=0.55]{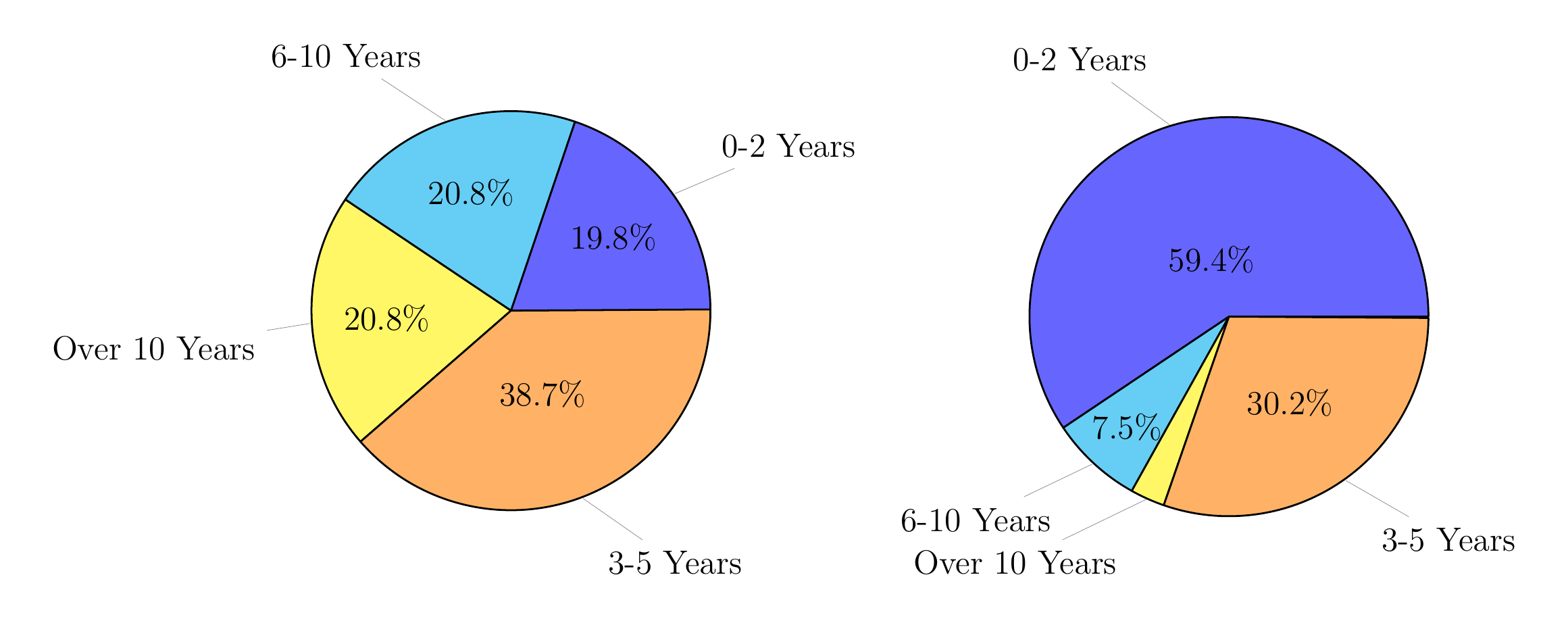}
  \caption{Experience of the respondents in the IT industry (\textbf{Left}) and the development of microservices systems (\textbf{Right})}
  \label{fig:SQ4-SQ5}
\end{figure*}

\textbf{Experience}: We asked the participants about their experiences in the IT industry (DQ4) and in the development of microservices systems (DQ5). Figure \ref{fig:SQ4-SQ5} (Left) shows that the majority of respondents (41 out of 106, 38.7\%) have worked in the IT industry for 3 to 5 years. On the other hand, Figure \ref{fig:SQ4-SQ5} (Right) shows that over half of the respondents (63 out of 106, 59.4\%) had little experience (from 0 to 2 years) in the development of microservices systems. We also received three responses in which the participants mentioned that they had more than 10 years of experience working on microservices. However, considering that the term “microservices” was first coined in 2011 \cite{comunytek}, which is around 9 years ago. We accept these results because (1) responses to all other survey questions from these three participants were valid, (2) the overall working experience of these participants in the IT industry was more than ten years and they were working as “C-level executives", and (3) we also found the blogs related to microservices (e.g., \cite{Jan2020}) in which practitioners claim \say{\textit{microservices have a very long history, not as short as many believe}}. For example, the similar term  “micro web services" was presented in 2005 \cite {Laura17}. 

\begin{figure*}[!htbp]
  \centering
  \includegraphics[scale=0.55]{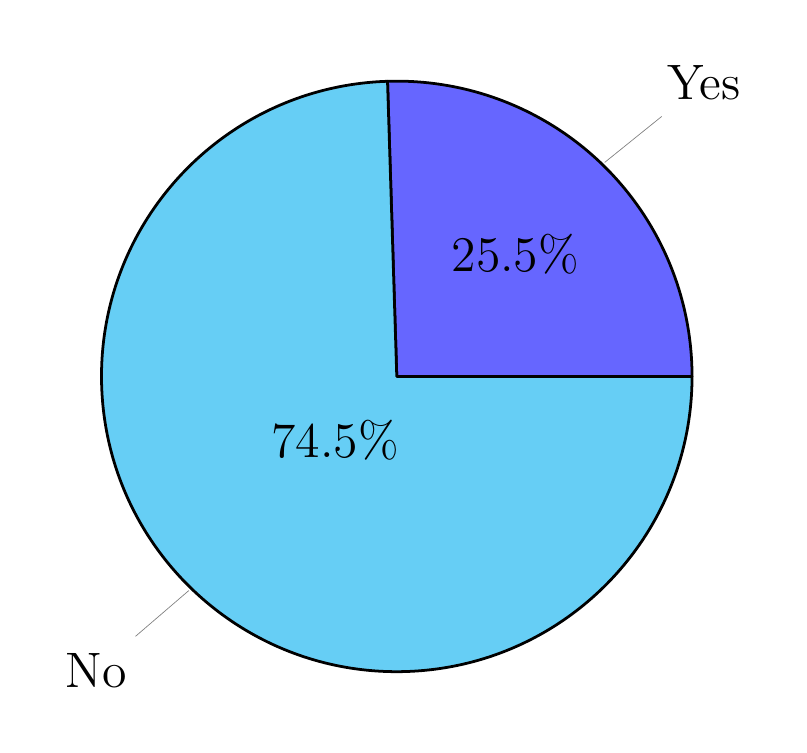}
  \caption{Participants professional training related to MSA}
  \label{fig:SQ6}
\end{figure*}

\textbf{Professional training}: Figure \ref{fig:SQ6} shows that 74.5\% (79 out of 106) of the respondents did not receive any professional training to develop microservices systems. Only 25.5\% (27 out of 106) of the respondents answered that they had such training. Note that we explicitly mentioned in DQ6 that professional training does not mean higher education (e.g., MSA course at universities).

\begin{figure*}[!htbp]
  \centering
  \includegraphics[scale=0.5]{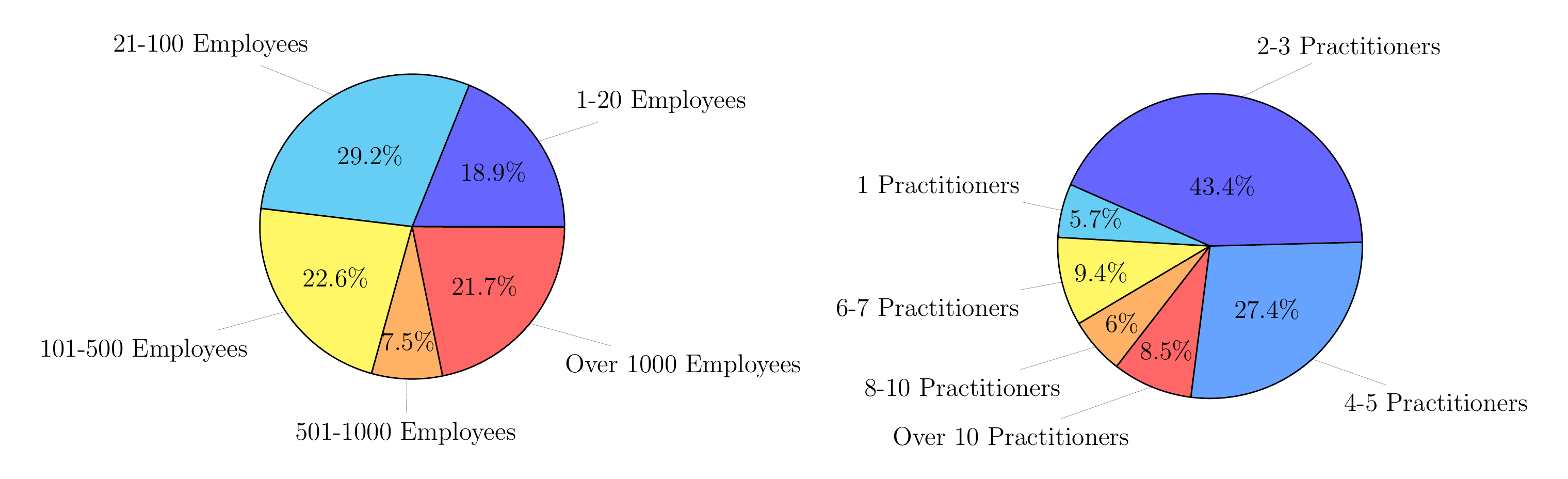}
  \caption{An overview on organization size (\textbf{Left}) and number of people working on one microservice (\textbf{Right})}
  \label{fig:SQ8-SQ9}
\end{figure*}

\textbf{Organization size and the number of people working on microservices}: The size of organizations varies from small (1-20 employees) to large (more than 1000 employees) (see Figure \ref{fig:SQ8-SQ9} (Left)). We found that 29.2\% (31 out of 106) of the respondents were from medium scale organizations (21-100 employees) and 21.7\% (23 out of 106) of the respondents were from large organizations.
Similarly, Figure \ref{fig:SQ8-SQ9} (Right) shows the number of persons working on one microservice.
43.4\% (46 out of 106) of the respondents stated that they were working in a group of 2 to 3 people, and 27.4\% (29 out of 106) answered that they were working in a group of 4 to 5 people.

\textbf{Application domains and development methods}: Figure \ref{fig:SQ7-SQ10} (Left) shows the domains of the participants' organizations where the survey participants worked. “E-commerce" (35 out of 106, 33.1\%), “Professional services" (24 out of 106, 22.6\%), and “Financial software" (23 out of 106, 21.6\%) are the most popular domains.  We also found that Scrum (63 out of 106, 59.4\%) and DevOps (53 out of 106, 50.0\%) are the two widely used development methods in the participants' organizations to develop microservices systems (see Figure \ref{fig:SQ7-SQ10} (Right)). Other popular development methods are iterative development and Rapid Application Development (RAD), which were mentioned by 25 and 14 participants, respectively. We also received 12 responses in which the participants indicated that they did not employ any development method. Note that one organization may adopt one or more development methods in their MSA projects.


\begin{figure}[H]
\begin{centering}
\footnotesize
\begin{tikzpicture}
\begin{axis}[
	xbar, 
	width=6.0cm, height=7.0cm, 
	enlarge y limits=0.02,
	enlargelimits=0.07,  
	symbolic y coords={Others, Real estate, Embedded systems, Telecommunication, Manufacturing, Insurance, Transportation, Education, Internet, Healthcare, Financial, Professional services, E-commerce},
	ytick=data,
	xmin=0,
	xtick distance=5,
	xmax=35,
	xlabel={\textbf{Num of participants}},
	nodes near coords, nodes near coords align={horizontal},
	every node near coord/.append 
]
\addplot coordinates {(35,E-commerce) (24,Professional services) (23,Financial) (21,Healthcare) (19,Internet) (17,Education) (11,Insurance) (11,Transportation) (9,Manufacturing) (7,Telecommunication) (6,Embedded systems) (4,Real estate) (4,Others)};
\end{axis}
\end{tikzpicture}  
\footnotesize
\begin{tikzpicture}
\begin{axis}[
	xbar, 
	width=6.0cm, height=6.9cm, 
	enlarge y limits=0.02,
	enlargelimits=0.07,  
	symbolic y coords={Not defined, Others, JAD, Lean, FDD, Spiral, Waterfall, XP, RAD, Iterative, DevOps, Scrum},
	ytick=data,
	ytick=data,
	xmin=0,
	xtick distance=10,
	xmax=70,
	xlabel={\textbf{Num of participants}},
	nodes near coords, nodes near coords align={horizontal},
	every node near coord/.append 
	]
\addplot coordinates {(63,Scrum) (53,DevOps) (16,Iterative) (14,RAD) (9,XP) (8,Waterfall) (7,FDD) (7,Spiral) (6,Lean) (4,JAD) (3,Others) (12,Not defined)};
\end{axis}
\end{tikzpicture}  
\caption{An overview on the domains of the organizations (\textbf{Left}) and the software development methods employed (\textbf{Right})}
\label{fig:SQ7-SQ10}
\end{centering}
\end{figure}
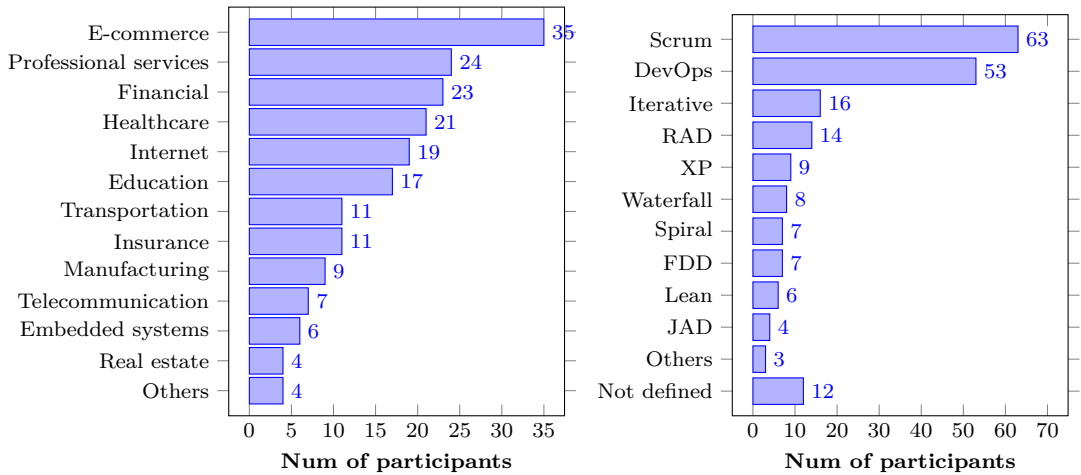
\begin{tcolorbox}[colback=gray!5!white,colframe=gray!75!black,title=Demographics]
\justify
\textbf{Finding 1.} The majority of participants did not receive any training to develop microservices systems.\\
\textbf{Finding 2.} The participants indicated that a team of 2-5 people often works on (develop, test, and deploy) each microservice.\\
\textbf{Finding 3.} Scrum and DevOps are the two most widely used development methods in MSA projects.
\end{tcolorbox}
\subsection{Design of microservices systems (RQ1)}
To realize how microservices systems are designed in the industry (RQ1), we asked the respondents to answer 17 survey questions (SQ11 to SQ27) (see Table \ref{tab:DesignofMSA} in Appendix B).

\subsubsection{Team structure, application decomposition, and implementation strategies}
We asked the survey respondents about whether a dedicated individual or team is responsible for creating and managing the architecture of microservices systems (SQ11) in their organizations. 36.7\% (39 out of 106) of the respondents said that their organizations do not have a dedicated person or team. We further investigated our results and found that most of these respondents are from small companies or small teams that possibly cannot afford a software architect. In contrast, 31.1\% (33 out of 106) answered that their organizations have such a team or person. Moreover, 32.1\% (34 out of 106) of the respondents shared that they have a dedicated team or person designing MSA, who has other responsibilities as well. 

\begin{figure*}[!ht]
  \centering
  \includegraphics[scale=0.55]{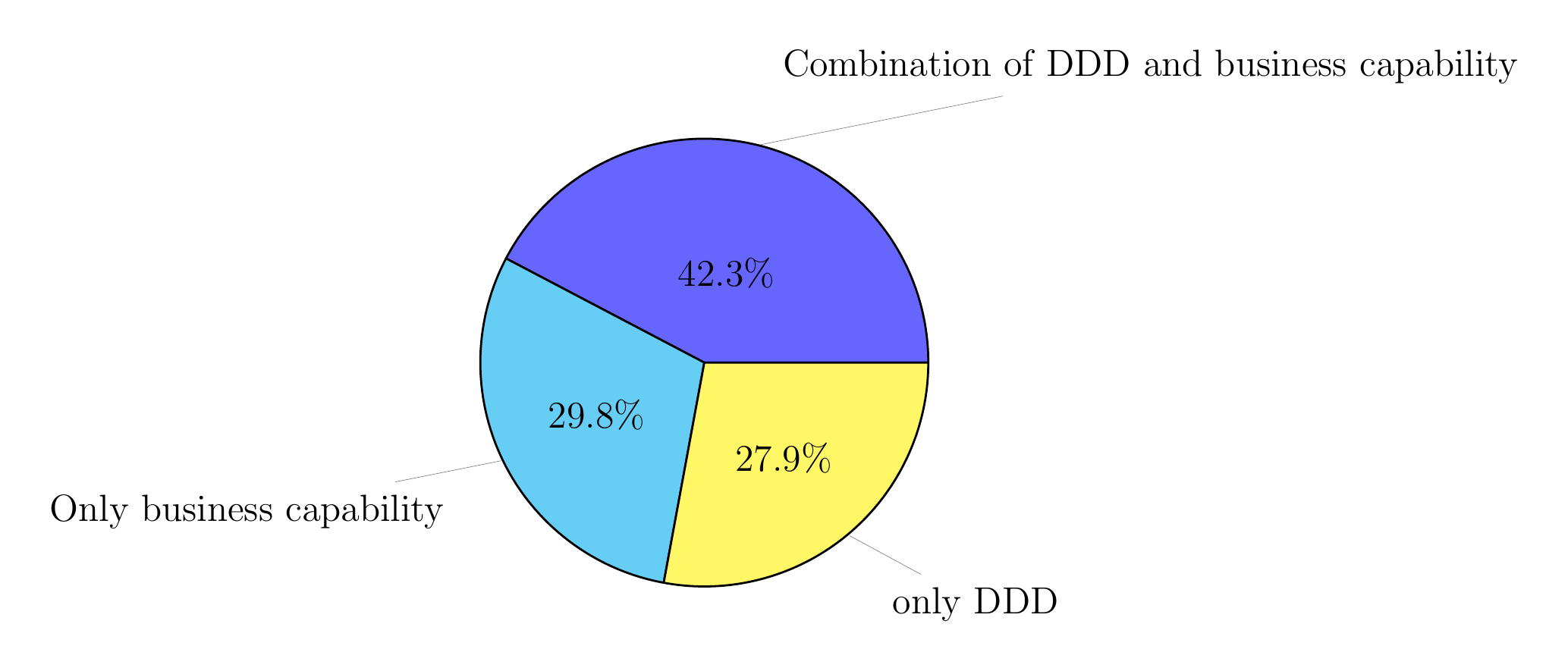}
  \caption{Strategies of application decomposition into microservices}
  \label{fig:SQ12}
\end{figure*}

We were also interested in understanding how an application is decomposed into microservices in the industry (SQ12) (see Figure \ref{fig:SQ12}). The number of valid responses to this survey question is 104 because two responses are meaningless. 42.3\% (44 out of 104) of the respondents answered that they used a combination of business capability and DDD to decompose an application into microservices. 29.8\% (31 out of 104) of the respondents decomposed an application into microservices by applying business capability. 27.9\% (29 out of 104) of the respondents used the DDD strategy for this purpose. For example, one respondent answered that ``\textit{we decompose an application into microservices by using the Banking Industry Architecture Network (BIAN)\footnote{\url{https://bian.org/}} domain model, i.e., the DDD strategy}'' \textbf{\textit{Architect (R14)}}.

On the other hand, all of the interviewees (IQ2.1.1 in Table \ref{tab:IQMsaDesign}) confirmed that the results of the survey reflect the actual use of application decomposition strategies (e.g., a combination of DDD and business capability) for microservices systems in the industry. Five out of six interviewees also confirmed that they are not familiar with any other approach that can be used for application decomposition. Four interviewees (i.e., P2, P3, P4, and P6) mentioned several reasons for using a combination of DDD and business capability, such as (i) a single strategy cannot meet the application decomposition goal in a good way, (ii) it supports the creative and iterative collaboration process between system stakeholders (e.g., technical expert, domain expert, and customers) to achieve the decomposition goals, (iii) it is suitable when implementing microservices in an agile or DevOps process, (iv) it provides the controlled flexibility for future changes as most of the changes remain around the particular domain, (v) it reduces the misunderstanding because requirements are gathered from a specific domain perspective, and (vi) it provides better team communication because all of the stakeholders use a shared set of the terminologies. Following is one representative quotation about the application decomposition strategies for microservices systems.


\faComment “\textit{Combination of DDD and business capability, helps to build microservices with clear responsibilities and thus increase the maintainability and evolvability of microservices systems}” \textbf{\textit{Architect and Application developer (P1)}}.

\begin{figure*}[hbt!]
  \centering
  \includegraphics[scale=0.5]{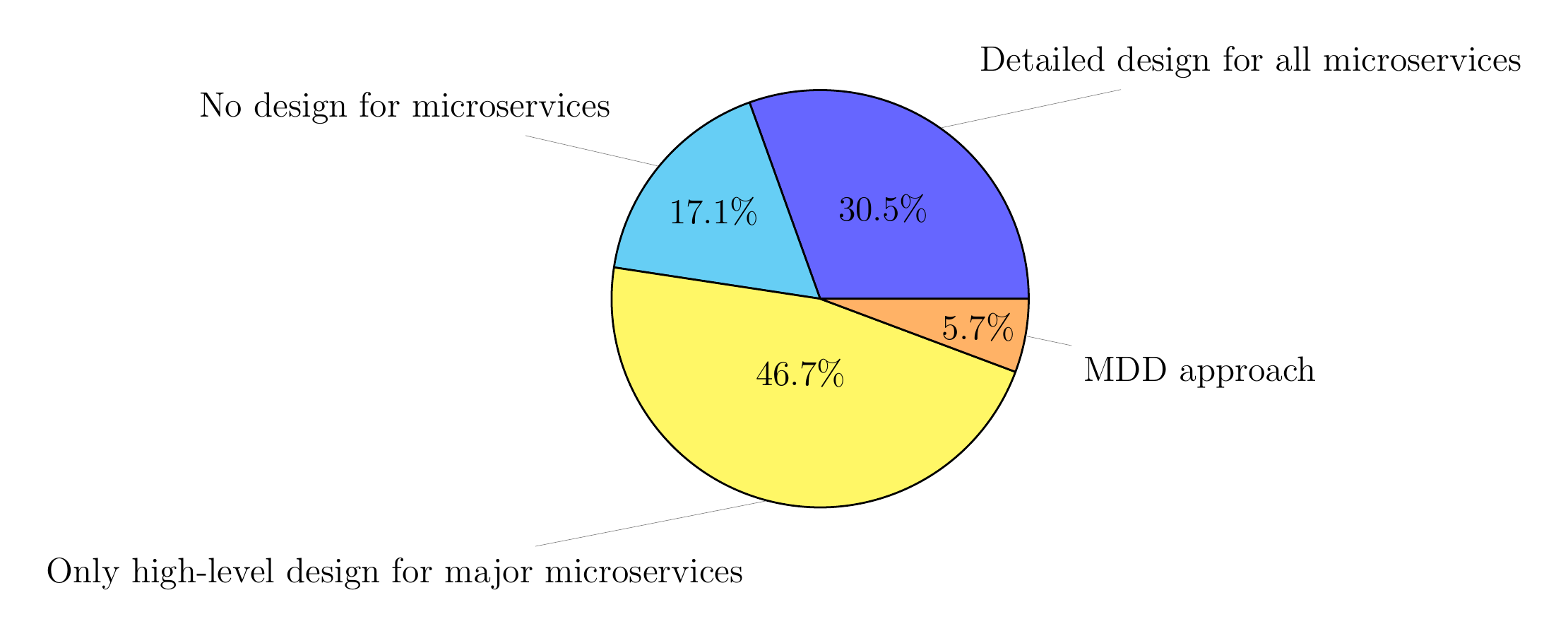}
  \caption{Strategies from design to implementation of microservices systems}
\label{fig:SQ13}
\end{figure*}

Figure \ref{fig:SQ13} shows the results of SQ13, in which the practitioners were asked to select the strategies to design and develop microservices systems. One response to SQ13 is invalid as the participant answered that “\textit{I have no idea}”, therefore, the number of valid responses is 105. From Figure \ref{fig:SQ13}, it shows that 46.7\% (49 out of 105) of the participants only modeled the high-level design (the major components) and 30.5\% (32 out of 105) of the participants modeled the detail design before writing the code. In the context of this finding, our interview demonstrates that practitioners have stated several reasons for using high-level design for microservices systems. For instance, four interviewees (i.e., P2, P4, P5, and P6) mentioned that high-level design is enough to (i) visualize the identified microservices and their major functionalities, (ii) show communication and orchestration style between microservices, (iii) indicate scalability points for each microservice, (vi) identify vulnerable and insecure points in microservices systems, and vii) present the boundaries of microservices. Another interviewee (i.e., P3) mentioned that the reasons for only using high-level design are (i) time pressure, which does not allow teams to spend time on low-level design, and (ii) having an experienced development team who does not need low-level design. Following is one representative quotation about the use of only high-level design for microservices systems.

\faComment “\textit{We always make the high-level design for the whole system (end to end) because it is enough to visualize the identified microservices and their fundamental functionalities, communication or orchestration style between microservices, and the boundaries of microservices.} \textbf{\textit{Architect and Application developer (P1)}}.

Moreover, a small percentage (6 out of 105, 5.7\%) of the practitioners applied the Model-Driven Development (MDD) approach for generating the code of microservices systems. Surprisingly, 17.1\% (18 out of 105) of the participants reported that they wrote the code for microservices systems without having either a high-level or detailed design.

We summarize some of the prominent risks from interviewees’ answers that many organizations may face the situation that there is no design or detailed design for microservices systems. The potential risks are (i) teams cannot correctly estimate the amount of work (e.g., person-hour) and time required to complete projects, (ii) implementing poor security standards and solutions, (iii) having no clear or shared vision for teams, (iv) flawed integration plan for microservices, (v) understandability issues for novice developers, (vi) system migration and evolution issues, and (vii) poor communication among the system stakeholders. Two interviewees (i.e., P1, P2) explicitly mentioned that

\faComment “\textit{Without a design of microservices systems, a system could become a nightmare for developers. Several kinds of risks can affect the microservices system as a whole, including poor work and time management, no clear and common vision about security standards, integration of microservices, and understandability of novice developers}" \textit{\textbf{Architect and Application developer (P3)}}.

\begin{tcolorbox}[colback=gray!5!white,colframe=gray!75!black,title=Key Findings of RQ1]
\justify
  \textbf{Finding 4.} Most of the practitioners used a combination of Domain-Driven Design (DDD) and business capabilities to break a monolith application into microservices.\\
  \textbf{Finding 5.} Most of the practitioners only created a high-level design for microservices systems before implementing them.
\end{tcolorbox}

\subsubsection{MSA architecting activities and design description}

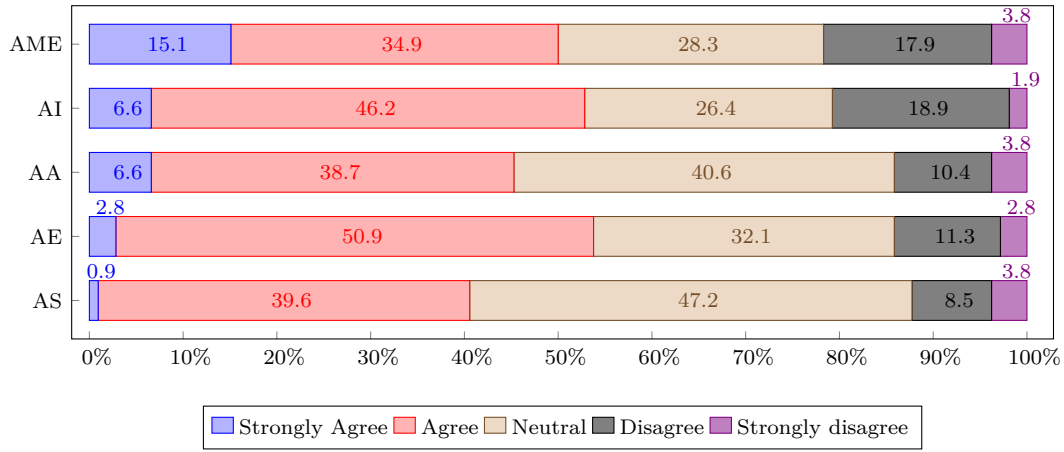
\begin{figure}[H]
\begin{centering}

\begin{tikzpicture}
  \begin{axis}[
     footnotesize,
      xbar stacked,
      width=14.5cm, height=6cm, 
      bar width=15pt,
      nodes near coords={
        \pgfkeys{/pgf/fpu=true}
        \pgfmathparse{\pgfplotspointmeta / 106 * 100}
        $\pgfmathprintnumber[fixed, precision=1]{\pgfmathresult}$
        \pgfkeys{/pgf/fpu=false}
      },
      nodes near coords custom/.style={
        every node near coord/.style={
          check for small/.code={
            \pgfkeys{/pgf/fpu=true}
            \pgfmathparse{\pgfplotspointmeta<#1}\%
            \pgfkeys{/pgf/fpu=false}
            \ifpgfmathfloatcomparison
              \pgfkeysalso{above=.5em}
            \fi
          },
          check for small,
        },
      },
      nodes near coords custom=6,
      xmin=-2, xmax=109,
      xtick={0, 10.6, ..., 106.1},
      ytick={1,...,5},
       yticklabels={AS, AE, AA, AI, AME},
      xtick pos=bottom,
      ytick pos=left,
      xticklabel={
        \pgfkeys{/pgf/fpu=true}
        \pgfmathparse{\tick / 106 * 100}
        $\pgfmathprintnumber[fixed, precision=1]{\pgfmathresult}\%$
        \pgfkeys{/pgf/fpu=false}
      },
      enlarge y limits=.15,
      legend style={at={(0.5,-0.20)}, anchor=north, legend columns=-1},
    	every node near coord/.append style={font=\footnotesize},
    ]
    
\addplot coordinates {(1,1) (3,2) (7,3) (7,4) (16,5)};
\addplot coordinates {(42,1) (54,2) (41,3) (49,4) (37,5)};
\addplot coordinates {(50,1) (34,2) (43,3) (28,4) (30,5)};
\addplot coordinates {(9,1) (12,2) (11,3) (20,4) (19,5)};
\addplot [color=violet, fill=violet!50] coordinates {(4,1) (3,2) (4,3) (2,4) (4,5)}; 

\legend{\strut Strongly Agree, \strut Agree, \strut Neutral, \strut Disagree, \strut Strongly disagree}
  \end{axis}
\end{tikzpicture}
\caption{MSA architecting activities used in industry (AA-Architectural Analysis, AS-Architectural Synthesis, AE-Architectural Evaluation, AI-Architectural Implementation, AME-Architectural Maintenance and Evolution}
\label{fig:SQ14-SQ18}
\end{centering}
\end{figure}

Five questions (SQ14-SQ18) were asked to understand to what extent the participants agree or disagree to use architecting activities (i.e., Architectural Analysis (AA), Architectural Synthesis (AS), Architectural Evaluation (AE), Architectural Implementation (AI), Architectural Maintenance and Evolution (AME)) in MSA projects. Figure \ref{fig:SQ14-SQ18} shows that 53.7\%  (54+3 out of 106) of the participants either “agree” or “strongly agree” with the use of AE (i.e., evaluating candidate architecture solutions when designing microservices systems to ensure design decisions are right). 52.8\% (49+7 out of 106) of the participants “agree” or “strongly agree” with the use of AI (i.e., refining coarse-grained architecture design of microservices systems into fine-grained detailed design with which developers can write code for microservices systems). 

Concerning to frequent use of AE activity, four interviewees (i.e., P1, P3, P5) mentioned that they use AE activity to (i) evaluate the functional and quality concerns (e.g., scalability, availability, performance, deployability, and security), ii) assess the architectural decisions along with their strengths and weaknesses, iii) get feedback about missing or additional architecturally significant requirements, and (iv) evaluate the business goals in the context of the designed architecture and current development processes. Three interviewees (i.e., P2, P4, P6) stated that they are unfamiliar with these architecting activities and their organizations have dedicated architects or teams to design microservices systems who are responsible for these activities. In the following, we provide one interviewee's response.

\faComment “\textit{We use AE to determine whether the microservices system's architecture is necessary, beneficial, and useful? Whether the current architecture or design requires continuous deployment? Whether the system requires greater availability and fault tolerance? Whether the system requires scalability?  Whether the architecture of the system has or adds complexity to the existing system?}” \textit{\textbf{Application developer (P5)}}.

Our results also indicate that the survey respondents pay more attention to architecting activities close to the solution space (i.e., AE, AI, AME) instead of the problem space (i.e., AA, AS). About this finding, four out of 6 interviewees (i.e., P1, P2, P3, P5) stated that they mainly focus on the solution space because their primary goal is to meet the customer requirements by providing them with working software. For instance, one participant explained his opinion in the following comment with an example.

\faComment “\textit{It is because, in the case of the solution space, scenarios are more comfortable to evaluate. For example, if we have a Zookeeper cluster located on two data centers. Losing two nodes within the data centers is an easy scenario to simulate when we have different architectures in the solution space. If we only focus on problem space and design an architecture resilient to the loss of one data center, it is more complicated. In the solution space, we have a set of proven reference architectures, and we use those and enrich them”} \textit{\textbf{Architect and Application developer (P3)}}.

It is interesting to note that a significant number of respondents took a “neutral” position about several MSA architecting activities, for example, AS (50 out of 106, 47.2\%), Architectural Analysis (43 out of 106, 40.6\%), and AE (34 out of 106, 32.1\%). A considerable number of participants “disagree” or “strongly disagree” on the usage of MSA architecting activities. For example, 21.7\% (19+4 out of 106) of the participants answered that they do not use AME, and 20.8\% (20+2 out of 106) of them responded that they do not use AI.


\begin{tcolorbox}[colback=gray!5!white,colframe=gray!75!black,title=Key Findings of RQ1]
\justify
\textbf{Finding 6.} Concerning MSA architecting activities, over half of the participants agreed to use Architecture Evaluation and Implementation when designing microservices systems.
\end{tcolorbox}

\begin{figure}[H]
\begin{centering}
\begin{tikzpicture}
  \begin{axis}[
        footnotesize,
      xbar stacked,
      width=13.0cm, height=5cm, 
      bar width=15pt,
      nodes near coords={
        \pgfkeys{/pgf/fpu=true}
        \pgfmathparse{\pgfplotspointmeta / 106 * 100}
        $\pgfmathprintnumber[fixed, precision=1]{\pgfmathresult}$
        \pgfkeys{/pgf/fpu=false}
      },
      nodes near coords custom/.style={
        every node near coord/.style={
          check for small/.code={
            \pgfkeys{/pgf/fpu=true}
            \pgfmathparse{\pgfplotspointmeta<#1}\%
            \pgfkeys{/pgf/fpu=false}
            \ifpgfmathfloatcomparison
              \pgfkeysalso{above=.5em}
            \fi
          },
          check for small,
        },
      },
      nodes near coords custom=6,
      xmin=-2, xmax=109,
      xtick={0, 10.6, ..., 106.1},
      ytick={1,...,4},
      yticklabels={ADL, DSL, UML, Boxes and Lines},
      xtick pos=bottom,
      ytick pos=left,
      xticklabel={
        \pgfkeys{/pgf/fpu=true}
        \pgfmathparse{\tick / 106 * 100}
        $\pgfmathprintnumber[fixed, precision=1]{\pgfmathresult}\%$
        \pgfkeys{/pgf/fpu=false}
      },
      enlarge y limits=.15,
      legend style={at={(0.5,-0.20)}, anchor=north, legend columns=-1},
    every node near coord/.append style={font=\footnotesize},
]
\addplot coordinates{(2,1) (8,2) (13,3) (29,4)};
\addplot coordinates{(23,1) (21,2) (20,3) (12,4)};
\addplot coordinates{(15,1) (23,2) (31,3) (25,4)};
\addplot coordinates{(25,1) (20,2) (29,3) (23,4)};
\addplot [color=violet, fill=violet!50] coordinates{(41,1) (34,2) (13,3) (17,4)};

\legend{\strut Very Often, \strut Often, \strut Sometimes, \strut Rarely, \strut Never}
\end{axis}
\end{tikzpicture}
\caption{Methods used for describing MSA in industry (UML-Unified Modeling Language, DSL-Domain Specific Languages, ADL-Architectural Description Language)}
\label{fig:SQ19}
\end{centering}
\end{figure}
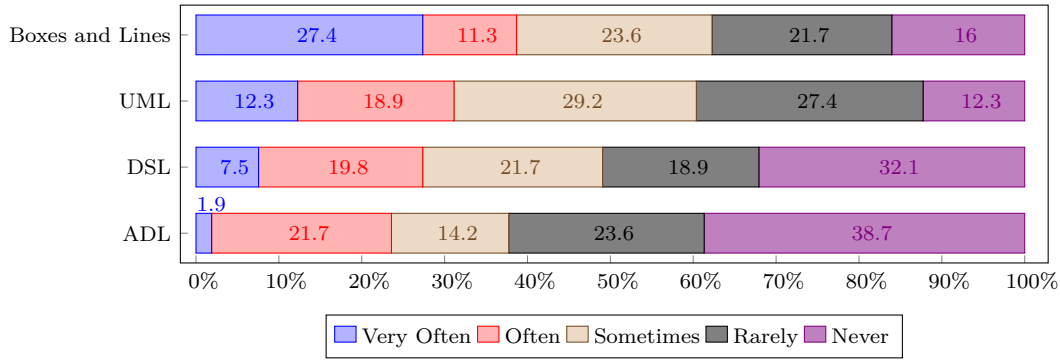

We also investigated about the MSA description methods in this survey. The results show that practitioners prefer to use informal Boxes and Lines for representing MSA. Figure \ref{fig:SQ19} shows that almost 51\% (29+25 out of 106) of the participants answered that they “very often” or “often” used Boxes and Lines for describing MSA. UML is reported as the second most used MSA description method, as 41.5\% (31+13 out of 106) of the participants answered that they used UML frequently (often or very often) for describing MSA. Regarding the frequent use of informal and semi-formal description methods, we summarize the responses of all the interviewees: (i) interactively visualize or describe several views of the system for stakeholders, (ii) provide visual representation and flexibility to changes, (iii) are low cost and easy to learn, (iv) are quick to implement and maintain the architecture documents, (v) are not limited to formal notations, (vi) effectively represent the communication between different parts of the system, vii) are easier to interpret, and (viii) have interactive tools support. One interview participant stated that

\faComment “\textit{We are not feeling the need for other approaches besides boxes and lines (informal approaches). Informal approaches are simpler to implement and effective to describe microservices architecture”} \textit{\textbf{Architect and Application developer (P3)}}.

A vast majority of the participants (89 out of 106, 84.7\%) responded that they never (41 out of 106, 38.7\%), sometimes (25 out of 106, 23.6\%), or rarely (23 out of 106, 21.6\%) used ADLs for describing MSA. DSLs are also not much popular in the industry, as only 29.2\% (8+23 out of 106) of the participants reported that they used (often or very often) DSLs for this purpose. According to interview participants, there are several reasons behind the limited use of ADLs and DSLs, including (i) more effort is required to learn ADLs and DSLs within a tight project delivery schedule, (ii) ADLs and DSLs cannot achieve all the architecting goals (e.g., visualization of architectural views), (iii) ADLs and DSLs lack of generalizability (most ADLs or DSLs are limited for a specific purpose), (iv) ADLs and DSLs have insufficient tooling support, (v) ADLs and DSLs cannot describe the complexity of microservices systems effectively, and (vi) ADLs and DSLs are not delivering benefits to working software. In the following, we provide one example comment from an interviewee about the use of ADLS and DSLs.

\faComment “\textit{I described architecture for hundreds of systems during my more than fifteen years of work experience from small to very large-scale companies (both monolithic and microservices). However, we did not use ADLs and DSLs either for monolithic or microservices systems because of their low business values and limited visualization to architectural views for the key system stakeholders}” \textit{\textbf{Architect and Application developer (P1)}}.



Overall, we identified that the use of formal approaches (e.g., ADLs and DSL) for describing MSA is not widespread. In contrast, practitioners prefer to use informal (e.g., Boxes and Lines) or semi-formal (e.g., UML) approaches for describing MSA.

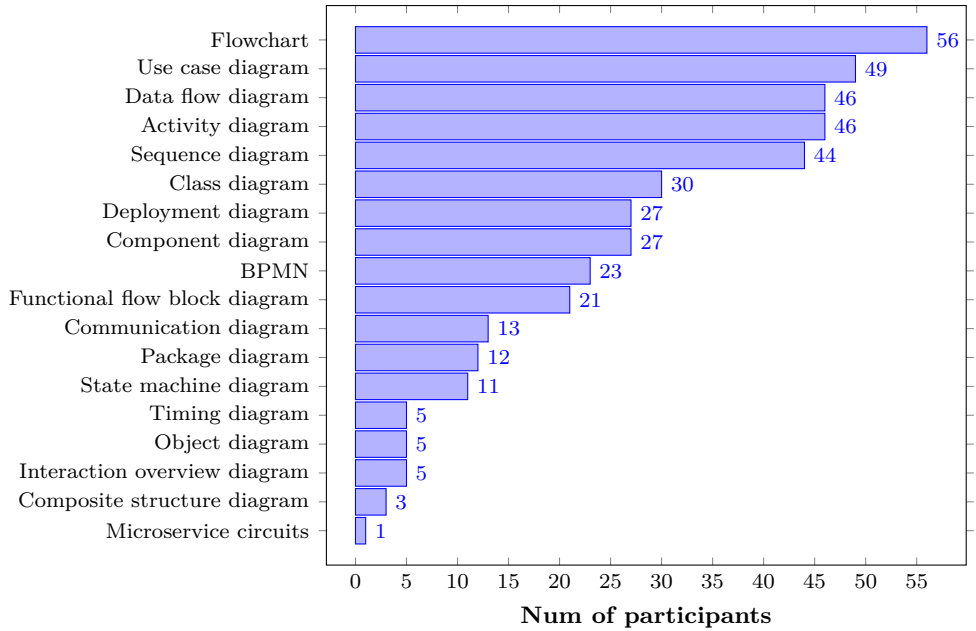
\begin{figure}[H]
\begin{centering}
\begin{tikzpicture}  
\begin{axis}[
	footnotesize,
	xbar, 
	width=10.0cm, height=9.0cm, 
	enlarge y limits=0.01,
	enlargelimits=0.07,  
	symbolic y coords={
	Microservice circuits, Composite structure diagram, Interaction overview diagram, Object diagram, Timing diagram, State machine diagram, Package diagram, Communication diagram, Functional flow block diagram, BPMN, Component diagram, Deployment diagram, Class diagram, Sequence diagram, Activity diagram, Data flow diagram, Use case diagram, Flowchart},
	ytick=data,
	xlabel={\textbf{Num of participants}},
	nodes near coords, nodes near coords align={horizontal},
	every node near coord/.append style={font=\footnotesize},
]
	\addplot coordinates {(56,Flowchart) (49,Use case diagram) (46,Activity diagram) (46,Data flow diagram) (44,Sequence diagram) (30,Class diagram) (27,Component diagram) (27,Deployment diagram) (23,BPMN) (21,Functional flow block diagram) (13,Communication diagram) (12,Package diagram) (11,State machine diagram) (5,Interaction overview diagram) (5,Object diagram) (5,Timing diagram) (3,Composite structure diagram) (1,Microservice circuits)};
\end{axis}
\end{tikzpicture}  
\caption{Diagrams for representing the design and architecture of microservices systems}
\label{fig:SQ20}
\end{centering}
\end{figure}

About the diagrams used to represent MSA, Figure \ref{fig:SQ20} shows the frequency of various types of diagrams used by the respondents to represent the design and architecture of microservices systems (SQ20). Flowchart (56 out of 106, 52.8\%) followed by Use case diagram (49 out of 106, 46.2\%), Data flow diagram (46 out 106, 43.3\%), and Activity diagram (46 out of 106, 43.3\%) are the widely used diagrams to represent MSA in the industry. Other popular diagrams are Sequence diagram (44 out of 106, 41.5\%), Class diagram (30 out of 106, 28.3\%), and Deployment diagram (27 out of 106, 25.4\%).
Interestingly, except Microservices circuits\footnote{\url{https://jolielang.gitbook.io/docs/architectural-composition/aggregation}}, all the diagrams above have been used for representing the scenario, process, logical, implementation, and physical view of monolithic architecture.

\begin{tcolorbox}[colback=gray!5!white,colframe=gray!75!black,title=Key Findings of RQ1]
\justify
\textbf{Finding 7.} The use of formal approaches (e.g., ADLs, DSLs) is not widespread in the industry to describe MSA. In contrast, practitioners prefer to use informal (e.g., Boxes and Lines) and semi-formal (e.g., UML) approaches for describing MSA.\\
\textbf{Finding 8.} Responses of the participants about the diagrams (e.g., flowchart, data flow, activity, sequence) for representing the design and architecture of microservices systems indicate that practitioners prefer to model the process view of MSA.
\end{tcolorbox}

\begin{figure}[!h]
\begin{centering}
\begin{tikzpicture}  
\begin{axis}[
	footnotesize,
	xbar, 
	width=11.0cm, height=4cm, 
	enlarge y limits=0.01,
	enlargelimits=0.07,  
	symbolic y coords={All the above, Storage, Process, Service interface, Domain, Services},
	xmin=20,
	xtick distance=5,
	xmax=50,
	ytick=data,
	xlabel={\textbf{Num of participants}},
	nodes near coords, nodes near coords align={horizontal},
	every node near coord/.append style={font=\footnotesize},
]
	\addplot coordinates {
	(48,Services) (27,Domain) (26,Service interface) (25,Process) (23,Storage) (37,All the above)};
\end{axis}
\end{tikzpicture}  
\caption{Design components of microservices systems}
\label{fig:SQ21}
\end{centering}
\end{figure}
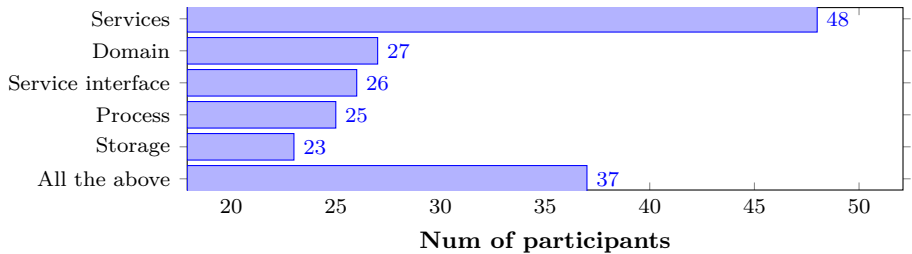

Another important aspect that we explored through this survey is MSA design components that can be used to describe MSA. We provided a list of potential architectural elements (e.g., domain, service, service interface, process, and storage) to the participants \cite{Sander}, and asked them to select which of these elements could be considered as MSA design components. Figure \ref{fig:SQ21} shows that 45.2\% (48 out of 106) of the respondents selected services (e.g., service gateway, service client) as MSA design components whereas 35.2\% (37 out of 106) of the participants said that they considered all the architectural elements in the list as MSA design components.

\subsubsection{Quality attributes and design patterns}
\label{sec:QADP}
We selected 15 QAs that are popular in the MSA and SOA context \cite{marquez2018actual, o2007quality, LI2020106449}. For each QA, the participants were asked to rank their importance for microservices systems using a Likert-scale question rated as “very important”, “important”, “somewhat important”, “not important”, and “not sure”. Table \ref{tab:MSAQAs} shows that 55.7\% (59 out of 106) of the respondents ranked Security as “very important”, followed by Availability (58 out of 106, 54.7\%), Performance (56 out of 106, 52.8\%), and Scalability (51 out of 106, 48.1\%). About these results, we asked the interview participants with IQ2.5.1 “\textit{what could be the reasons that practitioners consider security, availability, performance, and scalability the most important quality attributes for microservices systems}”. All the interviewees mentioned that they consider Security an important QA because (i) microservices systems are distributed in nature, ii) microservices systems are complex and have security vulnerabilities that an intruder can exploit. Two of the interviewees (P1, P4) also stated that “\textit{Security is equally crucial for microservices or any other type of distributed applications}”. Following is one representative quotation about the importance of Security in the context of microservices systems. 

\faComment “\textit{We have to keep up with the various security patches. However, in my understanding, providing security to microservices systems is relatively easy than monolithic applications because normally, we need to secure API gateway from intruders. Also, the isolated nature of microservices helps to limit the access of intruders to only vulnerable services}” \textit{\textbf{Architect and Application developer (P1)}}.

About availability, three interviewees (i.e., P1, P2, P3) stated that end-users usually do not compromise on the non-availability of the systems. One interviewee (i.e., P1) mentioned that end-users could compromise on the system performance to some extent in the peak time of system usage. However, it is not acceptable for the end-users that the system goes down for some time. Moreover, DDD plays a critical role in the availability of microservices systems because applications are decomposed into microservices according to subdomains of the system. If any problem occurs, it remains limited to specific microservice of subdomain and does not propagate in the whole system. We also investigated the reasons for considering Performance as an important QA during the design of microservices systems, and four interviewees (i.e., P1, P2, P3, P5) mentioned that (i) performance should not be degraded due to frequent communication calls and (ii) effective load balancing techniques can be introduced (e.g., creating multiple instances of one microservice and using a load balancer to improve the performance). Regarding Scalability, three interviewees (i.e., P2, P4, P6) stated that it is because scalability is one of the key features of microservices systems that provides a simplified way to scale a particular microservice instance by having a minimal effect on the availability of the system to end-users. Following is one representative quotation about the importance of Availability in the context of microservices systems.

\faComment “\textit{To achieve reliability for microservices systems, practitioners want to make their systems available 24/7. It is also essential when the system has several competitors or millions of users}” \textit{\textbf{Architect and Application developer (P5)}}.

On the other hand, only 26.4\% (28 out of 106) and 23.5\% (25 out of 106) of the respondents ranked Resilience and Portability as “very important” QAs, respectively. We also found several QAs that were ranked as “somewhat important” by the participants. Such QAs are Reusability (25 out of 106, 23.6\%), Compatibility (24 out of 106, 22.6\%), and Monitorability (23 out of 106, 21.7\%).
Overall, the results show that QAs are an essential architectural concern to design microservices systems.
{\renewcommand{\arraystretch}{1}
\footnotesize
\begin{longtable}{|l|l|c|c|c|c|c|}
\caption{Importance of QAs (in \%) for designing microservices systems (VI-Very Important, I-Important, SI-Somewhat Important, NI-Not Important, NS-Not Sure)}
\label{tab:MSAQAs}
\\\hline
\multicolumn{1}{|c|}{\textbf{ID}} &
  \textbf{Quality   attributes} &
  \textbf{VI} &
  \textbf{I} &
  \textbf{SI} &
  \textbf{NI} &
  \textbf{NS} \\ \hline
QA1  & Security                 & \cellcolor[HTML]{70AD47}55.7 & \cellcolor[HTML]{A9D08E}27.4                         & \cellcolor[HTML]{E2EFDA}9.4  & \cellcolor[HTML]{E2EFDA}3.8 & \cellcolor[HTML]{E2EFDA}3.8  \\ \hline
QA2  & Availability & \cellcolor[HTML]{70AD47}54.7 & \cellcolor[HTML]{A9D08E}29.2                         & \cellcolor[HTML]{E2EFDA}14.2 & \cellcolor[HTML]{E2EFDA}1.9 & \cellcolor[HTML]{E2EFDA}0.0  \\ \hline
QA3  & Performance & \cellcolor[HTML]{70AD47}52.8 & \cellcolor[HTML]{A9D08E} 33.0 & \cellcolor[HTML]{E2EFDA}9.4  & \cellcolor[HTML]{E2EFDA}2.8 & \cellcolor[HTML]{E2EFDA}1.9 \\ \hline
QA4  & Scalability              & \cellcolor[HTML]{70AD47}48.1 & \cellcolor[HTML]{A9D08E}38.7 & \cellcolor[HTML]{E2EFDA}6.6  & \cellcolor[HTML]{E2EFDA}3.8 & \cellcolor[HTML]{E2EFDA}2.8  \\ \hline
QA5  & Reliability              & \cellcolor[HTML]{70AD47}40.6                         & \cellcolor[HTML]{A9D08E}38.7 & \cellcolor[HTML]{E2EFDA}12.3 & \cellcolor[HTML]{E2EFDA}2.8 & \cellcolor[HTML]{E2EFDA}5.7  \\ \hline

QA6  & Maintainability          & \cellcolor[HTML]{A9D08E}35.8                         & \cellcolor[HTML]{70AD47} 42.5                         & \cellcolor[HTML]{A9D08E}20.8                         & \cellcolor[HTML]{E2EFDA}0.0 & \cellcolor[HTML]{E2EFDA}0.9  \\ \hline

QA7  & Usability& \cellcolor[HTML]{A9D08E} 35.8                         &  \cellcolor[HTML]{70AD47}40.6 & \cellcolor[HTML]{E2EFDA}14.2 & \cellcolor[HTML]{E2EFDA}2.8 & \cellcolor[HTML]{E2EFDA}6.6  \\ \hline

QA8  & Compatibility & \cellcolor[HTML]{A9D08E} 34.0& \cellcolor[HTML]{A9D08E} 34.0 & \cellcolor[HTML]{A9D08E} 22.6                         & \cellcolor[HTML]{E2EFDA}1.9 & \cellcolor[HTML]{E2EFDA}7.5  \\ \hline
QA9  & Testability & \cellcolor[HTML]{A9D08E} 33.0                         &  \cellcolor[HTML]{70AD47}41.5 & \cellcolor[HTML]{E2EFDA}19.8& \cellcolor[HTML]{E2EFDA}0.9 & \cellcolor[HTML]{E2EFDA}4.7  \\ \hline
QA10 & Monitorability& \cellcolor[HTML]{A9D08E} 32.1                         & \cellcolor[HTML]{A9D08E}38.7 & \cellcolor[HTML]{A9D08E}21.7 & \cellcolor[HTML]{E2EFDA}1.9 & \cellcolor[HTML]{E2EFDA}5.7  \\ \hline
QA11 & Functional   suitability & \cellcolor[HTML]{A9D08E} 30.2& \cellcolor[HTML]{70AD47} 42.5 & \cellcolor[HTML]{E2EFDA}15.1 & \cellcolor[HTML]{E2EFDA}4.7 & \cellcolor[HTML]{E2EFDA}7.5  \\ \hline
QA12 & Reusability & \cellcolor[HTML]{A9D08E} 30.2                         & \cellcolor[HTML]{A9D08E} 35.8 & \cellcolor[HTML]{A9D08E}23.6& \cellcolor[HTML]{E2EFDA}5.7 & \cellcolor[HTML]{E2EFDA}4.7  \\ \hline
QA13 & Resilience& \cellcolor[HTML]{A9D08E}26.4                         & \cellcolor[HTML]{A9D08E} 30.2 & \cellcolor[HTML]{A9D08E}22.6 & \cellcolor[HTML]{E2EFDA}5.7 & \cellcolor[HTML]{E2EFDA}15.1 \\ \hline
QA14 & Portability& \cellcolor[HTML]{A9D08E}23.6                         & \cellcolor[HTML]{A9D08E}35.8 & \cellcolor[HTML]{A9D08E}26.4                         & \cellcolor[HTML]{E2EFDA}7.5 & \cellcolor[HTML]{E2EFDA}6.6  \\ \hline
QA15 & Interoperability & \cellcolor[HTML]{A9D08E}21.7& \cellcolor[HTML]{A9D08E}42.5 & \cellcolor[HTML]{E2EFDA}17.0 & \cellcolor[HTML]{E2EFDA}5.7 & \cellcolor[HTML]{E2EFDA}13.2 \\ \hline
\end{longtable}}

\begin{tcolorbox}[colback=gray!5!white,colframe=gray!75!black,title=Key Findings of RQ1]
\justify
\textbf{Finding 9.} Over 80\% of the participants indicated that security, availability, performance, and scalability are very important QAs for the design of microservices systems.
\end{tcolorbox}

The participants were also asked to indicate how often they use patterns when designing microservices systems. The majority of MSA design patterns are decently (i.e., very often, often, sometimes) used in practice, and only some of them are not used or used less often (e.g., rarely, never) to design microservices systems. The results show that the following four MSA design patterns are used most often: API gateway (26.4\% very often, 42.5\% often, 11.3\% sometimes), Backend for frontend (17.9\% very often, 35.8\% often, 15.1\% sometimes), Access token (17.9\% very often, 34.9\% often, 16.0\% sometimes), and Database per service (12.3\% very often, 36.8\% often, 21.7\% sometimes). These four patterns are used to manage external APIs, application security, and databases for microservices systems. Similarly, we identified some other MSA design patterns that are frequently used for data management, deployment, and communication of microservices. The examples of such patterns are API composition (14.2\% very often, 33.0\% often, 35.8\% sometimes), Service instance per container (16.6\% very often, 30.3\% often, 24.5\% sometimes), and Messaging (14.2\% very often, 32.1\% often, 24.5\% sometimes). Overall, the combined usage (e.g., very often, often, Sometimes) reported by the participants for each MSA design pattern is more than 50.0\%.

We also asked IQ2.6.1 to the interviewees about the leading MSA design patterns used in industry. Four out of six interviewees (i.e., P1, P2, P4, P5) provided the following reasons for the frequent use of API gateway and Backend for frontend patterns: (i) API gateway and Backend for frontend are the key patterns for request routing, API composition, and protocol translation for microservices systems, (ii) these patterns have naturally bound with microservices architecture, (iii) API gateway offers a single entry-point for a specific group of microservices, (iv) Backend for frontend is a variant of the API gateway pattern that can be used to separate the frontend from the backend of the applications, and (v) these two patterns enhance the loose coupling property of MSA. About the frequent use of security pattern Access token, four interviewees (P1, P2, P4, P6) mentioned several reasons, such as (i) token-based authentication allowing an application to access an API, (ii) one-time login, iii) no session is persevered on the server side (stateless), iv) servers do not need to track and keep the information of users, (v) the security token is short-lived and can be revoked easily, (vi) the security token helps to save the server memory consumption, and (vii) if the service goes down for some reason, the security token is still valid because it is also used on the client side. Regarding the frequent use of Database per service pattern, six interviewees (P1, P2, P3, P4, P5, P6) provided four reasons that the Database per service pattern (i) helps to design and develop the loosely coupled applications, (ii) provides support for selecting databases according to the requirement of each microservice, (iii) helps to control the changes or failure by limiting them to a specific microservice or database, and (iv) provides freedom of using specific databases according to the team’s skills. In the following, we provide one representative quotation about the use of API gateway pattern for microservices systems.

\faComment “\textit{In my understanding, the reasons behind using these patterns are rooted in microservices systems. For instance, instead of the API gateway pattern, we did not use any other communication pattern, and it is almost considered a ditto standard for communication in the industry}” \textit{\textbf{Application developer (P4)}}. 

On the other hand, some MSA design patterns, such as Sagas (14.2\% rarely, 34.0\% never), Event sourcing (21.7\% rarely, 23.6\% never), and Self-registration (20.8\% rarely, 22.6\% never), are rarely or never used. More details about the use of MSA design patterns can be found in Table \ref{tab:MSApatterns}.


{\renewcommand{\arraystretch}{1}
\centering
\footnotesize
\begin{longtable}{|l|l|c|c|c|c|c|c|}
\caption{MSA design patterns (in \%) used in the industry (VO-Very Often, O-Often, S-Sometimes, R-Rarely, N-Never)}
\label{tab:MSApatterns}
\\\hline
\multicolumn{1}{|c|}{\textbf{ID}} &
  \multicolumn{1}{c|}{\textbf{MSA patterns}} &
  \textbf{VO} &
  \textbf{O} &
  \textbf{S} &
  \textbf{R} &
  \textbf{N} 
 \\ \hline
MSP1 &
  API gateway &
  \cellcolor[HTML]{A9D08E}26.4 &
  \cellcolor[HTML]{70AD47}42.5 &
  \cellcolor[HTML]{E2EFDA}11.3 &
  \cellcolor[HTML]{E2EFDA}10.4 &
  \cellcolor[HTML]{E2EFDA}9.4 \\ \hline
MSP2 &
   Backend for frontend &
  \cellcolor[HTML]{E2EFDA}17.9 &
  \cellcolor[HTML]{70AD47}35.8 &
  \cellcolor[HTML]{E2EFDA}15.1 &
  \cellcolor[HTML]{E2EFDA}17.9 &
  \cellcolor[HTML]{E2EFDA}13.2 \\ \hline
MSP3 &
  Access token &
  \cellcolor[HTML]{E2EFDA}17.9 &
  \cellcolor[HTML]{70AD47}34.9 &
  \cellcolor[HTML]{E2EFDA}16.0 &
  \cellcolor[HTML]{E2EFDA}17.0 &
  \cellcolor[HTML]{E2EFDA}14.2 \\ \hline
MSP4 &
  Service instance per   container &
  \cellcolor[HTML]{E2EFDA}16.0 &
  \cellcolor[HTML]{70AD47}30.2 &
  \cellcolor[HTML]{A9D08E}24.5 &
  \cellcolor[HTML]{E2EFDA}15.1 &
  \cellcolor[HTML]{E2EFDA}14.2 \\ \hline
MSP5 &
  Service registry &
  \cellcolor[HTML]{E2EFDA}16.0 &
  \cellcolor[HTML]{70AD47}29.2 &
  \cellcolor[HTML]{A9D08E}23.6 &
  \cellcolor[HTML]{E2EFDA}12.3 &
  \cellcolor[HTML]{E2EFDA}18.9 \\ \hline
MSP6 &
  API composition &
  \cellcolor[HTML]{E2EFDA}14.2 &
  \cellcolor[HTML]{70AD47}33.0 &
  \cellcolor[HTML]{70AD47}35.8 &
  \cellcolor[HTML]{E2EFDA}10.4 &
  \cellcolor[HTML]{E2EFDA}6.6 \\ \hline
MSP7 &
  Messaging &
  \cellcolor[HTML]{E2EFDA}14.2 &
  \cellcolor[HTML]{70AD47}32.1 &
  \cellcolor[HTML]{C6E0B4}24.5 &
  \cellcolor[HTML]{E2EFDA}17.9 &
  \cellcolor[HTML]{E2EFDA}11.3 \\ \hline
MSP8 &
  Remote procedure   invocation &
  \cellcolor[HTML]{E2EFDA}13.2 &
  \cellcolor[HTML]{70AD47}32.1 &
  \cellcolor[HTML]{E2EFDA}17.0 &
  \cellcolor[HTML]{E2EFDA}17.9 &
  \cellcolor[HTML]{E2EFDA}19.8 \\ \hline
MSP9 &
  Database per service &
  \cellcolor[HTML]{E2EFDA}12.3 &
  \cellcolor[HTML]{70AD47}36.8 &
  \cellcolor[HTML]{A9D08E}21.7 &
  \cellcolor[HTML]{E2EFDA}17.9 &
  \cellcolor[HTML]{E2EFDA}11.3 \\ \hline
MSP10 &
  Circuit breaker &
  \cellcolor[HTML]{E2EFDA}11.3 &
  \cellcolor[HTML]{70AD47}30.2 &
  \cellcolor[HTML]{E2EFDA}18.9 &
  \cellcolor[HTML]{E2EFDA}19.8 &
  \cellcolor[HTML]{E2EFDA}19.8 \\ \hline
MSP11 &
  Client-side   discovery &
  \cellcolor[HTML]{E2EFDA}10.4 &
  \cellcolor[HTML]{70AD47}34.9 &
  \cellcolor[HTML]{E2EFDA}19.8 &
  \cellcolor[HTML]{E2EFDA}19.8 &
  \cellcolor[HTML]{E2EFDA}15.1 \\ \hline
MSP12 &
  Server-side   discovery &
  \cellcolor[HTML]{E2EFDA}10.4 &
  \cellcolor[HTML]{A9D08E}24.5 &
  \cellcolor[HTML]{A9D08E}28.3 &
  \cellcolor[HTML]{E2EFDA}17.9 &
  \cellcolor[HTML]{E2EFDA}{\color[HTML]{333333} 18.9} \\ \hline
MSP13 &
  Serverless   deployment &
  \cellcolor[HTML]{E2EFDA}9.4 &
  \cellcolor[HTML]{A9D08E}25.5 &
  \cellcolor[HTML]{E2EFDA}18.9 &
  \cellcolor[HTML]{E2EFDA}17.9 &
  \cellcolor[HTML]{A9D08E}28.3 \\ \hline
MSP14 &
  Service instance per   host &
  \cellcolor[HTML]{E2EFDA}8.5 &
  \cellcolor[HTML]{A9D08E}29.2 &
  \cellcolor[HTML]{A9D08E}24.5 &
  \cellcolor[HTML]{E2EFDA}18.9 &
  \cellcolor[HTML]{E2EFDA}18.9 \\ \hline
MSP15 &
  Self-registration &
  \cellcolor[HTML]{E2EFDA}8.5 &
  \cellcolor[HTML]{A9D08E}25.5 &
  \cellcolor[HTML]{A9D08E}22.6 &
  \cellcolor[HTML]{A9D08E}20.8 &
  \cellcolor[HTML]{A9D08E}22.6 \\ \hline
MSP16 &
  Polling   publisher &
  \cellcolor[HTML]{E2EFDA}8.5 &
  \cellcolor[HTML]{C6E0B4}17.9 &
  \cellcolor[HTML]{A9D08E}29.2 &
  \cellcolor[HTML]{E2EFDA}16.0 &
  \cellcolor[HTML]{A9D08E}28.3 \\ \hline
MSP17 &
  Service instance per   VM &
  \cellcolor[HTML]{E2EFDA}7.5 &
  \cellcolor[HTML]{70AD47}30.2 &
  \cellcolor[HTML]{C6E0B4}18.9 &
  \cellcolor[HTML]{C6E0B4}17.0 &
  \cellcolor[HTML]{C6E0B4}26.4 \\ \hline
MSP18 &
  Application events &
  \cellcolor[HTML]{E2EFDA}7.5 &
  \cellcolor[HTML]{C6E0B4}24.5 &
  \cellcolor[HTML]{70AD47}33.0 &
  \cellcolor[HTML]{E2EFDA}18.9 &
  \cellcolor[HTML]{E2EFDA}16.0 \\ \hline
MSP19 &
  Sagas &
  \cellcolor[HTML]{E2EFDA}7.5 &
  \cellcolor[HTML]{C6E0B4}24.5 &
  \cellcolor[HTML]{E2EFDA}19.8 &
  \cellcolor[HTML]{E2EFDA}14.2 &
  \cellcolor[HTML]{548235}34.0 \\ \hline
MSP20 &
  Service deployment   platform &
  \cellcolor[HTML]{E2EFDA}6.6 &
  \cellcolor[HTML]{548235}32.1 &
  \cellcolor[HTML]{A9D08E}23.6 &
  \cellcolor[HTML]{E2EFDA}17.0 &
  \cellcolor[HTML]{A9D08E}20.8 \\ \hline
MSP21 &
  Multiple service   instance per host &
  \cellcolor[HTML]{E2EFDA}6.6 &
  \cellcolor[HTML]{70AD47}31.1 &
  \cellcolor[HTML]{A9D08E}22.6 &
  \cellcolor[HTML]{A9D08E}24.5 &
  \cellcolor[HTML]{E2EFDA}15.1 \\ \hline
MSP22 &
  Shared database &
  \cellcolor[HTML]{E2EFDA}6.6 &
  \cellcolor[HTML]{70AD47}31.1 &
  \cellcolor[HTML]{A9D08E}27.4 &
  \cellcolor[HTML]{E2EFDA}19.8 &
  \cellcolor[HTML]{E2EFDA}15.1 \\ \hline
MSP23 &
  Event sourcing &
  \cellcolor[HTML]{E2EFDA}5.7 &
  \cellcolor[HTML]{A9D08E}20.8 &
  \cellcolor[HTML]{A9D08E}28.3 &
  \cellcolor[HTML]{A9D08E}21.7 &
  \cellcolor[HTML]{A9D08E}23.6 \\ \hline
MSP24 &
  CQRS &
  \cellcolor[HTML]{E2EFDA}5.7 &
  \cellcolor[HTML]{E2EFDA}17.0 &
  \cellcolor[HTML]{A9D08E}32.1 &
  \cellcolor[HTML]{A9D08E}20.8 &
  \cellcolor[HTML]{A9D08E}24.5 \\ \hline
MSP25 &
  Transaction log   tailing &
  \cellcolor[HTML]{E2EFDA}4.7 &
  \cellcolor[HTML]{70AD47}30.2 &
  \cellcolor[HTML]{E2EFDA}19.8 &
  \cellcolor[HTML]{A9D08E}24.5 &
  \cellcolor[HTML]{A9D08E}20.8 \\ \hline
MSP26 &
  Domain-specific   protocol &
  \cellcolor[HTML]{E2EFDA}3.8 &
  \cellcolor[HTML]{548235}33.0 &
  \cellcolor[HTML]{E2EFDA}18.9 &
  \cellcolor[HTML]{A9D08E}23.6 &
  \cellcolor[HTML]{A9D08E}20.8 \\ \hline
\end{longtable}}

The participants also mentioned several other patterns via the ``Other'' field in SQ23. Such patterns are Model-view-controller (5 responses), Two-phase commit\footnote{\url{https://tinyurl.com/yyr3sl7q}} (3 responses), Facade (1 response), Singleton (1 response), Repository (1 response), API key\footnote{\url{https://tinyurl.com/y2sm2jqz}} (1 response), Rate limit\footnote{\url{https://tinyurl.com/yxroch9s}} (1 response), Service decorator\footnote{\url{https://tinyurl.com/y5rphfsx}} (1 response), Session memory (1 response), and Tolerant reader (1 response).

\begin{tcolorbox}[colback=gray!5!white,colframe=gray!75!black,title=Key Findings of RQ1]
\justify
\textbf{Finding 10.} API gateway and Backend for frontend as communication patterns, Database per service as data management pattern, and Access token as security pattern are reported as the most often used MSA design patterns in practice.
\end{tcolorbox}
\subsubsection{Skills} 
\begin{figure}[H]
\begin{centering}
\begin{tikzpicture}  
\begin{axis}[
	footnotesize,
	xbar, 
	width=9.0cm, height=4.5cm, 
	enlarge y limits=0.01,
	enlargelimits=0.07,  
	symbolic y coords={Knowledge of container technologies, Working in a dynamic team structure, Knowledge of securing microservices, Knowledge of DevOps, Architectural knowledge, Skills to implement MSA design patterns, Breaking down a complex coding task},
	ytick=data,
	nodes near coords, nodes near coords align={horizontal},
	every node near coord/.append style={font=\footnotesize},
]
	\addplot coordinates {(77,Breaking down a complex coding task) (60,Skills to implement MSA design patterns) (57,Architectural knowledge) (44,Knowledge of DevOps) (42,Knowledge of securing microservices) (35,Working in a dynamic team structure) (32,Knowledge of container technologies)};
\end{axis}
\end{tikzpicture}  
\caption{Skills required to design and implement MSA}
\label{fig:SQ28}
\end{centering}
\end{figure}
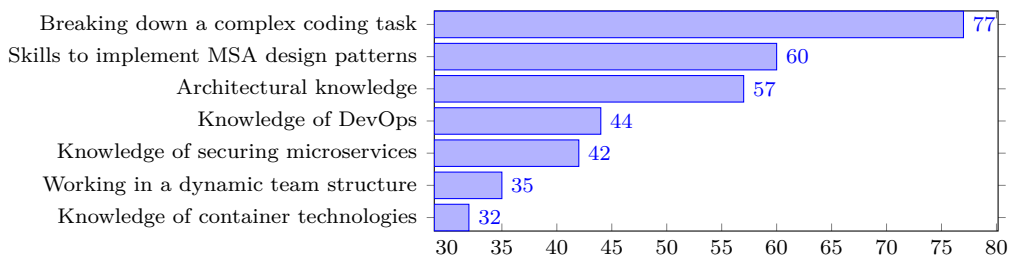

We asked the participants about the skills (including hard and soft skills) required to design and implement MSA (SQ27). We identified these skills from the literature, e.g., \citep{xia2019practitioners, Bartosz, li2015makes}. 72.6\% (77 out of 106) of the participants believed that practitioners should have the skills to break down a complex coding task into smaller tasks, and 56.6\% (60 out of 106) of the participants mentioned that practitioners should have skills to implement the functionality of a microservice by using suitable MSA design patterns.
On the other hand, only 30.1\% (32 out of 106) of the participants answered that practitioners should know container technologies (see Figure \ref{fig:SQ28} for more details).

\subsubsection{Challenges and solutions}
\label{designCha&sol}
\textbf{Challenges}: The participants were asked about architectural challenges that they may encounter during the design of microservices systems (SQ25). We provided a list of challenges identified in a recent literature review \cite{waseemMSAdevops}. The results show that \textit{DC1: clearly defining the boundaries of microservices} is the most common challenge in MSA design (74 out of 106, 69.8\%). The other two top-ranked challenges in MSA design are \textit{DC2: addressing security concerns} (49 out of 106, 46.2\%) and \textit{DC3: managing microservices complexity at the design level} (37 out of 106, 34.9\%).

Concerning these design challenges, we asked IQ2.7.1 to the interviewees about their reasons. All the interviewees agreed that these are valid challenges. Regarding the reasons for the challenge DC1, four out of six interviewees (i.e., P1, P2, P5, P6) mentioned that DC1 occurs due to (i) failure in deciding about the responsibilities or functionalities of each microservice, (ii) failure to locate the overlapping and shared functionalities between microservices, (iii) lack of business knowledge when creating a new application, and (iv) no iterative approaches (e.g., sprint meetings) being employed to refine the design of microservices systems. About the reasons for the challenge DC2, three out of six interviewees (i.e., P1, P3, P4) provided the following reasons: (i) insecure communication between microservices, (ii) vulnerable security standards and tools, (iii) insecure cloud infrastructure for deploying and scaling microservices, and (iv) poor design and introducing insecure code for microservices systems. About the reasons for the challenge DC3, four out of six interviewees (i.e., P1, P2, P3, P6) provided the following reasons: (i) hard to identify end-to-end microservices and decide about their communication patterns, (ii) defining microservices for every system feature or activity (too fine-grained), (iii) poorly defining the boundaries of microservices (i.e., DC1), and (iv) too many interactions and dependencies between microservices. Following is one representative quotation about the challenge DC3 during the design of microservices systems. 

\faComment “\textit{We don’t have any silver bullet to reduce the complexity. We need to balance the complexity of the system while defining the boundaries of microservices. Complexity cannot be mitigated entirely, but we need to bring it to the level where it does not severely affect the system’s quality (performance, availability, security)}” \textit{\textbf{Architect and Application developer (P3)}}.

On the other hand, for each design challenge, the participants were further asked to indicate if these challenges impact monitoring and testing of microservices systems. The fourth and fifth columns of Table \ref{tab:designChallenges} show how many respondents believed that certain challenge has an impact on monitoring and testing of microservices systems. More than half of the respondents felt that the challenge of \textit{DC1: clearly defining the boundaries of microservices} impacts both monitoring and testing of microservices systems. Similarly, 61.9\% (63 out of 106) and 59.4\% (58 out of 106) of the respondents agreed that \textit{DC2: Addressing security concerns} can affect testing and monitoring of microservices systems, respectively. More details can be seen in Table \ref{tab:designChallenges}.

\begin{table}[H]
\footnotesize
\caption{Challenges faced during the design of microservices systems and their impact on monitoring and testing of microservices systems (in \%)}
\label{tab:designChallenges}
\resizebox{\textwidth}{!}{%
\begin{tabular}{|l|l|c|c|c|}
\hline
\multicolumn{1}{|c|}{} &
  \multicolumn{1}{c|}{} &
   &
  \multicolumn{2}{c|}{\textbf{Impact of Challenges on}} \\ \cline{4-5} 
\multicolumn{1}{|c|}{\multirow{-2}{*}{\textbf{ID}}} &
  \multicolumn{1}{c|}{\multirow{-2}{*}{\textbf{MSA Design   Challenge}}} &
  \multirow{-2}{*}{\textbf{Response}} &
  \textbf{Monitoring} &
  \textbf{Testing} \\ \hline
DC1 &
  Clearly defining the boundaries of microservices &
  \cellcolor[HTML]{70AD47}69.8 &
  \cellcolor[HTML]{A9D08E}59.4 &
  \cellcolor[HTML]{A9D08E}54.7 \\ \hline
DC2 &
  Addressing security concerns &
  \cellcolor[HTML]{C6E0B4}46.2 &
  \cellcolor[HTML]{A9D08E}59.4 &
  \cellcolor[HTML]{70AD47}61.3 \\ \hline

DC3 &
  Managing microservices complexity at the design level &
  \cellcolor[HTML]{E2EFDA}34.9 &
  \cellcolor[HTML]{A9D08E}57.5 &
  \cellcolor[HTML]{A9D08E}52.8 \\ \hline
DC4 &
  Recovering MSA from existing code &
  \cellcolor[HTML]{E2EFDA}34.9 &
  \cellcolor[HTML]{A9D08E}57.5 &
  \cellcolor[HTML]{A9D08E}50.9 \\ \hline

DC5 &
  Reaching scalability &
  \cellcolor[HTML]{E2EFDA}26.4 &
  \cellcolor[HTML]{70AD47}60.4 &
  \cellcolor[HTML]{A9D08E}51.9 \\ \hline
DC6 &
  Finding appropriate modelling abstractions for microservices &
  \cellcolor[HTML]{E2EFDA}23.6 &
  \cellcolor[HTML]{C6E0B4}49.1 &
  \cellcolor[HTML]{C6E0B4}48.1 \\ \hline
DC7 &
  Separating functional and operational concerns &
  \cellcolor[HTML]{E2EFDA}23.6 &
  \cellcolor[HTML]{70AD47}67.0 &
  \cellcolor[HTML]{C6E0B4}49.1 \\ \hline
DC8 &
  Addressing microservices communication at the design level &
  \cellcolor[HTML]{E2EFDA}22.6 &
  \cellcolor[HTML]{70AD47}62.3 &
  \cellcolor[HTML]{A9D08E}50.0 \\ \hline
DC9 &
  Addressing data management at the design level &
  \cellcolor[HTML]{E2EFDA}21.7 &
  \cellcolor[HTML]{70AD47}60.4 &
  \cellcolor[HTML]{A9D08E}50.9 \\ \hline
DC10 &
  Aligning team structure with the architecture of MSA-based system &
  \cellcolor[HTML]{E2EFDA}19.8 &
  \cellcolor[HTML]{A9D08E}50.9 &
  \cellcolor[HTML]{70AD47}60.4 \\ \hline
\end{tabular}%
}
\end{table}

 \textbf{Solutions for general challenges – from survey results}: Through an open-ended survey question (SQ26), we asked the participants what solutions they used to address or mitigate the challenges in MSA design. We received 36 answers, and after evaluating each answer, only 23 responses were found as valid. We excluded those solutions that were incomplete, inconsistent, or do not adequately address the MSA design challenges. We classified the valid answers into two categories (see Table \ref{tab:DesignSol}) as presented below:
 
 \begin{itemize}
   \item \textbf{Strategy, patterns, and DSLs}: 
   Most of the participants recommended the DDD strategy (10 responses) to address the challenge of \textit{DC1: clearly defining the boundaries of microservices}. This strategy can help in determining the boundaries of microservices through concepts like bounded context \cite{haselbock2018expert}. To give an example, one participant explicitly mentioned that \say{\textit{we built a micro services-oriented platform (i.e., Dolittle\footnote{\url{https://dolittle.io/contributing/guidelines/domain\_driven\_design/}}) that helps in the identification of MSA boundaries by using the DDD strategy}} \textbf{\textit{C-level executive (R5)}}. However, one respondent questioned the usefulness of DDD in determining the boundaries of microservices and said that \say{\textit{DDD is not always the right option for identification of MSA boundaries as it does not always address problems of scalability and architectural issues for microservices systems. For instance, some boundaries that defined through DDD may not fit with the business model or may not even help to achieve the scalability}} \textbf{\textit{Application developer (R45)}}. Vertical decomposition pattern was deemed to achieve appropriate granularity, scalability, and reliability in microservices systems
   \cite{hasselbring2016microservices,hasselbring2017microservice}. One participant stated that \say{\textit{we use vertical decomposition pattern for identification of microservices boundaries}} \textbf{\textit{System analyst (R33)}}. One other participant mentioned that \say{\textit{Circuit breaker pattern was considered useful to control the flow of traffic and APIs calls between the services, as well as to protect the system from both transient failures and buggy code}} \textbf{\textit{Architect (R19)}}. We also received a few answers in which the participants recommended patterns and DSL as solutions to mitigate the challenges related to the design of MSA, such as microservice API patterns\footnote{\url{https://microservice-api-patterns.org}}, structural design pattern\footnote{\url{https://csrc.nist.gov/publications/detail/sp/800-204/draft}} (e.g., re-deployable Jolie Circuit breaker), and DSL Jolie\footnote{\url{https://jolie-lang.org}}. 
  
  \item\textbf{Adoption of software development methods}: We received one response in which a participant suggested that \say{\textit{developers who have good knowledge of microservices should be involved during the architecting process of the MSA in order to effectively deal with challenges that may arise on design level}} \textbf{\textit{Consultant (R102)}}. Furthermore, some participants also proposed to use DevOps (3 responses), Scrum (2 responses), Feature-driven development (1 response), and Team sprints (1 response) to address the challenges (DC1 and DC10 in Table \ref{tab:designChallenges}) related to the design of microservices systems. However, none of the participants provided detailed descriptions about how software development methods (e.g., DevOps) can address the challenges of MSA design (e.g., security concerns). 
 \end{itemize}

 {\renewcommand{\arraystretch}{1}
 \footnotesize
 \begin{longtable}{|p{2.5cm}|p{5.5cm}|l|c|}
    \caption{Proposed solutions to address the MSA design challenges}
    \label{tab:DesignSol}
    \\ \hline
\textbf{Type} & \textbf{Proposed   Solution} & \textbf{\begin{tabular}[c]{@{}l@{}}Challenge ID\\Addressed\end{tabular}} & \multicolumn{1}{l|}{\textbf{Count}} \\ \hline
\multirow{5}{*}{\begin{tabular}[c]{@{}l@{}}Strategy, patterns, \\ and DSL\end{tabular}} & DDD & DC1 & \multirow{5}{*}{15} \\ \cline{2-3}
 & Vertical   decomposition pattern & DC1, DC5 &  \\ \cline{2-3}
 & Microservice   API patterns (e.g., public API, community API ) & DC2, DC5 &  \\ \cline{2-3}
 & Structural   design pattern & DC3 &  \\ \cline{2-3}
 & Circuit   breaker, Jolie (DSL) & DC8 &  \\ \hline
Software   development methods & DevOps,   Scrum, FDD, and Team sprints & DC1, D10 & 8 \\ \hline
    
    \end{longtable}
    
    }

\textbf{Solutions for top three challenges - from interview results}: We summarized the solutions from the interview results for the leading three challenges, i.e., \textit{DC1: clearly defining the boundaries of microservices}, \textit{DC2: addressing security concerns}, and \textit{DC3: managing microservices complexity at the design level}.
\begin{itemize}
    \item \textbf{Solutions for clearly defining the boundaries of microservices}: The interviewees proposed several measures to address this challenge: (i) gain a deeper understanding of business requirements, application domains, technologies, business processes, and development processes, (ii) introduce and implement the concept of miniservices \cite{GuptaMiniservice} for defining the boundaries of microservices, (iii) identify possible contexts or subdomains of the application by using certain strategies (e.g., DDD), (iv) discover the common and isolated functionality of each subdomain, (v) find the suitable granularity level for microservices, and (vi) ensure the visibility of defined boundaries of microservices.
    \item \textbf{Solutions for addressing security concerns}: Microservices systems can be secured on the network, load balancing, firewall, and application level by simply buying the updated security solutions from cloud providers. For example, AWS provides their own way of securing microservices (e.g., API facade pattern), Kubernetes offers Role-Based Access Control component to limit the access of specific microservice to authorized users. It is suggested that end-users constantly update their firewalls and other security standards to address several security challenges. One of the practitioners (i.e., P2) also recommended employing DevSecOps\footnote{\url{https://www.devsecops.org/}} and microservices together as a solution to address security challenges, because both technologies have a natural alliance to protect the applications.
    \item \textbf{Solutions for managing microservices complexity at the design level}: There is no silver bullet for addressing the complexity of microservices. However, it can be reduced by taking these measures: (i) do not design or develop microservices for every system’s feature, (ii) define a manageable set of responsibilities for each microservice, (iii) understand the operating environment of microservices (e.g., deployment and monitoring infrastructure), and (iv) clearly define the boundaries of microservices.
\end{itemize}

\begin{tcolorbox}[colback=gray!5!white,colframe=gray!75!black,title=Key Findings of RQ1]
\justify\textbf{Finding 11.} \textit{Clearly defining the boundaries of microservices} and \textit{addressing security concerns} are the most reported challenges during the design of microservices systems. Moreover, it has been identified that MSA design challenges have a significant impact on monitoring and testing of microservices systems.\\
\textbf{Finding 12.} Practitioners indicated that they could better understand and define the boundaries of microservices by using Domain-Driven Design (DDD) strategy.
\end{tcolorbox}

\subsubsection{Differences between the subpopulations of the study for MSA design}
\label{GapsMSAdesign}

We analyzed the answers to Likert scale survey questions in Table \ref{tab:SigDiffArchDscrip}, Table \ref{tab:SigDiffQAs}, and Table \ref{tab:SigDifPatterns}. To better understand practitioners' perspectives on the design of microservices systems, we divided the survey respondents into three groups (i.e., \textit{Experience $\le$ 2 years vs. Experience \textgreater{} 2 years}, \textit{MSA style vs. No MSA style}, \textit{Employees size $\le$ 100 vs. Employees size \textgreater{} 100}). We calculated the P-value to indicate statistically significant difference for each survey question statement and the Effect size to indicate the magnitude of difference between two groups. The dark grey color in table cells shows that the first group (e.g., Experience $\le$ 2 years) is more likely to agree with the survey question statements. In contrast, the light grey color shows that the second group (e.g., Experience \textgreater{} 2 years) is more likely to agree with the survey question statements. In the following, we report the statistical comparison for MSA architecting activities and description approaches, quality attributes, and MSA design patterns.


\textbf{MSA architecting activities and description approaches} (see Table \ref{tab:SigDiffArchDscrip}):
\textit{Experience $\le$ 2 years vs. Experience \textgreater{} 2 years}: There are two survey question statements (i.e., \faWrench{} {\textbf{Architectural maintenance and evolution}} and \faWrench{} {\textbf{Domain specific languages}}) with statistically significant differences between Experience $\le$ 2 years and Experience \textgreater{} 2 years groups. The Experience \textgreater{} 2 years group is more likely to agree on using Architectural maintenance and evolution activity in MSA projects than the Experience $\le$ 2 years group. On the other hand, the Experience $\le$ 2 years group is more likely to use Domain Specific Languages for describing MSA than the Experience \textgreater{} 2 years group. \textit{MSA style vs. No MSA style}: There are three survey question statements (i.e., \faWrench{} {\textbf{Architectural maintenance and evolution}}, \faGears{} {\textbf{Unified modeling language}}, \faGears{} {\textbf{Domain specific languages}}) with statistically significant differences between MSA style and No MSA style groups. The MSA style group is more likely to agree on using Architectural maintenance and evolution activity in MSA projects than the No MSA style group. Moreover, the MSA style group is also more likely to use Unified modeling language and Domain specific languages for describing the architecture of microservices systems than the MSA style group. \textit{Employees size $\le$ 100 vs. Employees size \textgreater{} 100}: We did not find any survey question statement with statistically significant difference between Employees size $\le$ 100 and Employees size \textgreater{} 100 groups.

\begin{table}[H]
\caption{Statistically significant differences on the survey results about MSA architecting activities and description approaches}
\label{tab:SigDiffArchDscrip}
\footnotesize

\resizebox{\textwidth}{!}{\begin{tabular}{|r|c|c|c|c|c|c|c|c|}
\hline 
& &\textbf{Likert Distro.} & \multicolumn{2}{c|}{\textbf{Exper. $\le$ 2 vs. \textgreater{} 2 years}} & \multicolumn{2}{c|}{\textbf{MSA style vs. No MSA style}} & \multicolumn{2}{c|}{\textbf{Empl. $\le$ 100 vs. \textgreater{} 100}}\tabularnewline \hline
\textbf{\textbf{Survey Question Statement}}& \textbf{SQ\#} & Mean value & P-value & Effect size & P-value & Effect size & P-value & Effect size\tabularnewline
\hline 
\multicolumn{9}{|c|}{\cellcolor{blue!10}\scriptsize{MSA architecting activities (Strongly agree (5), Agree (4), Neutral
(3), Disagree (2), Strongly disagree (1))}}\tabularnewline
\hline 
Architectural analysis & SQ14 & 3.36 & 0.55 & 0.29 & 0.12 & -0.19 & 0.93 & -0.13\tabularnewline
\hline 
Architectural synthesis & SQ15 & 3.26 & 0.63 & 0.31 & 0.23 &-0.03 & 0.96 & 0.01\tabularnewline
\hline 
Architectural evaluation & SQ16 & 3.39 & 0.59 & -0.05 & 0.23 & 0.13 & 0.96 &-0.23\tabularnewline
\hline 
Architectural implementation & SQ17 & 3.38 & 0.52 & -0.16 & 0.13 & -0.32 & 0.95 & 0.09\tabularnewline
\hline 
Architectural maintenance and evolution & SQ18 & 3.40 & \faWrench{} \textbf{0.03} & \cellcolor{gray!20}-0.11 & \faGears{} \textbf{0.04} & \cellcolor{gray!50}0.21 & 0.89 & -0.17\tabularnewline
\hline 
\multicolumn{9}{|c|}{\cellcolor{blue!10}\scriptsize{Approaches used for describing the architecture of microservices systems
(Very often (5), Often(4), Sometimes (3), Rarely (2), Never (1))}}\tabularnewline
\hline 
Boxes and lines & \multirow{4}{*}{SQ19} & 3.35 & 0.33 & -1.05 & 0.12 & 1.10 & 0.91 & -0.91\tabularnewline
\cline{1-1} \cline{3-9} 
Unified modeling language &  & 3.12 & 0.15 & 0.13 & \faGears{} \textbf{0.01} & \cellcolor{gray!50}0.42 & 0.90 & 0.27\tabularnewline
\cline{1-1} \cline{3-9} 
Domain specific languages &  & 2.57 & \faWrench{} \textbf{0.04} & \cellcolor{gray!50}0.70 & \faGears{} \textbf{0.00} & \cellcolor{gray!50}0.09 & 0.91 & 0.50\tabularnewline
\cline{1-1} \cline{3-9} 
Architectural description languages &  & 2.22 & 0.46 & 0.58 & 0.06 & -0.59 & 0.94 & 0.45\tabularnewline
\hline 
\end{tabular}}

\end{table}

\textbf{Quality attributes}: \textit{Experience $\le$ 2 years vs. Experience \textgreater{} 2 years} (see Table \ref{tab:SigDiffQAs}): There are three survey question statements (i.e., \faWrench{} {\textbf{ Usability}}, \faWrench{} {\textbf{ Resilience}}, \faWrench{} {\textbf{Interoperability}}) with statistically significant differences between Experience $\le$ 2 years and Experience \textgreater{} 2 years groups. The Experience \textgreater{} 2 years group is more likely to agree on importance of Usability and Resilience for designing MSA than the Experience $\le$ 2 years group. On the other hand, the Experience $\le$ 2 years group is more likely to agree on importance of Interoperability for designing MSA than the Experience \textgreater{} 2 years group. \textit{MSA style vs. No MSA style}: There are seven survey question statements (i.e., \faGears{} {\textbf{Security}}, \faGears{} {\textbf{Availability}}, \faGears{} {\textbf{Scalability}}, \faGears{} {\textbf{Reliability}}, \faGears{} {\textbf{Maintainability}}, \faGears{} {\textbf{Compatibility}}, \faGears{} {\textbf{Resilience}}) with statistically significant differences between MSA style and No MSA style groups. The MSA style group is more likely to agree on importance of Security, Availability, Scalability, Reliability, Maintainability, and Compatibility for designing MSA than the No MSA style group. On the other hand, the No MSA style group is more likely to agree on importance of Resilience for designing MSA than the MSA style group. \textit{Employees size $\le$ 100 vs. Employees size \textgreater{} 100}: We identified two  survey question statements (i.e., \faGroup{} {\textbf{Functional suitability}}, \faGroup{} {\textbf{Resilience}}) with statistically significant differences between Employees size $\le$ 100 and Employees size \textgreater{} 100 groups. The Employees size $\le$ 100 group is more likely to agree on the importance of the Functional suitability and Resilience quality attributes for designing MSA than the Employees size \textgreater{} 100 group. On the other hand, we did not find any survey question statement about the importance of quality attributes with statistically significant difference for the Employees size \textgreater{} 100 group.

\begin{table}[H]
\caption{Statistically significant differences on the survey results about quality attributes in MSA design}
\hrule
\begin{centering}
\resizebox{\textwidth}{!}{\begin{tabular}{{|r|c|c|c|c|c|c|c|c|}}
\hline 
& &\textbf{Likert Distro.} & \multicolumn{2}{c|}{\textbf{Exper. $\le$ 2 vs. \textgreater{} 2 years}} & \multicolumn{2}{c|}{\textbf{MSA style vs. No MSA style}} & \multicolumn{2}{c|}{\textbf{Empl. $\le$ 100 vs. \textgreater{} 100}}\tabularnewline \hline
\textbf{\textbf{Survey Question Statement}}& \textbf{SQ\#} & Mean value & P-value & Effect size & P-value & Effect size & P-value & Effect size\tabularnewline
\hline 
\multicolumn{9}{|c|}{\cellcolor{blue!10}\scriptsize{Quality attribute when designing MSA-based systems (Very important
(5), Important(4), Somewhat important (3), Important (2), Not sure
(1))}}\tabularnewline
\hline 
Security & \multirow{15}{*}{SQ22} & 3.87 & 0.50 & -0.17 & \faGears{} \textbf{0.14} & \cellcolor{gray!50}0.10 & 0.80 & 0.00\tabularnewline
\cline{1-1} \cline{3-9} 
Availability &  & 3.87 & 0.14 & -0.40 & \faGears{} \textbf{0.15} & \cellcolor{gray!50}0.50 & 0.87 & -0.12\tabularnewline
\cline{1-1} \cline{3-9} 
Performance &  & 3.80 & 0.41 & -0.36 & 0.14 & 0.59 & 0.54 & -0.23\tabularnewline
\cline{1-1} \cline{3-9} 
Scalability &  & 3.69 & 0.38 & -0.21 & \faGears{} \textbf{0.09} & \cellcolor{gray!50}0.25 & 0.93 & -0.07\tabularnewline
\cline{1-1} \cline{3-9} 
Reliability &  & 3.52 & 0.29 & -0.45 & \faGears{} \textbf{0.05} & \cellcolor{gray!50}0.50 & 0.92 & -0.36\tabularnewline
\cline{1-1} \cline{3-9} 
Maintainability &  & 3.51 & 0.87 & -0.03 & \faGears{} \textbf{0.03} & \cellcolor{gray!50}0.18 & 0.78 & 0.00\tabularnewline
\cline{1-1} \cline{3-9} 
Usability &  & 3.43 & \faWrench{} \textbf{0.03} & \cellcolor{gray!20}-0.05 & 0.15 & 0.09 & 0.78 & -0.01\tabularnewline
\cline{1-1} \cline{3-9} 
Compatibility &  & 3.42 & 0.75 & 0.36 & \faGears{} \textbf{0.01} & \cellcolor{gray!50}0.36 & 0.88 & -0.05\tabularnewline
\cline{1-1} \cline{3-9} 
Testability &  & 3.42 & 0.90 & -0.20 & 0.05 & 0.53 & 0.97 & 0.06\tabularnewline
\cline{1-1} \cline{3-9} 
Monitorability &  & 3.40 & 0.32 & -0.25 & 0.06 & 0.02 & 0.62 & -0.11\tabularnewline
\cline{1-1} \cline{3-9} 
Functional suitability &  & 3.36 & 0.34 & 0.15 & 0.12 & -0.05 & \faGroup{} 0.04 & \cellcolor{gray!50}0.14\tabularnewline
\cline{1-1} \cline{3-9} 
Reusability &  & 3.41 & 0.19 & -0.09 & 0.07 & \cellcolor{gray!50}0.36 & 0.72 & -0.07\tabularnewline
\cline{1-1} \cline{3-9} 
Resilience &  & 3.20 & \faWrench{} \textbf{0.04} & \cellcolor{gray!20} -0.04 & \faGears{} \textbf{0.00} & \cellcolor{gray!20} -0.07 & \faGroup{} \textbf{0.03} & \cellcolor{gray!50}0.17\tabularnewline
\cline{1-1} \cline{3-9} 
Portability &  & 3.28 & 0.18 & 0.14 & 0.09 & 0.47 & 0.56 & 0.16\tabularnewline
\cline{1-1} \cline{3-9} 
Interoperability &  & 3.13 & \faWrench{} \textbf{0.03} & \cellcolor{gray!50}0.35 & 0.05 & 0.32 & 0.59 & 0.04\tabularnewline
\hline 
\end{tabular}}
\par\end{centering}
\label{tab:SigDiffQAs}
\end{table}

\textbf{MSA design patterns}: \textit{Experience $\le$ 2 years vs. Experience \textgreater{} 2 years} (see Table \ref{tab:SigDifPatterns}): There are four survey question statements (i.e., \faWrench{} {\textbf{Service registry}}, \faWrench{} {\textbf{Serverless deployment}}, \faWrench{} {\textbf{Polling publisher}}, \faWrench{} {\textbf{Shared database}}) with statistically significant differences between Experience $\le$ 2 years and Experience \textgreater{} 2 years groups. The Experience \textgreater{} 2 years group is more likely to use Service registry, Serverless deployment, Polling publisher, and Shared database patterns for designing MSA than the Experience $\le$ 2 years group. On the other hand, we did not find any survey question statement about the use of MSA design patterns with statistically significant differences for Experience $\le$ 2 years group. \textit{MSA style vs. No MSA style}: There are sixteen survey question statements (e.g., \faGears{} {\textbf{Backend for frontend}}, \faGears{} {\textbf{Access token}}, \faGears{} {\textbf{Service registry}}) with statistically significant differences between MSA style and No MSA style groups. We identified that the No MSA style group is more likely to use patterns (e.g., Backend for frontend, Access token, Service registry) for designing MSA than the MSA style group. On the other hand, we did not find any survey question statement about using MSA design patterns with statistically significant differences for the MSA style group. \textit{Employees size $\le$ 100 vs. Employees size \textgreater{} 100}: We identified three survey question statements (i.e., \faGroup{} {\textbf{Remote procedure invocation}}, \faGroup{} {\textbf{Polling publisher}}, \faGroup{} {\textbf{Multiple service instance per host}}) with statistically significant differences between Employees size $\le$ 100 and Employees size \textgreater{} 100 groups. The Employees size \textgreater{} 100 group is more likely to use Polling publisher and Multiple service instance per host  patterns for designing MSA than the Employees size $\le$ 100 group. Moreover, the Employees size $\le$ 100 group is more likely to use Remote procedure invocation pattern for designing MSA than the Employees size \textgreater{} 100.
\begin{table}[H]
\caption{Statistically significant differences on the survey results about MSA design patterns}

\label{tab:SigDifPatterns}
\hrule
\resizebox{\textwidth}{!}{\begin{tabular}{|r|c|c|c|c|c|c|c|c|}
\hline 
& &\textbf{Likert Distro.} & \multicolumn{2}{c|}{\textbf{Exper. $\le$ 2 vs. \textgreater{} 2 years}} & \multicolumn{2}{c|}{\textbf{MSA style vs. No MSA style}} & \multicolumn{2}{c|}{\textbf{Empl. $\le$ 100 vs. \textgreater{} 100}}\tabularnewline \hline
\textbf{\textbf{Survey Question Statement}}& \textbf{SQ\#} & Mean value & P-value & Effect size & P-value & Effect size & P-value & Effect size\tabularnewline
\hline 
\multicolumn{9}{|c|}{\cellcolor{blue!10}\scriptsize{MSA design patterns when designing MSA-based systems (Very important
(5), Important (4), Somewhat important (3), Important (2), Not sure
(1))}}\tabularnewline
\hline 
API gateway & \multirow{26}{*}{SQ23} & 3.67 & 0.48 & -0.37 & 0.07 & 0.26 & 0.94 & -0.53\tabularnewline
\cline{1-1} \cline{3-9} 
Backend for frontend &  & 3.29 & 0.23 & 0.02 & \faGears{} \textbf{0.00} & \cellcolor{gray!20}-0.54 & 0.90 & 0.05\tabularnewline
\cline{1-1} \cline{3-9} 
Access token &  & 3.26 & 0.31 & 0.27 &  \faGears{} \textbf{0.00} & \cellcolor{gray!20}-0.22 & 0.90 & -0.41\tabularnewline
\cline{1-1} \cline{3-9} 
Service instance per VM &  & 2.75 & 0.21 & 0.03 & 0.06 & -0.60 & 0.95 & -0.12\tabularnewline
\cline{1-1} \cline{3-9} 
Service registry &  & 3.13 &  \faWrench{} 0.01 & \cellcolor{gray!20}-0.15 &  \faGears{} \textbf{0.00} & \cellcolor{gray!20}-0.37 & 0.95 & -0.60\tabularnewline
\cline{1-1} \cline{3-9} 
API composition &  & 3.42 & 0.45 & -0.11 & 0.06 & -0.03 & 0.93 & -0.37\tabularnewline
\cline{1-1} \cline{3-9} 
Messaging &  & 3.21 & 0.32 & -0.20 & 0.06 & -0.19 & 0.90 & -0.07\tabularnewline
\cline{1-1} \cline{3-9} 
Remote procedure invocation &  & 3.03 & 0.19 & 0.12 &  \faGears{} \textbf{0.00} & \cellcolor{gray!20}-0.31 & \faGroup{} \textbf{0.02} & \cellcolor{gray!50}0.06\tabularnewline
\cline{1-1} \cline{3-9} 
Database per service &  & 3.26 & 0.40 & 0.06 &  \faGears{} \textbf{0.01} & \cellcolor{gray!20}-0.09 & 0.96 & -0.13\tabularnewline
\cline{1-1} \cline{3-9} 
Circuit breaker &  & 2.97 & 0.17 & -0.60 & 0.06 & -0.40 & 0.95 & -0.82\tabularnewline
\cline{1-1} \cline{3-9} 
Client-side discovery &  & 3.07 & 0.29 & 0.12 & \faGears{} \textbf{0.01} & \cellcolor{gray!20}-0.12 & 0.91 & -0.21\tabularnewline
\cline{1-1} \cline{3-9} 
Server-side discovery  &  & 2.92 & 0.21 & -0.42 &  \faGears{} \textbf{0.00} & \cellcolor{gray!20}-0.21 & 0.94 & -0.20\tabularnewline
\cline{1-1} \cline{3-9} 
Serverless deployment &  & 2.70 & \faWrench{} \textbf{0.01} & \cellcolor{gray!20} -0.16 &  \faGears{}  \textbf{0.00} & \cellcolor{gray!20}-0.52 & 0.93 & 0.21\tabularnewline
\cline{1-1} \cline{3-9} 
Service instance per host &  & 2.94 & 0.24 & 0.01 & 0.05 & 0.09 & 0.89 & -0.20\tabularnewline
\cline{1-1} \cline{3-9} 
Self-registration &  & 2.79 & 0.14 & 0.28 & \faGears{} \textbf{0.00} & \cellcolor{gray!20}-0.53 & 0.94 & \cellcolor{gray!50}0.02\tabularnewline
\cline{1-1} \cline{3-9} 
Polling publisher &  & 2.64 & \faWrench{} \textbf{0.02} & \cellcolor{gray!20} -0.29 & 0.24 & -0.68 &\faGroup{} 0.02 & -0.38\tabularnewline
\cline{1-1} \cline{3-9} 
Application events &  & 2.92 & 0.30 & 0.10 & \faGears{} \textbf{0.01} & -0.13 & 0.91 & -0.21\tabularnewline
\cline{1-1} \cline{3-9} 
Sagas &  & 2.60 & 0.29 & 0.07 & 0.22 & -0.81 & 0.96 & -0.22\tabularnewline
\cline{1-1} \cline{3-9} 
Service deployment platform &  & 2.88 & 0.40 & 0.22 & \faGears{} \textbf{0.01} & -0.15 & 0.95 & 0.06\tabularnewline
\cline{1-1} \cline{3-9} 
Multiple service instance per host &  & 2.90 & 0.22 & -0.06 & 0.32 & -0.17 & \faGroup{} \textbf{0.03} & \cellcolor{gray!20}-0.43\tabularnewline
\cline{1-1} \cline{3-9} 
Shared database &  & 2.93 & \faWrench{} \textbf{0.03} & -0.24 & \faGears{} \textbf{0.01} & -0.13 & 0.95 & 0.12\tabularnewline
\cline{1-1} \cline{3-9} 
Event sourcing &  & 2.65 & 0.26 & 0.18 & \faGears{} \textbf{0.01} & \cellcolor{gray!20} -0.32 & 0.89 & 0.12\tabularnewline
\cline{1-1} \cline{3-9} 
Command query responsibility segregation &  & 2.60 & 0.25 & -0.02 & 0.20 & -0.48 & 0.92 & -0.38\tabularnewline
\cline{1-1} \cline{3-9} 
Transaction log tailing &  & 2.75 & 0.26 & 0.28 & \faGears{} \textbf{0.01} & \cellcolor{gray!20}-0.55 & 0.95 & -0.16\tabularnewline
\cline{1-1} \cline{3-9} 
Service instance per container &  & 3.22 & 0.20 & -0.36 & \faGears{} \textbf{0.00} & \cellcolor{gray!20}-0.03 & 0.95 & -0.15\tabularnewline
\cline{1-1} \cline{3-9} 
Domain-specific protocol &  & 2.80 & 0.33 & 0.10 & \faGears{} \textbf{0.01} & \cellcolor{gray!20}-0.49 & 0.92 & -0.45\tabularnewline
\hline 
\end{tabular}}
\end{table}

\subsection{Monitoring of microservices systems (RQ2)}
\subsubsection{Monitoring metrics} 
We asked about thirteen monitoring metrics (SQ28) identified from the grey literature, e.g., \cite{Susan17, DaveSwersky2018monitoring, Apurva16}. According to participants' responses, the most popular monitoring metric is resource usage (e.g., CPU and memory usage) (71 out of 106, 66.9\%), followed by load balancing (57 out of 106, 53.7\%) and availability (56 out of 106, 52.8\%).

To understand the reasons behind using monitoring metrics, we asked IQ3.1.1 (see Table \ref{tab:IQmonitoring}). All the interviewees were familiar with the monitoring metrics. The interviewees have acknowledged that they mainly use resource usage, load balancing, and availability as the monitoring metrics. They argued that monitoring metrics are used to (i) provide a better customer experience (e.g., avoiding failure and improving performance), (ii) estimate the budget (e.g., payment according to the use of cloud infrastructure, such as Microsoft Azure, Amazon AWS), (iii) assess application behavior on high traffic time (e.g., Christmas), (iv) monitor the overall health of host infrastructure (e.g., memory/CPU utilization), (v) count the number of hosts and pods, (vi) detect problems with multi-threading, (vii) identify degraded states or failure points before the system entirely fails, (viii) improve the performance of microservices, and (ix) monitor the availability state (percentage) of a system for end-users. Following is one representative quotation about the monitoring metrics for microservices systems.

\faComment “\textit{The use of monitoring metrics allows us to get first-hand information about each microservice’s behaviour. Monitoring metrics can help to understand, visualize, and isolate failures points. We can also use monitoring metrics to check the efficiency of microservices systems}” \textit{\textbf{Application developer (P5)}}.

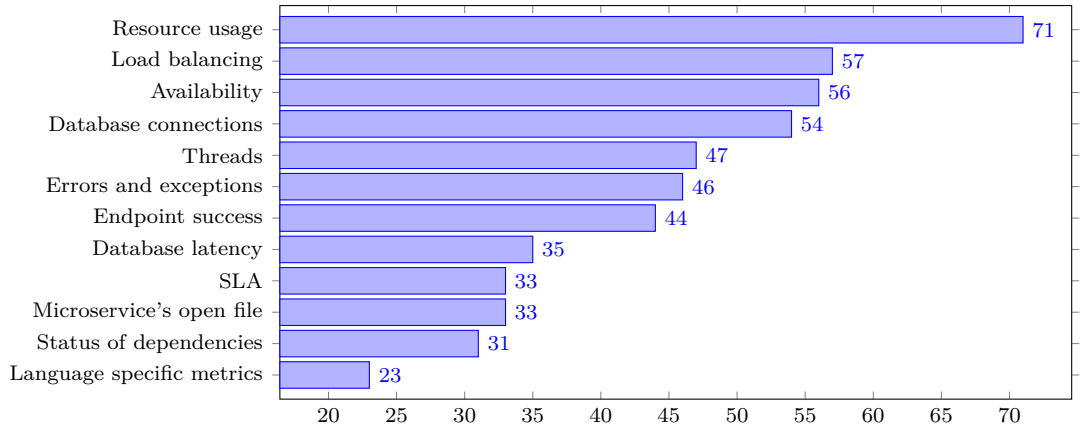
\begin{figure}[H]
\begin{centering}
\begin{tikzpicture}  
\begin{axis}[
	footnotesize,
	xbar, 
	width=12.0cm, height=6.8cm, 
	enlarge y limits=0.01,
	enlargelimits=0.07,  
	symbolic y coords={Language specific metrics, Status of dependencies, Microservice’s open file, SLA, Database latency, Endpoint success, Errors and exceptions, Threads, Database connections, Availability, Load balancing, Resource usage},
	xmin=20,
	xtick distance=5,
	xmax=71,
	ytick=data,
	nodes near coords, nodes near coords align={horizontal},
	every node near coord/.append style={font=\footnotesize},
]
	\addplot coordinates {(71,Resource usage) (57,Load balancing) (56,Availability) (54,Database connections) (47,Threads) (46,Errors and exceptions) (44,Endpoint success) (35,Database latency) (33,Microservice’s open file) (33,SLA) (31,Status of dependencies) (23,Language specific metrics)};
\end{axis}
\end{tikzpicture}  
\caption{Metrics used for monitoring microservices systems}
\label{fig:SQ29}
\end{centering}
\end{figure}

The other popular monitoring metrics reported by the participants are database connections (54 out of 106, 50.9\%), threads (47 out of 106, 44.3\%), errors and exceptions (46 out of 106, 43.3\%), and endpoints success (44 out of 106, 41.5\%) (see Figure \ref{fig:SQ29} for more details).

\subsubsection{Monitoring practices} 
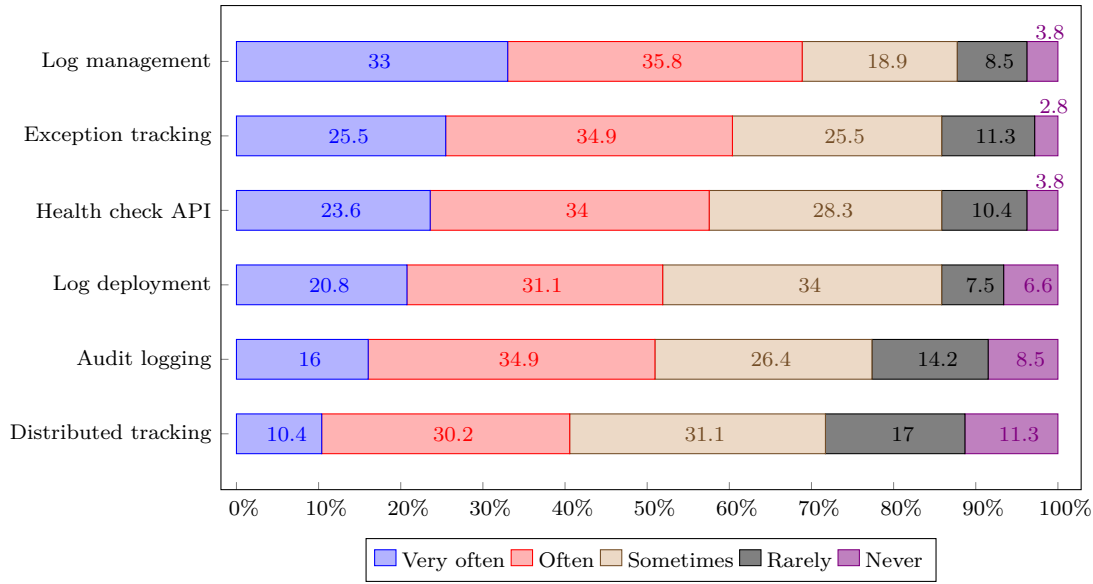
\begin{figure}[H]
\begin{centering}
\begin{tikzpicture}
  \begin{axis}[
      footnotesize,
      xbar stacked,
      width=12.9cm, height=8cm, 
      bar width=15pt,
      nodes near coords={
        \pgfkeys{/pgf/fpu=true}
        \pgfmathparse{\pgfplotspointmeta / 106 * 100}
        $\pgfmathprintnumber[fixed, precision=1]{\pgfmathresult}$
        \pgfkeys{/pgf/fpu=false}
      },
      nodes near coords custom/.style={
        every node near coord/.style={
          check for small/.code={
            \pgfkeys{/pgf/fpu=true}
            \pgfmathparse{\pgfplotspointmeta<#1}\%
            \pgfkeys{/pgf/fpu=false}
            \ifpgfmathfloatcomparison
              \pgfkeysalso{above=.5em}
            \fi
          },
          check for small,
        },
      },
      nodes near coords custom=6,
      xmin=-2, xmax=109,
      xtick={0, 10.6, ..., 106.1},
      ytick={1,...,7},
      yticklabels={Distributed tracking, Audit logging, Log deployment, Health check API, Exception tracking, Log management},
      xtick pos=bottom,
      ytick pos=left,
      xticklabel={
        \pgfkeys{/pgf/fpu=true}
        \pgfmathparse{\tick / 106 * 100}
        $\pgfmathprintnumber[fixed, precision=1]{\pgfmathresult}\%$
        \pgfkeys{/pgf/fpu=false}
      },
      enlarge y limits=.15,
      legend style={at={(0.5,-0.10)}, anchor=north, legend columns=-1},
    every node near coord/.append style={font=\footnotesize},
    ]
	
\addplot coordinates{(11,1) (17,2)  (22,3) (25,4) (27,5) (35,6)};
\addplot coordinates{(32,1) (37,2)  (33,3) (36,4) (37,5) (38,6)};
\addplot coordinates{(33,1) (28,2)  (36,3) (30,4) (27,5) (20,6)};
\addplot coordinates{(18,1) (15,2)  (8,3) (11,4) (12,5) (9,6)};
\addplot [color=violet, fill=violet!50] coordinates{(12,1) (9,2) (7,3) (4,4) (3,5) (4,6)};
\legend{\strut Very often, \strut Often, \strut Sometimes, \strut Rarely, \strut Never}
\end{axis}
\end{tikzpicture}
\caption{Practices used for monitoring microservices systems}
\label{fig:monitoringpractices}
\end{centering}
\end{figure}

We asked the participants which practices they use to monitor microservices systems (SQ29) about six monitoring practices that have been identified from grey literature (e.g., \cite{Carstensen_2019, DaveSwersky2018monitoring, rajput2018hands}), (see Figure~\ref{fig:monitoringpractices}). The participants mentioned the usage of the following three monitoring practices most often: log management (33.0\% very often, 35.8\% often, 18.9\% sometimes), exception tracking (25.5\% very often, 34.9\% often, 25.5\% sometimes), and health check API (23.6\% very often, 34.0\% often, 28.3\% sometimes).

To understand the reasons behind using the top three monitoring practices, we asked IQ3.2.1. We summarized the collected reasons for using the top three monitoring practices according to the responses from five out of six interviewees (i.e., P1, P2, P3, P4, P5): monitoring practices (i) provide the ability to track user activities for preventing, detecting, or minimizing the impact of a data breach, (ii) help to continue analyzing the stability of applications, (iii) help to perform the root cause analysis for specific issues, (iv) help to track the causes of errors and analyze security issues, (v) are used to track and notify the exceptions for developers, and (vi) show the operational status of the service and its interconnection with other services. Following is one representative quotation about the reason of using the top three monitoring practices.

\faComment “\textit{In my opinion, these three practices complement each other to efficiently manage the quality of each microservice in different operational environments. Besides that, the Health Check API shows the operational status of the service and its interconnection with other microservices.}” \textit{\textbf{Application developer (P5)}}.

On the other hand, distributed tracking (17.0\% rarely, 11.3\% never) and audit logging (14.2\% rarely, 8.5\% never) are less used (see Figure \ref{fig:monitoringpractices} for more details).



\subsubsection{Tools}
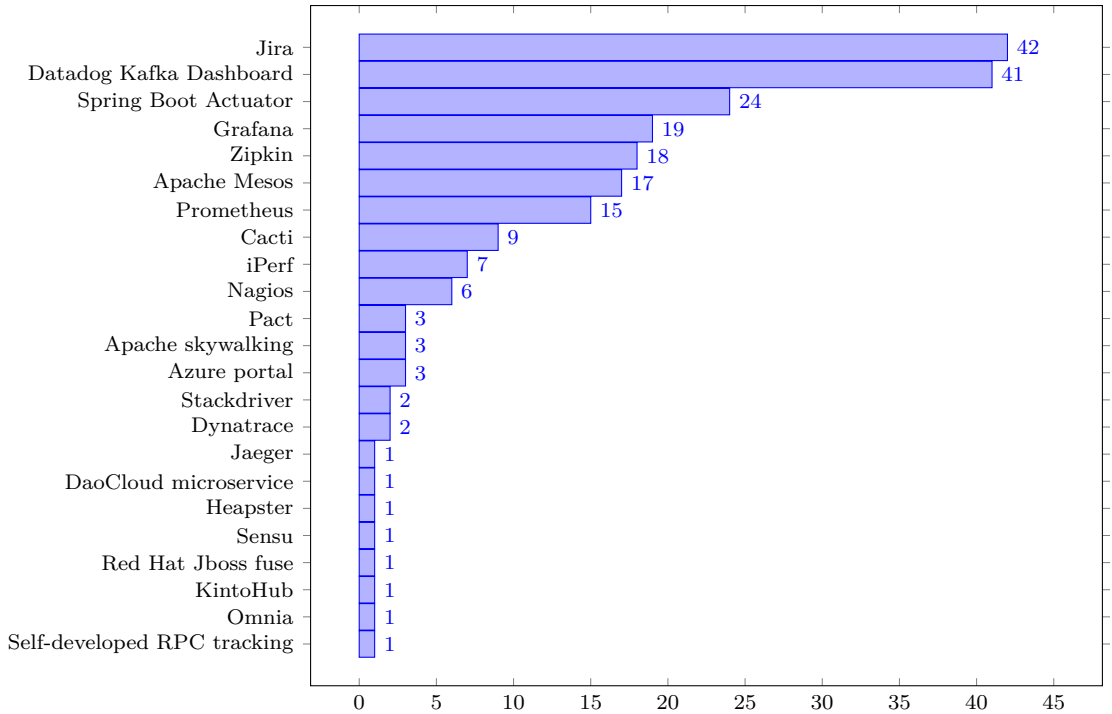
\begin{figure}[H]
\begin{centering}
\begin{tikzpicture}  
\begin{axis}[
	footnotesize,
	xbar, 
	width=12.0cm, height=10.6cm, 
	enlarge y limits=0.01,
	enlargelimits=0.07,  
	symbolic y coords={Self-developed RPC tracking, Omnia, KintoHub, Red Hat Jboss fuse, Sensu, Heapster, DaoCloud microservice, Jaeger, Dynatrace, Stackdriver, Azure portal, Apache skywalking, Pact, Nagios, iPerf, Cacti, Prometheus, Apache Mesos, Zipkin, Grafana, Spring Boot Actuator, Datadog Kafka Dashboard, Jira},
	ytick=data,
	xmin=0,
	xtick distance=5,
	xmax=45,
	nodes near coords, nodes near coords align={horizontal},
	every node near coord/.append style={font=\footnotesize},
]
	\addplot coordinates {(42,Jira) (41,Datadog Kafka Dashboard) (24,Spring Boot Actuator) (19,Grafana) (18,Zipkin) (17,Apache Mesos) (15,Prometheus) (9,Cacti) (7,iPerf) (6,Nagios) (3,Pact) (3,Apache skywalking) (3,Azure portal) (2,Stackdriver) (2,Dynatrace) (1,Jaeger) (1,DaoCloud microservice) (1,Heapster) (1,Sensu) (1,Red Hat Jboss fuse) (1,KintoHub) (1,Omnia) (1,Self-developed RPC tracking)};

\end{axis}
\end{tikzpicture}  
\caption{Tools used for monitoring microservices systems}
\label{fig:monitoringtools}
\end{centering}
\end{figure}

A vast majority of the participants indicated Jira (42 out of 106, 39.6\%) and Datadog Kafka Dashboard (41 out of 106, 38.6\%) as their monitoring tools, which are followed by Spring Boot Actuator (24 out of 106, 22.6\%), Grafana (19 out of 106, 17.9\%), and Zipkin (18 out of 106, 16.9\%) (see Figure \ref{fig:monitoringtools} for details). These tools are mainly used to monitor network performance (e.g., speed, bandwidth), resource usage (e.g., CPU, memory), and errors and exceptions for microservices systems. We also received 17 short answers in the free text field of survey question SQ30, where the participants provided 11 monitoring tools that they used to monitor microservices systems. 
These tools are Apache skywalking\footnote{\url{https://skywalking.apache.org/}} (3 responses), Azure portal\footnote{\url{https://azure.microsoft.com/en-us/features/azure-portal/}} (3 responses), Stackdriver\footnote{\url{https://cloud.google.com/stackdriver/}} (2 responses), Dynatrace\footnote{\url{https://www.dynatrace.com/}} (2 responses), Jaeger\footnote{\url{https://www.jaegertracing.io/}} (1 response), DaoCloud microservice platform\footnote{\url{http://blog.daocloud.io/dmp/}} (1 response), Heapster\footnote{\url{https://github.com/kubernetes-retired/heapster}} (1 response), Sensu\footnote{\url{https://sensu.io/}} (1 response), Red Hat Jboss fuse\footnote{\url{https://www.redhat.com/en/technologies/jboss-middleware/fuse}} (1 response), KintoHub\footnote{\url{https://www.kintohub.com/}} (1 response), and self-developed RPC tracking tool (1 response).

Besides that, we also asked IQ3.3.1 about the use of Jira as a monitoring tools. Out of the six interviewees, three interviewees stated that Jira could be integrated with Integrated Development Environment (IDE) to monitor microservices systems. Two other interviewees considered issue tracking as part of monitoring of microservices systems and other types of systems, and they said that Jira as an issue tracking tool can be considered as a monitoring tool. One interviewee was not sure about the use of Jira for monitoring purposes, and he mentioned that “\textit{we are using Dynatrace for monitoring microservices systems in our company}”.

\begin{tcolorbox}[colback=gray!5!white,colframe=gray!75!black,title=Key Findings of RQ2]
\justify
\textbf{Finding 13.} Resource usage, load balancing, and availability are the most frequently used monitoring metrics for microservices systems.\\
\textbf{Finding 14.} Log management, exception tracking, and health check API are the most frequently employed monitoring practices for microservices systems.
\end{tcolorbox}

\subsubsection{Challenges and solutions} 
\textbf{Challenges}: The respondents were asked to indicate what monitoring challenges they faced (i.e., SQ31). The monitoring challenges listed as answers of SQ31 were collected from the academic and grey literature \cite{waseemMSAdevops, Neependra17, heinrich2017performance}. The third column of Table \ref{tab:monitoringChallenges} shows the percentage of the practitioners' responses for each challenge. The three main challenges that the participants faced during the monitoring of microservices systems are \textit{collection of monitoring metrics data and logs from containers} (58 out of 106, 54.7\%), \textit{distributed tracing} (48 out of 106, 45.2\%), and \textit{having many components to monitor (complexity)} (44 out of 106, 41.5\%). Other highly ranked challenges are \textit{performance monitoring} (43 out of 106, 40.5\%), \textit{analyzing the collected data} (37 out of 106, 34.9\%), and \textit{failure zone detection} (36 out of 106, 33.9\%).
    
{\renewcommand{\arraystretch}{1}
\footnotesize
\centering
\begin{longtable}{|c|l|c|}
\caption{Challenges faced during the monitoring of microservices systems (in \%)}
\label{tab:monitoringChallenges}
\\\hline
    & \multicolumn{1}{c|}{}                                        &                              \\
\multirow{-2}{*}{\textbf{ID}} & \multicolumn{1}{c|}{\multirow{-2}{*}{\textbf{Monitoring Challenges}}}   & \multirow{-2}{*}{\textbf{Responses}} \\ \hline
MC1 & Collection of monitoring metrics data and logs from containers   & \cellcolor[HTML]{548235}54.7 \\ \hline
MC2 & Distributed tracing                                          & \cellcolor[HTML]{548235}45.3 \\ \hline
MC3 & Having many components to monitor (complexity)               & \cellcolor[HTML]{548235}41.5 \\ \hline
MC4 & Performance monitoring                                       & \cellcolor[HTML]{548235}40.6 \\ \hline
MC5 & Analysing the collected data                                 & \cellcolor[HTML]{A9D08E}34.9 \\ \hline
MC6 & Failure zone detection                                       & \cellcolor[HTML]{A9D08E}34.0 \\ \hline
MC7 & Availability of the monitoring tools                         & \cellcolor[HTML]{C6E0B4}21.7 \\ \hline
MC8 & Monitoring of application running inside containers           & \cellcolor[HTML]{C6E0B4}21.7 \\ \hline
MC9 & Maintaining monitoring infrastructures                       & \cellcolor[HTML]{E2EFDA}17.0 \\ \hline
\end{longtable}}


We asked IQ3.4.1 about the reasons for the top three monitoring challenges. Four out of six interviewees (i.e., P1, P2, P3, P5) provided the following reasons about the monitoring challenges: (i) communication between hundreds of microservices, (ii) runtime monitoring of microservices systems often does not have a standardized infrastructure, (iii) different languages, databases, and frameworks for developing microservices, and (iv) logs and dataflows in different formats. Following is one representative quotation about the reason for the monitoring challenges.

\faComment “\textit{Indeed, these are significant challenges. However, I think this is due to the complex nature of microservices systems}” \textit{\textbf{Architect and Application developer (P2)}}.


{\renewcommand{\arraystretch}{1}
\centering
\footnotesize
\begin{longtable}{|c|c|c|c|c|c|}
\caption{The severity of the challenges on the monitoring of microservices systems (in \%)}
\label{tab:monitoringSeverityImpact}
\\\hline
 &
  \multicolumn{5}{c|}{\textbf{Severity Scale}} \\ \cline{2-6} 
\multirow{-2}{*}{\textbf{ID}} &
  \textbf{Catastrophic} &
  \textbf{Major} &
  \textbf{Moderate} &
  \textbf{Minor} &
  \textbf{Insignificant} \\ \hline
MC1 &
  \cellcolor[HTML]{E2EFDA}3.8 &
  \cellcolor[HTML]{E2EFDA}17.9 &
  \cellcolor[HTML]{70AD47}48.1 &
  \cellcolor[HTML]{C6E0B4}23.6 &
  \cellcolor[HTML]{E2EFDA}6.6 \\ \hline
MC2 &
  \cellcolor[HTML]{E2EFDA}3.8 &
  \cellcolor[HTML]{C6E0B4}25.5 &
  \cellcolor[HTML]{A9D08E}38.7 &
  \cellcolor[HTML]{C6E0B4}24.5 &
  \cellcolor[HTML]{E2EFDA}7.5 \\ \hline
MC3 & \cellcolor[HTML]{E2EFDA}5.7 & \cellcolor[HTML]{A9D08E}34.0 & \cellcolor[HTML]{A9D08E}33.0 & \cellcolor[HTML]{E2EFDA}17.0 & \cellcolor[HTML]{E2EFDA}10.4 \\ \hline
MC4 &
  \cellcolor[HTML]{E2EFDA}6.6 &
  \cellcolor[HTML]{A9D08E}35.8 &
  \cellcolor[HTML]{A9D08E}32.1 &
  \cellcolor[HTML]{C6E0B4}20.8 &
  \cellcolor[HTML]{E2EFDA}4.7 \\ \hline
MC5 &
  \cellcolor[HTML]{E2EFDA}6.6 &
  \cellcolor[HTML]{A9D08E}38.7 &
  \cellcolor[HTML]{A9D08E}38.7 &
  \cellcolor[HTML]{E2EFDA}12.3 &
  \cellcolor[HTML]{E2EFDA}3.8 \\ \hline
MC6 &
  \cellcolor[HTML]{E2EFDA}9.4 &
  \cellcolor[HTML]{A9D08E}31.1 &
  \cellcolor[HTML]{A9D08E}36.8 &
  \cellcolor[HTML]{E2EFDA}16.0 &
  \cellcolor[HTML]{E2EFDA}6.6 \\ \hline
MC7 &
  \cellcolor[HTML]{E2EFDA}3.8 &
  \cellcolor[HTML]{A9D08E}35.8 &
  \cellcolor[HTML]{A9D08E}35.8 &
  \cellcolor[HTML]{E2EFDA}18.9 &
  \cellcolor[HTML]{E2EFDA}5.7 \\ \hline
MC8 &
  \cellcolor[HTML]{E2EFDA}5.7 &
  \cellcolor[HTML]{E2EFDA}19.8 &
  \cellcolor[HTML]{70AD47}43.4 &
  \cellcolor[HTML]{C6E0B4}24.5 &
  \cellcolor[HTML]{E2EFDA}6.6 \\ \hline
MC9 &
  \cellcolor[HTML]{E2EFDA}7.5 &
  \cellcolor[HTML]{C6E0B4}24.5 &
  \cellcolor[HTML]{70AD47}44.3 &
  \cellcolor[HTML]{E2EFDA}17.0 &
  \cellcolor[HTML]{E2EFDA}6.6 \\ \hline
\end{longtable}
}

\textbf{Solutions for monitoring challenges - from interview and survey results}: We received 24 distinct answers as the solutions for addressing monitoring challenges (SQ33). After checking the answers, we found only 11 valid solutions. We excluded those solutions that were incomplete, inconsistent, or did not adequately address the challenges of monitoring microservices systems. The participants recommended ten tools and one in-house developed framework to address the challenges related to monitoring of microservices systems.
 \begin{itemize}
     \item \textbf{Monitoring tools and a self-developed framework}: We received five responses in which the participants named the tools and provided the purposes of using tools. An respondent said that \say{\textit{we use Prometheus and Spring Actuator for collecting, aggregates, and visualize the monitoring data}}  \textbf{\textit{Operational staff (R6)}}. Similarly, another participant shared that \say{\textit{our team uses ELK (Elasticsearch, Logstash, and Kibana) Stack to address the challenges related to a different kind of monitoring, for instance, application performance and log monitoring}} \textbf{\textit{DevOps engineer (R11)}}. In another example, a participant pointed out that \say{\textit{we use health check API tools to detect the health of microservices (e.g., failure zone detection)}} \textbf{\textit{DevOps engineer (R91)}}. Some participants only recommended a few tools without explaining the reasons behind using them in monitoring microservices systems. For example, Grafana\footnote{\url{https://grafana.com/}} (1 response), Zipkin\footnote{\url{https://zipkin.io/}} (2 responses), and Raygun APM\footnote{\url{https://raygun.com/platform/apm}} (2 responses). However, we identified that these tools could be used to address several challenges. For instance, MC1, MC2, MC4, and MC5 (see Table \ref{tab:MonitoringSol}). On the other hand, only one participant referred to their in-house developed framework to address the challenges of \textit{MC1: collection of monitoring metrics data}, \textit{MC6: failure zone detection}, and \textit{MC8: monitoring of application running inside containers}. However, this participant did not provide any reference for his developed framework. We also asked IQ3.4.2 about the solutions of monitoring challenges. Three out of six interviewees (i.e., P1, P3, P4) only suggested the names of several tools that they were using in their organizations and these tools were successfully used to address most of the monitoring challenges. The tool names are (i) Dynatrace, (ii) Elastic Stack, (iii) Amazon CloudWatch, (iv) Jaeger, (v) Graphana Dashboard, and (vi) Prometheus.
\end{itemize}
     
{\renewcommand{\arraystretch}{1}
 \footnotesize
 \begin{longtable}{|p{2.5cm}|p{5.5cm}|l|c|}
    \caption{Proposed solutions to address the monitoring challenges}
    \label{tab:MonitoringSol}
    \\ \hline
\hline
\multicolumn{1}{|l|}{\textbf{Type}} &
  \textbf{Proposed   Solution} &
  \textbf{\begin{tabular}[c]{@{}l@{}}Challenge ID\\Addressed\end{tabular}} &
  \textbf{Count} \\ \hline
\multirow{7}{*}{\begin{tabular}[c]{@{}c@{}}Monitoring tools \\ and frameworks\end{tabular}} &
  ELK   (Elasticsearch, Logstash, and Kibana) Stack &
  MC1, MC4, MC5 &
  \multirow{7}{*}{11} \\ \cline{2-3}
 & Prometheus   and Spring Boot Actuator  & MC1 &  \\ \cline{2-3}
 & Grafana                  & MC1 &  \\ \cline{2-3}
 & Zipkin                   & MC2 &  \\ \cline{2-3}
 & Raygun APM               & MC4 &  \\ \cline{2-3}
 & Health check   API tools & MC6 &  \\ \cline{2-3}
 & In-house   framework     & MC6 &  \\ \hline
    \end{longtable}
    }

\begin{tcolorbox}[colback=gray!5!white,colframe=gray!75!black,title=Key Findings of RQ2]
\justify
\textbf{Finding 15.} The most prominent challenges of monitoring microservices systems are collection of monitoring metrics data and logs from containers, and distributed tracing.
\end{tcolorbox}

\subsubsection{Differences between the subpopulations of the study for monitoring of microservices systems}
\label{GapMSAmonitoring}
We analyzed the answers to Likert scale survey questions related to monitoring microservices in Table \ref{tab:SigDifMonitorPractices} and Table \ref{tab:SigDifMonitorChallenges}. In the following, we report the statistical comparison for monitoring practices and challenges of microservices systems.

\textbf{Monitoring practices} (see Table \ref{tab:SigDifMonitorPractices}): \textit{Experience $\le$ 2 years vs. Experience \textgreater{} 2 years}: There are two survey question statements (i.e., \faWrench{} {\textbf{Health check API}}, \faWrench{} {\textbf{Log deployment and changes}}) with statistically significant differences between Experience $\le$ 2 years and Experience \textgreater{} 2 years groups. The Experience \textgreater{} 2 years group is more likely to use Health check API and Log deployment and changes for monitoring microservices systems than the Experience $\le$ 2 years group. On the other hand, we did not find any survey question statement about using monitoring practices with statistically significant differences for Experience $\le$ 2 years group. \textit{MSA style vs. No MSA style}: There are two survey question statements (i.e., \faGears{} {\textbf{Audit logging}}, \faGears{} {\textbf{Distributed tracking}}) with statistically significant differences between MSA style and No MSA style groups. The No MSA style group is more likely to use Audit logging and Distributed tracking for monitoring microservices systems than the MSA style group. However, we did not find any survey question statement about using monitoring practices with statistically significant differences for the MSA style group. \textit{Employees size $\le$ 100 vs. Employees size \textgreater{} 100}: We identified one survey question statement (i.e., \faGroup{} {\textbf{Distributed tracking}}) with statistically significant difference between Employees size $\le$ 100 and Employees size \textgreater{} 100 groups. The Employees size \textgreater{} 100 group is more likely to use Distributed tracking for monitoring microservices systems than the Employees size $\le$ 100 group. Moreover, we did not find any survey question statement about using monitoring practices with statistically significant differences for the Employees size $\le$ 100 group.


\begin{table}[H]
    \centering
\caption{Statistically significant differences on the survey results about monitoring practices}

\label{tab:SigDifMonitorPractices}
\resizebox{\textwidth}{!}{\begin{tabular}{|r|c|c|c|c|c|c|c|c|}
\hline 
& &\textbf{Likert Distro.} & \multicolumn{2}{c|}{\textbf{Exper. $\le$ 2 vs. \textgreater{} 2 years}} & \multicolumn{2}{c|}{\textbf{MSA style vs. No MSA style}} & \multicolumn{2}{c|}{\textbf{Empl. $\le$ 100 vs. \textgreater{} 100}}\tabularnewline \hline
\textbf{\textbf{Survey Question Statement}}& \textbf{SQ\#} & Mean value & P-value & Effect size & P-value & Effect size & P-value & Effect size\tabularnewline
\hline 
\multicolumn{9}{|c|}{\cellcolor{blue!10}\scriptsize{Practices to monitor microservices systems (Very important
(5), Important (4), Somewhat important (3), Important (2), Not sure
(1))}}\tabularnewline
\hline 
Log management & \multirow{7}{*}{SQ29} & 3.89 & 0.51 & -0.42 & 0.10 & 0.93 & 0.94 & -0.54\tabularnewline
\cline{1-1} \cline{3-9} 
Audit logging &  & 3.42 & 0.43 & -0.42 & \faGears{} \textbf{0.02} & \cellcolor{gray!20}-0.09 & 0.92 & -0.54\tabularnewline
\cline{1-1} \cline{3-9} 
Distributed tracking &  & 3.13 & 0.53 & -0.47 & \faGears{} \textbf{0.02} & \cellcolor{gray!20}-0.18 & \faGroup{} \textbf{0.02} & \cellcolor{gray!20}-0.38\tabularnewline
\cline{1-1} \cline{3-9} 
Exception tracking &  & 3.73 & 0.60 & -0.31 & 0.07 & 0.41 & 0.93 & -0.01\tabularnewline
\cline{1-1} \cline{3-9} 
Health check API &  & 3.64 & \faWrench{} \textbf{0.02} & \cellcolor{gray!20}-0.40 & 0.05 &0.24 & 0.89 & 0.00\tabularnewline
\cline{1-1} \cline{3-9} 
Log deployment and changes &  & 3.54 & \faWrench{}  \textbf{0.03} & \cellcolor{gray!20}-0.17 & 0.05 & 0.17 & 0.94 & 0.37\tabularnewline
\hline 
\end{tabular}}
\end{table}
\textbf{Monitoring challenges} (see Table \ref{tab:SigDifMonitorChallenges}): 
\textit{Experience $\le$ 2 years vs. Experience \textgreater{} 2 years}: There are two survey question statements (i.e., \faWrench{} {\textbf{MC3}}, \faWrench{} {\textbf{MC8}}) with statistically significant differences between Experience $\le$ 2 years and Experience \textgreater{} 2 years groups. The Experience $\le$ 2 years group is more likely to agree that MC3 (i.e., having many components to monitor) challenge has a severe impact on monitoring of microservices systems than the Experience \textgreater{} 2 years group. On the other hand, the Experience \textgreater{} 2 years group is more likely to agree that MC8 (i.e., monitoring of application running inside containers) challenge has a critical impact on monitoring of microservices systems than Experience $\le$ 2 years group. \textit{MSA style vs. No MSA style}: There are two survey question statements (i.e., \faGears{} {\textbf{MC4}}, \faGears{} {\textbf{MC6}}) with statistically significant differences between MSA style and No MSA style groups. The MSA style group is more likely to agree that MC4 (i.e., performance monitoring) and MC6 (i.e., failure zone detection) challenges have severe impact on monitoring of microservices systems than the No MSA style group. However, we did not find any survey question statement about monitoring challenges with statistically significant differences for the No MSA style group. \textit{Employees size $\le$ 100 vs. Employees size \textgreater{} 100}: We identified one survey question statement (i.e., \faGroup{} {\textbf{MC5}}) with statistically significant difference between Employees size $\le$ 100 and Employees size \textgreater{} 100 groups. The Employees size \textgreater{} 100 group is more likely to agree that MC5 (i.e., analyzing the collected data) challenge has a severe impact on monitoring of microservices systems than the Employees size $\le$ 100 group. Moreover, we did not find any survey question statement about monitoring challenges with statistically significant difference for the Employees size $\le$ 100 group.

\begin{table}[H]
    \centering
\caption{Statistically significant differences on the survey results about monitoring challenges}

\label{tab:SigDifMonitorChallenges}
\resizebox{\textwidth}{!}{\begin{tabular}{|r|c|c|c|c|c|c|c|c|}
\hline 
& &\textbf{Likert Distro.} & \multicolumn{2}{c|}{\textbf{Exper. $\le$ 2 vs. \textgreater{} 2 years}} & \multicolumn{2}{c|}{\textbf{MSA style vs. No MSA style}} & \multicolumn{2}{c|}{\textbf{Empl. $\le$ 100 vs. \textgreater{} 100}}\tabularnewline \hline
\textbf{\textbf{Survey Question Statement}}& \textbf{SQ\#} & Mean value & P-value & Effect size & P-value & Effect size & P-value & Effect size\tabularnewline
\hline 
\multicolumn{9}{|c|}{\cellcolor{blue!10}\scriptsize{Severity of each challenge related to the monitoring of MSA-based
systems (Catastrophic (5), Major (4), Moderate (3), Minor (2), Insignificant(1))}}\tabularnewline
\hline 
MC1 & \multirow{9}{*}{SQ32} & 3.15 & 0.16 & -0.03 & 0.09 & -0.02 & 0.94 & -0.13\tabularnewline
\cline{1-1} \cline{3-9} 
MC2 &  & 2.94 & 0.44 & -0.21 & 0.06 & 0.03 & 0.94 & 0.23\tabularnewline
\cline{1-1} \cline{3-9} 
MC3 &  & 3.09 & \faWrench{}  \textbf{0.04} & \cellcolor{gray!50}0.21 & 0.08 & 0.99 & 0.94 & -0.25\tabularnewline
\cline{1-1} \cline{3-9} 
MC4 &  & 3.20 & 0.14 & 0.04 & \faGears{}  \textbf{0.03} & \cellcolor{gray!50}-0.43 & 0.94 & -0.20\tabularnewline
\cline{1-1} \cline{3-9} 
MC5 &  & 3.33 & 0.19 & -0.09 & 0.14 & -0.06 & \faGroup{}  \textbf{0.04} & \cellcolor{gray!20}-0.01\tabularnewline
\cline{1-1} \cline{3-9} 
MC6 &  & 3.24 & 0.12 & -0.12 & \faGears{}  \textbf{0.04} & \cellcolor{gray!50}-0.08 & 0.97 & 0.10\tabularnewline
\cline{1-1} \cline{3-9} 
MC7 &  & 2.89 & 0.19 & -0.04 & 0.12 & 0.26 & 0.95 & -0.16\tabularnewline
\cline{1-1} \cline{3-9} 
MC8 &  & 2.94 & \faWrench{}  \textbf{0.02} & \cellcolor{gray!20}-0.06 & 0.07 & -0.04 & 0.94 & 0.04\tabularnewline
\cline{1-1} \cline{3-9} 
MC9 &  & 3.09 & 0.15 & 0.08 & 0.08 & 0.41 & 0.94 & 0.04\tabularnewline
\hline 
\end{tabular}}
\end{table}


\subsection{Testing of microservices systems (RQ3)}

\subsubsection{Testing strategies}
We asked SQ34 to collect information relevant to the use of testing strategies for microservices systems \cite{waseemtestingMSA, Renz2016, JakeLumetta2018MSATesting, clemson2014}. The top three most frequently used testing strategies are: Unit testing (34.0\% very often, 29.2\% often, and 10.4\% sometimes), E2E testing (26.4\% very often, and 29.2\% often, 10.4\% sometimes), and Integration testing (23.6\% very often, 35.8\% often, and 6.6\% sometimes). 

We asked IQ4.1.1 (see Table \ref{tab:IQMsatesting}) about the reasons for using the top three testing strategies. All of the interviewees were familiar with these testing strategies. They also acknowledged that they frequently use these testing strategies to test microservices systems. According to the interviewees, these are fundamental testing strategies that can be used for all type of systems (e.g., MSA-based systems, SOA-based systems). In the following, we summarized the collected reasons. Concerning the reasons of using Unit testing, all the interviewees mentioned that they use this strategy because Unit testing (i) allows isolated testing of the design components of microservices systems (e.g., services, domains, service interfaces), (ii) makes sure each microservice working correctly and delivering the required functionalities, (iii) helps to identify the problems with individual microservice in the early stage of system development. About the reasons of using integration testing, four out of six interviewees (i.e., P2, P3, P5, P6) stated that this strategy integrates microservices together to verify that (i) microservices are working collaboratively to achieve the business goals, (ii) communication paths between microservices are working correctly, and (iii) third-party APIs and microservices are working together as expected. About the reasons of using E2E testing, three out of six interviewees (i.e., P2, P5, P6) stated that they typically perform E2E testing for testing the whole microservices systems, and the purposes of using E2E testing are to (i) ensure the correctness of the whole microservices systems and (ii) understand how the customers will use the developed systems. Following is one representative quotation about the reason of using testing strategies for microservices systems.

\faComment “\textit{The reasons for using these testing strategies are the same for microservices and monolithic systems. To the best of my knowledge. There is no specific testing technique that is designed or used in the industry specifically to test microservices systems}” \textit{\textbf{Architect and Application developer (P1)}}.

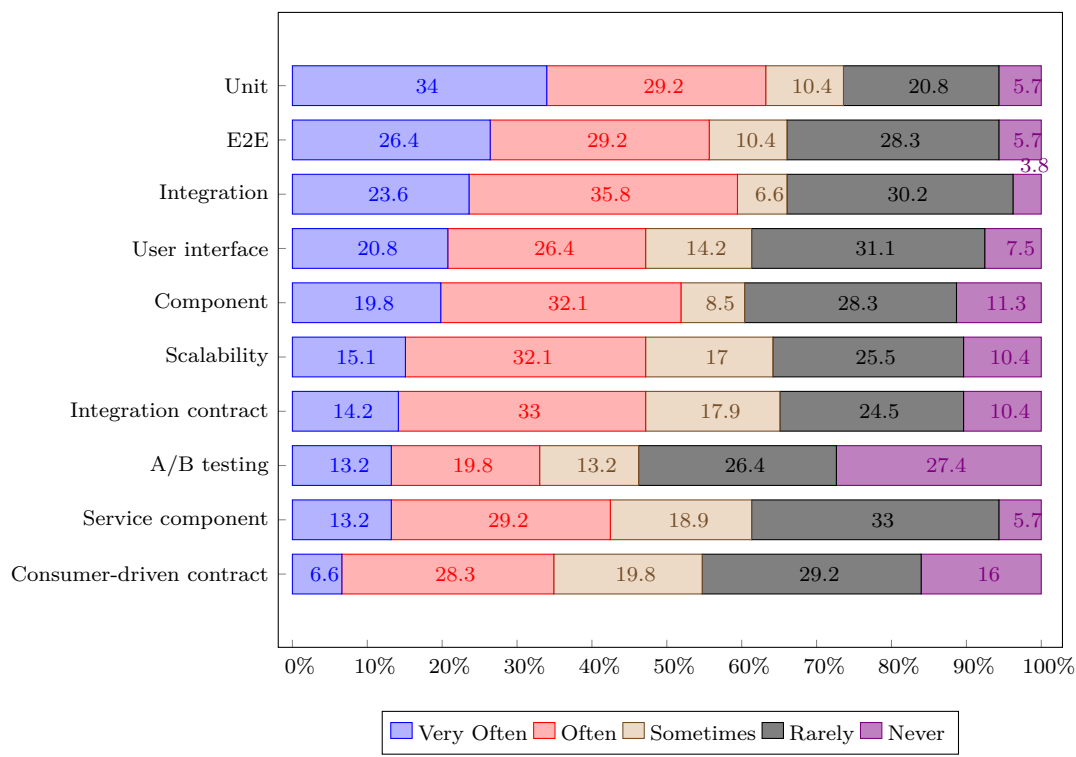
\begin{figure}[H]
\begin{centering}
\begin{tikzpicture}
\begin{axis}[
      footnotesize,
      xbar stacked,
      width=11.9cm, height=10cm, 
      bar width=15pt,
      nodes near coords={
        \pgfkeys{/pgf/fpu=true}
        \pgfmathparse{\pgfplotspointmeta / 106 * 100}
        $\pgfmathprintnumber[fixed, precision=1]{\pgfmathresult}$
        \pgfkeys{/pgf/fpu=false}
      },
      nodes near coords custom/.style={
        every node near coord/.style={
          check for small/.code={
            \pgfkeys{/pgf/fpu=true}
            \pgfmathparse{\pgfplotspointmeta<#1}\%
            \pgfkeys{/pgf/fpu=false}
            \ifpgfmathfloatcomparison
              \pgfkeysalso{above=.5em}
            \fi
          },
          check for small,
        },
      },
      nodes near coords custom=6,
      xmin=-2, xmax=109,
      xtick={0, 10.6, ..., 106.1},
      ytick={1,...,10},
      yticklabels={Consumer-driven contract, Service component, A/B testing, Integration contract, Scalability, Component, User interface, Integration, E2E, Unit},
      xtick pos=bottom,
      ytick pos=left,
      xticklabel={
        \pgfkeys{/pgf/fpu=true}
        \pgfmathparse{\tick / 106 * 100}
        $\pgfmathprintnumber[fixed, precision=1]{\pgfmathresult}\%$
        \pgfkeys{/pgf/fpu=false}
      },
      enlarge y limits=.15,
      legend style={at={(0.5,-0.10)}, anchor=north, legend columns=-1},
    every node near coord/.append style={font=\footnotesize},
    ]
\addplot coordinates{(7,1) (14,2) (14,3) (15,4) (16,5) (21,6) (22,7)(25,8) (28,9) (36,10)};
\addplot coordinates{(30,1) (31,2) (21,3) (35,4) (34,5) (34,6) (28,7)(38,8) (31,9) (31,10)};
\addplot coordinates{(21,1) (20,2) (14,3) (19,4) (18,5) (9,6) (15,7)(7,8) (11,9) (11,10)};
\addplot coordinates{(31,1) (35,2) (28,3) (26,4) (27,5) (30,6) (33,7)(32,8) (30,9) (22,10)};
\addplot [color=violet, fill=violet!50] coordinates{(17,1) (6,2) (29,3) (11,4) (11,5) (12,6) (8,7)(4,8) (6,9) (6,10)};
\legend{\strut Very Often, \strut Often, \strut Sometimes, \strut Rarely, \strut Never}
\end{axis}
\end{tikzpicture}
\caption{Testing strategies used to test microservices systems}
\label{fig:SQ30}
\end{centering}
\end{figure}

On the other hand, the following three testing strategies are used sparingly: A/B testing (26.4\% rarely, 27.4\% never), Consumer-driven contract testing (29.2\% rarely, 16.0\% never), and Service component testing (33.0\% rarely, 5.7\% never) (see Figure \ref{fig:SQ30} for details).

\subsubsection{Tools}
 
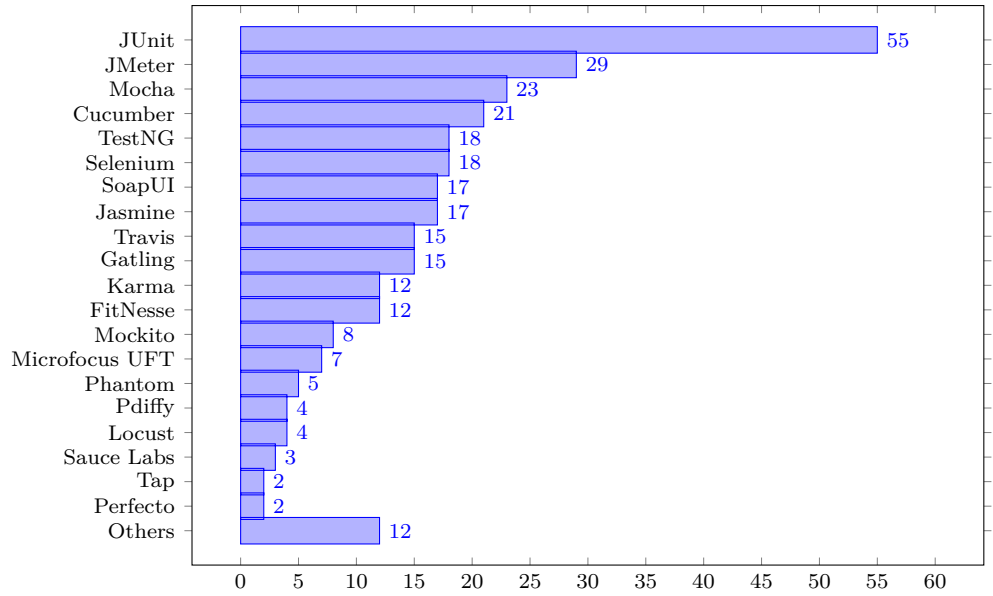
\begin{figure}[H]
\begin{centering}
\begin{tikzpicture}  
\begin{axis}[
	footnotesize,
	xbar, 
	width=12.0cm, height=9.0cm, 
	enlarge y limits=0.01,
	enlargelimits=0.07,  
	symbolic y coords={Others, Perfecto, Tap, Sauce Labs, Locust, Pdiffy, Phantom, Microfocus UFT, Mockito, FitNesse, Karma, Gatling, Travis, Jasmine, SoapUI, Selenium, TestNG, Cucumber, Mocha, JMeter, JUnit
},
	ytick=data,
	xmin=0,
	xtick distance=5,
	xmax=60,
	nodes near coords, nodes near coords align={horizontal},
	every node near coord/.append style={font=\footnotesize},
]
	\addplot coordinates {(55,JUnit) (29,JMeter) (23,Mocha) (21,Cucumber) (18,Selenium) (18,TestNG) (17,Jasmine) (17,SoapUI) (15,Gatling) (15,Travis) (12,FitNesse) (12,Karma) (8,Mockito) (7,Microfocus UFT) (5,Phantom) (4,Locust) (4,Pdiffy) (3,Sauce Labs) (2,Perfecto) (2,Tap) (12,Others)
};
\end{axis}
\end{tikzpicture}  
\caption{Tools used to test microservices systems}
\label{fig:SQ36}
\end{centering}
\end{figure}

We also asked SQ36 about the tools used for testing microservices systems. A large number of participants responded that they used JUnit (55 out of 106, 51.9\%), JMeter (29 out of 106, 27.3\%), and Mocha (23 out of 106, 21.7\%) (see Figure \ref{fig:SQ36} for details). 

We also asked IQ4.2.1 to understand the reasons for using the top three testing tools. Regarding JUnit and JMeter, all the interviewees agreed that these are fundamental testing tools that are required to test Java-based systems (including microservices systems). Moreover, three out of six interviewees (i.e., P2, P3, P5) mentioned that these tools also provide efficiency, simplicity, and agility in testing of microservices systems.

In the “Other" field of this survey question, the participants also mentioned several other tools. We classified these tools into the category of “Others" in Figure \ref{fig:SQ36}, including Nock (2 responses), Nunit (2 responses), Xunit (2 responses), Moq (2 responses), Jest (2 responses), Fluent assert (1 response), and Pact (1 response). We also received one response in which the participant mentioned that \textit{\say{We manually test microservices systems}}.

\begin{tcolorbox}[colback=gray!5!white,colframe=gray!75!black,title=Key Findings of RQ3]
\justify
\textbf{Finding 16.} The most commonly used testing strategies to test microservices systems are unit testing, E2E testing, and integration testing. These testing strategies are also used to test monolith and SOA-based systems. \\
\textbf{Finding 17.} Our results show that the most of the practitioners used Junit, JMeter, and Mocha tools to test microservices systems.
\end{tcolorbox}


\subsubsection{Skills}
\begin{figure}[H]
\begin{centering}
\begin{tikzpicture}  
\begin{axis}[
	footnotesize,
	xbar, 
	width=10.0cm, height=4.9cm, 
	enlarge y limits=0.01,
	enlargelimits=0.07,  
	symbolic y coords={Knowledge about Linux commands, Knowledge about test management tools, Knowledge about test automation tools, Analytically and logically thinking, Knowledge about multiple databases, Writing good unit test cases, Writing good integration test cases},
	ytick=data,
	xmin=30,
	xtick distance=5,
	xmax=70,
	nodes near coords, nodes near coords align={horizontal},
	every node near coord/.append style={font=\footnotesize},
]
	\addplot coordinates {(70,Writing good integration test cases) (63,Writing good unit test cases) (61,Knowledge about multiple databases)(58,Analytically and logically thinking) (54,Knowledge about test automation tools) (41,Knowledge about test management tools) (38,Knowledge about Linux commands)};
\end{axis}
\end{tikzpicture}  
\caption{Skills required to test microservices systems}
\label{fig:SQ38-skills}
\end{centering}
\end{figure}
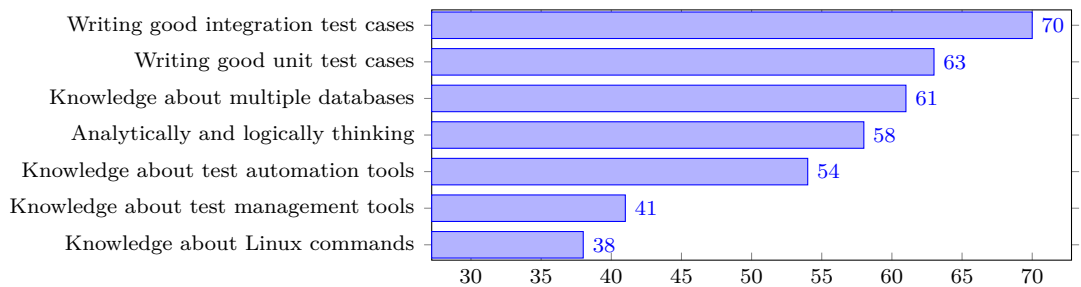

This survey also reports the skills required to test microservices systems properly (SQ38). Seven skills related to testing of microservices systems were collected after reviewing the relevant literature, e.g., \cite{xia2019practitioners, Bartosz, Tejaswini}. According to results of the survey, the top three skills that are required to test microservices systems are \textit{writing good integration test cases} (70 out of 106, 66\%), \textit{writing good unit test cases} (63 out of 106, 59.4\%), and having \textit{knowledge about multiple databases} (61 out of 106, 57.5\%).
Other highly-ranked skills are having \textit{analytically and logically thinking} (58 out of 106, 54.7\%) and \textit{knowledge about test automation tools} (54 out of 106, 50.9\%) (see Figure \ref{fig:SQ38-skills} for more details).

\subsubsection{Challenges and solutions}
\label{sec:ChallengeAndSolutions}
\textbf{Challenges}: Similar to SQ24 and SQ31, we asked the participants to select the challenges of testing microservices systems from a given list in SQ36. The list of the testing microservices challenges was prepared by reviewing the literature, e.g., \cite{rajput2018hands, waseemtestingMSA, JakeLumetta2018MSATesting, sotomayor2019comparison}. The third column of Table \ref{tab:TestingChallenges} shows percentage of responses for each challenge. 
Exactly half (53 out of 106, 50.0\%) of the participants reported that \textit{TC1: creating and implementing manual tests for microservices is challenging}. Other prominent challenges are \textit{TC2: Integration testing}(46 out of 106, 43.3\%), \textit{TC3: Debugging of microservices on container platforms} (45 out of 106, 42.4\%), and \textit{TC4: Creating and implementing automated tests} (41 out of 106, 38.6\%).  Moreover, some of the participants (16 out of 106, 15.0\%) felt that \textit{TC10: Polyglot technologies} could pose a challenge for testing of microservices systems.

{\renewcommand{\arraystretch}{1}
\footnotesize
\centering
\begin{longtable}{|l|p{8.7cm}|c|}
\caption{Challenges faced during the testing of microservices systems (in \%)}
\label{tab:TestingChallenges}
\\\hline
    & \multicolumn{1}{c|}{}                                          &                              \\
\multirow{-2}{*}{\textbf{ID}} & \multicolumn{1}{c|}{\multirow{-2}{*}{\textbf{Testing Challenges}}}                    & \multirow{-2}{*}{\textbf{Responses}} \\ \hline
TC1 & Creating and implementing manual tests & \cellcolor[HTML]{548235}50.0 \\ \hline
TC2 & Integration testing of microservices& \cellcolor[HTML]{548235}43.4 \\ \hline
TC3 & Debugging of microservices on container platforms         & \cellcolor[HTML]{548235}42.5 \\ \hline
TC4 & Creating and implementing automated tests& \cellcolor[HTML]{A9D08E}38.7 \\ \hline
TC5 & Independent microservices & \cellcolor[HTML]{A9D08E}32.1 \\ \hline
TC6 & Complexity and size of microservices systems                   & \cellcolor[HTML]{A9D08E}31.1 \\ \hline
TC7 & Understanding about each testable microservice                 & \cellcolor[HTML]{A9D08E}30.2 \\ \hline
TC8 & Performance testing of microservices& \cellcolor[HTML]{A9D08E}30.2 \\ \hline
TC9 & Many independent teams& \cellcolor[HTML]{C6E0B4}20.8 \\ \hline
TC10 & Polyglot technologies in microservices systems                & \cellcolor[HTML]{C6E0B4}15.1         \\ \hline
\end{longtable}}

Furthermore, we investigated the severity of each challenge on testing of microservices systems by asking the participants to rate them through a five-point Likert-scale: “catastrophic”, “major”, “moderate”, “minor”, and “insignificant” (see Table \ref{tab:testingchallengesimpact}). The Likert-scale points “catastrophic” and “major” are used to record the “major long-term impact” and “major short-term impact” of the challenges on testing of microservices systems. The Likert-scale point “moderate” is used to record “significant impact” of the challenges. In contrast, the Likert-scale points “minor” and “insignificant” are used to record the “short-term impact” and “minimal impact” of the challenges.
According to the results, all the challenges have “significant” or “major short-term impact” on testing of microservices systems, for instance, more than 56.6\% of the participants rated all the testing challenges as “major” or “moderate” on the severity scale, and a few (less than 10\%) of the participants reported that all the challenges have a “major long-term impact” on testing of microservices systems (see Table \ref{tab:testingchallengesimpact}).

We asked IQ4.3.1 about the reasons for the top three testing challenges. All the interviewees were agreed that these are valid challenges. Regarding the reasons of the challenge TC1, the interviewees stated that manual testing could become a problem due to the number of microservices and communication between them. Regarding the reasons of the challenge TC2, three interviewees (i.e., P2, P3, P5) mentioned that integration testing is challenging because of (i) multiple connecting points, (ii) limited knowledge of testers about all the microservices, and (iii) manually analyzing logs across multiple microservices. Regarding the reasons of the challenge TC3, two interviewees (i.e., P1, P2) pointed out that debugging of microservices on container platforms is challenging due to the lack of automatic debugging tool support. Following is one representative quotation about the reason for the testing challenges.

\faComment “\textit{Our major challenge is to write effective test cases for the integration testing of microservices systems, which requires extensive knowledge about each microservice and application domain.}” \textit{\textbf{Architect and Application developer (P1)}}.

On the other hand, many participants reported that the testing challenges have “short-term” or “minimal” impact on testing of microservices systems. For example, \textit{TC1: creating and implementing manual tests} (26.4\% minor, 11.3\% insignificant), \textit{TC5: independent microservices} (26.4\% minor, 9.4\% insignificant), and \textit{TC7: understanding about each testable microservice} (21.7\% minor, 8.5\% insignificant). Overall, we received contradictory results about severity of the challenges on testing of microservices systems.

{\renewcommand{\arraystretch}{1}
\footnotesize
\begin{longtable}{|c|c|c|c|c|c|}
\caption{The severity of the challenges on testing of microservices systems (in \%)}
\label{tab:testingchallengesimpact}
\\\hline
 &
  \multicolumn{5}{c|}{\textbf{Severity Scale}} \\ \cline{2-6} 
\multirow{-2}{*}{\textbf{ID}} &
  \textbf{Catastrophic} &
  \textbf{Major} &
  \textbf{Moderate} &
  \textbf{Minor} &
  \textbf{Insignificant} \\ \hline
TC1 &
  \cellcolor[HTML]{E2EFDA}5.7 &
  \cellcolor[HTML]{A9D08E}20.8 &
  \cellcolor[HTML]{70AD47}35.8 &
  \cellcolor[HTML]{A9D08E}26.4 &
  \cellcolor[HTML]{C6E0B4}11.3 \\ \hline
TC2 &
  \cellcolor[HTML]{E2EFDA}2.8 &
  \cellcolor[HTML]{70AD47}31.1 &
  \cellcolor[HTML]{70AD47}39.6 &
  \cellcolor[HTML]{A9D08E}17.9 &
  \cellcolor[HTML]{E2EFDA}8.5 \\ \hline
TC3 &
  \cellcolor[HTML]{E2EFDA}6.6 &
  \cellcolor[HTML]{70AD47}34.0 &
  \cellcolor[HTML]{70AD47}32.1 &
  \cellcolor[HTML]{A9D08E}18.9 &
  \cellcolor[HTML]{E2EFDA}8.5 \\ \hline
TC4 &
  \cellcolor[HTML]{E2EFDA}1.9 &
  \cellcolor[HTML]{70AD47}30.2 &
  \cellcolor[HTML]{70AD47}37.7 &
  \cellcolor[HTML]{A9D08E}22.6 &
  \cellcolor[HTML]{E2EFDA}7.5 \\ \hline
TC5 &
  \cellcolor[HTML]{E2EFDA}4.7 &
  \cellcolor[HTML]{A9D08E}21.7 &
  \cellcolor[HTML]{70AD47}37.7 &
  \cellcolor[HTML]{A9D08E}26.4 &
  \cellcolor[HTML]{E2EFDA}9.4 \\ \hline
TC6 &
  \cellcolor[HTML]{E2EFDA}7.5 &
  \cellcolor[HTML]{70AD47}34.0 &
  \cellcolor[HTML]{70AD47}36.8 &
  \cellcolor[HTML]{C6E0B4}16.0 &
  \cellcolor[HTML]{E2EFDA}5.7 \\ \hline
TC7 &
  \cellcolor[HTML]{E2EFDA}4.7 &
  \cellcolor[HTML]{A9D08E}22.6 &
  \cellcolor[HTML]{70AD47}42.5 &
  \cellcolor[HTML]{A9D08E}21.7 &
  \cellcolor[HTML]{E2EFDA}8.5 \\ \hline
TC8 &
  \cellcolor[HTML]{E2EFDA}6.6 &
  \cellcolor[HTML]{A9D08E}24.5 &
  \cellcolor[HTML]{70AD47}46.2 &
  \cellcolor[HTML]{C6E0B4}17.9 &
  \cellcolor[HTML]{E2EFDA}4.7 \\ \hline
TC9 &
  \cellcolor[HTML]{E2EFDA}9.4 &
  \cellcolor[HTML]{A9D08E}23.6 &
  \cellcolor[HTML]{70AD47}37.7 &
  \cellcolor[HTML]{A9D08E}20.8 &
  \cellcolor[HTML]{E2EFDA}8.5 \\ \hline
TC10 &
  \cellcolor[HTML]{E2EFDA}4.7 &
  \cellcolor[HTML]{A9D08E}25.5 &
  \cellcolor[HTML]{70AD47}43.4 &
  \cellcolor[HTML]{C6E0B4}19.8 &
  \cellcolor[HTML]{E2EFDA}6.6 \\ \hline
\end{longtable}}

\textbf{Solutions for testing challenges - from interview and  survey results}: We received 25 responses from practitioners for addressing testing challenges, in which only 13 responses are valid after our checking (see Table \ref{tab:TestingSol}). The participants recommended the following testing strategies, and approaches and guidelines to address the testing challenges. 
\begin{itemize}
    \item \textbf{Testing strategies}: Two participants \textbf{\textit{(DevOps engineer (R69) and Software quality engineer (R89))}} suggested that there should be test automation for alleviating the challenge of \textit{TC1: creating and implementing manual tests in microservices systems}. According to the participants, testing the whole microservices system is tricky and difficult in a manual way. They extended their answers by suggesting automation for testing strategies (e.g., unit, integration, E2E, and consumer-driven contract testing) with the help of DevOps. One \textbf{\textit{DevOps engineer}} also said that \say{\textit{manual testing has several issues with distributed systems like microservices, for example, test case coverage, testing the whole system, domain knowledge about the system under test, and many others. I must suggest enterprises should use test case and test management software tool like Testrail\footnote{\url{https://www.gurock.com/testrail/}} \textit{ \textbf{DevOps engineer (R69)}}}}.
    
    \item \textbf{Approaches and guidelines}: 
    We received three answers in which the participants \textbf{\textit{(System analyst (R24), Architect (R54), and Software quality engineer (R19))}} stressed upon the development of microservices systems through the Behavior-Driven Development (BDD) approach. According to the participants, the BDD approach can be used to address the challenges like \textit{implementing automated tests and integration testing}. Concerning to BDD, one \textbf{\textit{Software quality engineer}} said that \say{\textit{we have focused very much on BDD and want to explore this even further. Right now, we are very much doing unit level testing through BDD but want to grow out of it to other testing strategies (e.g., integration, end to end) with DevOps infrastructure (R19)}}.
    About addressing the testing challenge of \textit{TC7: understanding about each testable microservice}, we received several answers. For example, one other participant suggested that \say{\textit{you must ensure the project has an explicit requirement, required test cases are consistent and have enough coverage rate}} \textbf{\textit{Software quality engineer (R31)}}. We asked IQ4.3.2 about the solutions for testing challenges. In the following, we report the summary of the solutions received from the interviewees: (i) start testing from individual microservice with technically sound staff, (ii) each microservice should have its own database, (iii) introduce DevOps or CI/CD for continuous testing of microservices systems, (iv) write effective test cases according to the granularity of each microservice, and (v) thoroughly test microservices systems through unit, integration, component, and E2E testing strategies.
    \end{itemize}
         {\renewcommand{\arraystretch}{1}
 \footnotesize
 \begin{longtable}{|l|p{5.5cm}|l|c|}
    \caption{Proposed solutions to address the testing challenges}
    \label{tab:TestingSol}
   \\\hline
   \textbf{Type} &
  \textbf{Proposed   Solution} &
  \textbf{\begin{tabular}[c]{@{}l@{}}Challenge ID\\Addressed\end{tabular}} &
  \textbf{Count} \\ \hline
\multicolumn{1}{|c|}{Testing strategies} &
  \begin{tabular}[c]{@{}l@{}} Unit, Integration, E2E,\\ and Consumer-driven contract testing\end{tabular} &
  TC1, TC2 &
  \multirow{3}{*}{13} \\ \cline{1-3}
\multirow{1}{*}{\begin{tabular}[c]{@{}l@{}}Approaches and \\ guidelines\end{tabular}} &
  Behavior-driven development &
  TC2, TC4 &
   \\ \cline{2-3}
 &
  \begin{tabular}[c]{@{}l@{}}Test cases should have \\ enough coverage rate.\end{tabular} &
  TC7 &
   \\ \hline
    \end{longtable}
    }




\begin{tcolorbox}[colback=gray!5!white,colframe=gray!75!black,title=Key Findings of RQ3]
\justify

\textbf{Finding 18.} Most of the practitioners considered creating and implementing manual tests, integration testing, and debugging of the microservices deployed on container platforms a challenging job.
\end{tcolorbox}

\subsubsection{Differences between the subpopulations of the study testing of microservices systems}
\label{GapsMSAtesting}

We analyzed the answers to Likert scale survey questions related to testing of microservices systems in Table \ref{tab:SigDiffTestingStrategies} and Table \ref{tab:SigDiffTestingchallenges}.   In  the  following,  we  report  the  statistical comparison for testing strategies and testing challenges of microservices systems.

\textbf{Testing strategies} (see Table \ref{tab:SigDiffTestingStrategies}): \textit{Experience $\le$ 2 years vs. Experience \textgreater{} 2 years}: There are three survey question statements (i.e., \faWrench{} {\textbf{Component testing}}, \faWrench{} {\textbf{User interface testing}}, \faWrench{} {\textbf{A/B testing}}) with statistically significant differences between Experience $\le$ 2 years and Experience \textgreater{} 2 years groups. The Experience $\le$ 2 years group is more likely to use User interface and A/B testing strategies to test microservices systems than the Experience \textgreater{} 2 years group. On the other hand, the Experience \textgreater{} 2 years group is more likely to use Component testing strategy to test microservices systems than  the Experience $\le$ 2 years group. \textit{MSA style vs. No MSA style}: There are six survey question statements (i.e., \faGears{} {\textbf{Integration contract testing}}, \faGears{} {\textbf{Consumer-driven contract testing}}, \faGears{} {\textbf{Component testing}}, \faGears{} {\textbf{Scalability testing}}, \faGears{} {\textbf{End to End testing}}, \faGears{} {\textbf{A/B testing}}) with statistically significant differences between MSA style and No MSA style groups. The MSA style group is more likely to use Integration contract, Component, and End to End testing strategies to test microservices systems than the No MSA style group. On the other hand, the No MSA style group is more likely to use Consumer-driven contract, Scalability, and A/B testing strategies to test microservices systems than the MSA style group. \textit{Employees size $\le$ 100 vs. Employees size \textgreater{} 100}: We identified only one survey question statement (i.e., \faGroup{} {\textbf{End to End testing}}) with statistically significant difference between Employees size $\le$ 100 and Employees size \textgreater{} 100 groups. The Employees size $\le$ 100 group is more likely to use End to End testing strategy to test microservices systems than the Employees size \textgreater{} 100 group. Moreover, we did not find any survey question statement about using testing strategies with statistically significant difference for the Employees size \textgreater{} 100 group.

\begin{table}[H]
\caption{Statistically significant differences on the survey results about testing strategies}
\label{tab:SigDiffTestingStrategies}
\hrule
\resizebox{\textwidth}{!}{\begin{tabular}{|r|c|c|c|c|c|c|c|c|}
\hline 
& &\textbf{Likert Distro.} & \multicolumn{2}{c|}{\textbf{Exper. $\le$ 2 vs. \textgreater{} 2 years}} & \multicolumn{2}{c|}{\textbf{MSA style vs. No MSA style}} & \multicolumn{2}{c|}{\textbf{Empl. $\le$ 100 vs. \textgreater{} 100}}\tabularnewline \hline
\textbf{\textbf{Survey Question Statement}}& \textbf{SQ\#} & Mean value & P-value & Effect size & P-value & Effect size & P-value & Effect size\tabularnewline
\hline 
\hline 
\multicolumn{9}{|c|}{\cellcolor{blue!10}\scriptsize{Strategies to test MSA-based systems (Very important (5), Important
(4), Somewhat important (3), Important (2), Not sure (1))}}\tabularnewline
\hline 
Unit testing & \multirow{10}{*}{SQ34} & 3.82 & 0.26 & -0.64 & 0.06 & 0.75 & 0.93 & -0.26\tabularnewline
\cline{1-1} \cline{3-9} 
Integration contract testing &  & 3.27 & 0.57 & -0.33 & \faGears{} 0.01 & \cellcolor{gray!50}0.37 & 0.90 & -0.16\tabularnewline
\cline{1-1} \cline{3-9} 
Consumer-driven contract testing &  & 2.92 & 0.23 & 0.13 & \faGears{} 0.01 & \cellcolor{gray!20}-0.03 & 0.95 & 0.06\tabularnewline
\cline{1-1} \cline{3-9} 
Integration testing &  & 3.75 & 0.27 & -0.13 & 0.09 & 0.45 & 0.94 & -0.20\tabularnewline
\cline{1-1} \cline{3-9} 
Component testing &  & 3.42 & \faWrench{} 0.03 & \cellcolor{gray!20}-0.16 & \faGears{} 0.02 & \cellcolor{gray!50}0.14 & 0.92 & -0.25\tabularnewline
\cline{1-1} \cline{3-9} 
Scalability testing &  & 3.29 & 0.25 & 0.16 & \faGears{} 0.01 & \cellcolor{gray!20}-0.13 & 0.90 & -0.12\tabularnewline
\cline{1-1} \cline{3-9} 
Service component testing &  & 3.26 & 0.25 & -0.01 & 0.12 & 0.33 & 0.92 & -0.26\tabularnewline
\cline{1-1} \cline{3-9} 
User Interface Testing &  & 3.41 & \faWrench{} 0.04 & \cellcolor{gray!50}0.42 & 0.08 & 0.02 & 0.96 & 0.38\tabularnewline
\cline{1-1} \cline{3-9} 
End to End Testing &  & 3.63 & 0.41 & 0.08 & \faGears{} 0.02 & \cellcolor{gray!50}0.12 & \faGroup{} 0.03 & \cellcolor{gray!50}0.22\tabularnewline
\cline{1-1} \cline{3-9} 
A/B Testing &  & 2.80 & \faWrench{} 0.02 & \cellcolor{gray!50}0.20 & \faGears{} 0.00 & \cellcolor{gray!20}-0.42 & 0.89 & 0.51\tabularnewline
\hline 
\end{tabular}}
\end{table}
\textbf{Testing challenges} (see Table \ref{tab:SigDiffTestingchallenges}): \textit{Experience $\le$ 2 years vs. Experience \textgreater{} 2 years}: There are two survey question statements (i.e., \faWrench{}  {\textbf{TC4}}, \faWrench{} {\textbf{TC7}}) with statistically significant differences between Experience $\le$ 2 years and Experience \textgreater{} 2 years groups. The Experience $\le$ 2 years group is more likely to agree that TC4 (i.e., creating and implementing automated tests) and TC7 (i.e., understanding about each testable microservice) challenges have a critical impact on testing of microservices systems than the Experience \textgreater{} 2 years group. However, we did not find any survey question statement about testing challenges with statistically significant difference for the Experience \textgreater{} 2 years group. \textit{MSA style vs. No MSA style}: There are two survey question statements (i.e., \faGears{} {\textbf{TC1}}, \faGears{} {\textbf{TC5}}) with statistically significant differences between MSA style and No MSA style groups. The MSA style group is more likely to agree that TC1 (i.e., creating and implementing manual tests) challenge has a severe impact on testing of microservices systems than the No MSA style group. On the other hand, the No MSA style group is more likely to agree that TC5 (i.e., independent microservices) challenge has a critical impact on testing of microservices systems than the MSA style group. \textit{Employees size $\le$ 100 vs. Employees size \textgreater{} 100}: We identified only one survey question statement (i.e., \faGroup{} {\textbf{TC6}}) with statistically significant difference between Employees size $\le$ 100 and Employees size \textgreater{} 100 groups. The Employees size $\le$ 100 group is more likely to agree that TC6 (i.e., complexity and size of microservices systems) challenge has a severe impact on testing of microservices systems than the Employees size \textgreater{} 100 group. Moreover, we did not find any survey question statement about testing challenges of microservices systems with statistically significant difference for the Employees size \textgreater{} 100 group.

\begin{table}[H]
\caption{Statistically significant differences on the survey results about testing challenges}
\label{tab:SigDiffTestingchallenges}
\hrule
\resizebox{\textwidth}{!}{\begin{tabular}{|r|c|c|c|c|c|c|c|c|}
\hline 
& &\textbf{Likert Distro.} & \multicolumn{2}{c|}{\textbf{Exper. $\le$ 2 vs. \textgreater{} 2 years}} & \multicolumn{2}{c|}{\textbf{MSA style vs. No MSA style}} & \multicolumn{2}{c|}{\textbf{Empl. $\le$ 100 vs. \textgreater{} 100}}\tabularnewline \hline
\textbf{\textbf{Survey Question Statement}}& \textbf{SQ\#} & Mean value & P-value & Effect size & P-value & Effect size & P-value & Effect size\tabularnewline
\hline 
\multicolumn{9}{|c|}{\cellcolor{blue!10}\scriptsize{Severity (impact) of each challenge related to testing of MSA-based
systems? (Catastrophic (5), Major (4), Moderate (3), Minor (2), Insignificant(1))}}\tabularnewline
\hline 
TC1 & \multirow{10}{*}{SQ37} & 2.87 & 0.45 & -0.16 & \faGears{} 0.03 & \cellcolor{gray!50}0.32 & 0.92 & 0.09\tabularnewline
\cline{1-1} \cline{3-9} 
TC2 &  & 3.06 & 0.27 & 0.00 & 0.09 & 0.23 & 0.94 & -0.11\tabularnewline
\cline{1-1} \cline{3-9} 
TC3 &  & 3.13 & 0.26 & -0.01 & 0.05 & 0.33 & 0.93 & 0.08\tabularnewline
\cline{1-1} \cline{3-9} 
TC4 &  & 2.95 &\faWrench{} 0.02 & \cellcolor{gray!50}0.22 & 0.08 & -0.02 & 0.94 & 0.00\tabularnewline
\cline{1-1} \cline{3-9} 
TC5 &  & 2.89 & 0.25 & -0.08 & \faGears{} 0.04 &  \cellcolor{gray!20} -0.05 & 0.93 & 0.13\tabularnewline
\cline{1-1} \cline{3-9} 
TC6 &  & 3.21 & 0.25 & -0.03 & 0.07 & 0.16 &  \faGroup{} 0.01 & \cellcolor{gray!50}0.09\tabularnewline
\cline{1-1} \cline{3-9} 
TC7 &  & 2.95 &  \faWrench{} 0.04 & \cellcolor{gray!50}0.06 & 0.07 & -0.12 & 0.94 & 0.09\tabularnewline
\cline{1-1} \cline{3-9} 
TC8 &  & 3.10 & 0.55 & 0.00 & 0.13 & -0.30 & 0.94 & 0.06\tabularnewline
\cline{1-1} \cline{3-9} 
TC9 &  & 3.08 & 0.24 & 0.07 & 0.05 & -0.18 & 0.92 & 0.23\tabularnewline
\cline{1-1} \cline{3-9} 
TC10 &  & 3.03 & 0.26 & 0.14 & 0.09 & -0.13 & 0.94 & 0.15\tabularnewline
\hline 
\end{tabular}}
\end{table}
\section{Discussion} \label{sec:discussion}
Our empirical study sought to unveil the state of practice of designing, testing, and monitoring microservices systems. In particular, we conducted an online survey with 106 practitioners from 29 countries and interviewed 6 practitioners from four countries. In the following, Section~\ref{analysis} discusses the main findings and implications for researchers. Section \ref{compMSAvsNonMSA} compares the key findings of our survey with other surveys conducted in the context of non-MSA-based systems, and Section~\ref{sub_sec:Implications} discusses the implications for practitioners.

\subsection{Analysis of the results} \label{analysis}

\subsubsection{Design of microservices systems}\label{subsec:discussion_designing}

    \textbf{Application decomposition strategies}:
    An application can be divided into small-scale microservices using various decomposition strategies. DDD and business capability strategies define services corresponding to subdomains and business objects, respectively \cite{richardson2018microservices}. Our survey participants indicated that the applications can be decomposed using DDD, business capability, and a combination of DDD and business capability, in which the combination of DDD and business capability is identified as the leading strategy (\textbf{Finding 4}). The interview participants also acknowledged this finding that adopting a combination of DDD and business capability (i) provides a better opportunity to achieve decomposition goals, (ii) is suitable when using microservices with DevOps or agile practices, (iii) helps to organize the development and operations teams, and (iv) helps to establish a flexible and scalable design for microservices systems. We have found several responses in which both the interviewees and the survey participants have confirmed that DDD and business capability strategies are complementary to each other and can also help to address the challenges of \textit{DC1: defining boundaries of microservices} and \textit{DC3: managing the complexity of microservices at the design level}. Overall, concerning application decomposition strategies, the result of our study partially confirms the finding of several existing studies (e.g., \cite{zhang2019microservice, Wang2020, richardson2018microservices, de2018net}) and the personal experience of the microservices practitioners (e.g., \cite{ADDMA, BDDMA, IDDMA}). We also found several application decomposition strategies for microservices (e.g., decomposed by verbs or use cases \cite{taibicloser19}, data flow-driven approach \cite{LI2019110380}, interface analysis \cite{baresi2017microservices}, service cutter \cite{Gysel2016}) in the literature. However, none of the survey and interview participants mentioned these or other strategies for decomposing applications into microservices in their responses.
\begin{leftrightbar}
\textbf{Research Implication 1}

We found that many organizations use the combination of DDD and business capability to decompose monolithic applications into microservices. We assert that future empirical studies should seek the guidelines that detail how to break a monolith into microservices using the DDD and business capability strategies together. Such guidelines would be beneficial for software teams (particularly inexperienced ones) to smoothly migrate from a monolithic system to a microservices system.

\end{leftrightbar}

\textbf{MSA architecting activities}:
The results of this study indicate that practitioners are more likely to agree on using Architectural Evaluation (AE) and Architectural Implementation (AI) activities when designing microservices systems (\textbf{Finding 6}). This finding was also confirmed by the interviewees, who mentioned several reasons for using AE and AI. These results are expected because practitioners focus on identifying, evaluating, and implementing QAs while designing the architecture of microservices systems. According to Bachmann et al. \cite{cojocaru2019attributes}, the minimum quality requirements to be evaluated during the design of microservices systems are granularity, cohesion, coupling, scalability, response time, security, health management, execution cost, and reusability.

Moreover, many participants confirmed the use of architectural implementation, in which a coarse-grained MSA design is refined into a detailed design. Detailed design (1) provides a clear vision for microservices implementation, such as each microservice's functionality and interface for communicating with other microservices and (2) allows developers to evaluate design options before implementing them in microservices systems. Furthermore, a good number of participants also considered Architectural Maintenance and Evolution (AME) during the design of microservices systems. This is because generally, microservices systems are developed using iterative development methods (e.g., SCRUM and DevOps, see \textbf{Finding 3}) in which changes (e.g., new requirements) are continuously accommodated, and AME helps to ensure the consistency and integrity of the MSA during the iterations. The results regarding using AE, AI, and AME indicate that practitioners pay more attention to the architecting activities close to the solutions space (i.e., AE, AI, and AME) instead of the problem space (i.e., AA and AS).

On the other hand, we received a significant amount of “neutral" responses from the participants about using AA and AS. This result indicates that many participants were undecided about using these two activities for architecting microservices systems. One potential reason is that microservices practitioners are not fully aware of the concepts or terminologies (e.g., AA, AS) used to describe architecting activities. Our results also show that about 40\% to 50\% of the participants either “strongly agree” or “agree” with using various architecting activities when designing a microservices system; on the other side, around 10\% to 20\% of the participants “strongly disagree” or “disagree” with using the architecting activities (see Figure \ref{fig:SQ14-SQ18}). Therefore, the opinions of the participants on using MSA architecting activities are contradictory.

\begin{leftrightbar}
\textbf{Research Implication 2}

We found a lack of understating about MSA architecting activities among the practitioners and some contradictory responses from the survey practitioners. Research is needed to empirically investigate the necessity, understandability, and actual use of the MSA architecting activities in the industry.
\end{leftrightbar}

\textbf{MSA description methods}: Architecture description methods are used to document architecture of software systems using specific notations (e.g., UML diagrams) in architecture views \cite{bachmann2011documenting}. The results show that formal approaches are not widely used in the industry to describe MSA, in contrast, the practitioners prefer to use informal and semi-formal methods for describing MSA (\textbf{Finding 7}). We identified several reasons about why the practitioners do not prefer formal approaches but tend to use informal and semi-formal methods for describing MSA (see Section 4.2.2). This finding has also been confirmed in several review studies on microservices architecture (e.g., \cite{di2019architecting, alshuqayran2016systematic, waseemMSAdevops}).

On the other hand, the results related to diagrams representing the design and architecture of MSA-based systems (see Figure \ref{fig:SQ20}) reflect that practitioners describe MSA in multiple views, which is common practice in documenting architecture \cite{bachmann2011documenting}. For instance, our results show that the leading diagrams for representing the design and architecture of microservices systems are flowchart, use case diagram, data flow diagram, activity diagram, and sequence diagram (\textbf{Finding 8}), which are used to describe the process (dynamic) view of microservices systems \cite{fowler2004uml}. According to Gerber et al. \cite{Anna2017}, microservices systems could have many dynamic operations, such as scaling up and scaling down services and deployment of services across data centers or cloud platforms. The behavioral diagrams (e.g., activity diagram, use case diagram) represent the dynamic operations of microservices systems by showing collaborations between microservices and changes to their internal states. We also asked about diagrams that can be used to illustrate the behavioral view of the microservice system. For instance, communication, state machine, timing, and interaction overview diagrams. We also noticed that a considerable number of the participants reported class, deployment, component, package, and object diagrams for representing the design and architecture of microservices systems, which are used to describe the structural (or Static) view of microservices systems. Moreover, our results indicate that the diagrams used by practitioners for describing MSA also follow the 4+1 architectural view model \cite{kruchten19954}, a commonly-used architecture framework for describing the architecture of traditional systems. For instance, use case diagram represents the scenario view; flowchart, activity, functional flow block, communication, and timing diagram represent the process view; class diagram, object diagram, package diagram, a composite structure diagram, and microservice circuits represent the logical view; component diagram represents the implementation view, and deployment diagram represents the deployment view.

\begin{leftrightbar}

\textbf{Research Implication 3}

Regarding MSA description methods, the survey results confirm the findings of our recent SMS on microservices architecture in DevOps \cite{waseemMSAdevops}, in which we identified that MSA is mostly described by using informal description methods (e.g., Boxes and Lines). Therefore, we suggest further research about the pros and cons of using informal description methods for describing MSA. Moreover, it is also interesting to investigate how formal description methods (e.g., DSLs, ADLs) can be adapted to describe MSA in industry.
\end{leftrightbar}

\textbf{Quality attributes}: Our study results show that security, availability, performance, and scalability are the most important QAs considered in MSA design (\textbf{Finding 9}). Previous research (e.g., \cite{LI2020106449,waseemMSAdevops}) also highlights the importance of these QAs. For instance, \textbf{Finding 9} partially confirms the study results by Li et al. \cite{LI2020106449} that scalability, performance, availability, monitorability, and security are identified as the most concerned QAs in MSA.
We reported several reasons collected from the interviews in the Results section regarding \textbf{Finding 9} (see Section \ref{sec:QADP}). Here, we further analyzed the top four QAs: (1) About security, unlike traditional monolithic applications, MSA-based systems are composed of independent services that work together to fulfill the business goals. Microservices provide public interfaces, use network-exposed APIs, and develop applications using polyglot technologies, thus open the doors to intruders, and consequently the security of messages, queues, and API endpoints requires advanced strategies to protect at the infrastructure and code level. According to the literature review conducted by Pereira-Vale et al., \cite{Pereira2019SecMec}, authentication, authorization, and credentials are the most widely used security strategies by microservices developers. 
(2) Regarding availability, one of the main benefits of MSA (and also one of the reasons that the industry adopts MSA) is that it provides a fault-tolerant architectural style \cite{newman2015building}. It also implies that practitioners must use virtualized resources (e.g., containers, Virtual Machines (VMs), data storage) to ensure a high availability of microservices systems \cite{toffetti2015architecture}. 
(3) Performance is a major concern for all types of systems. With MSA, an application is decomposed into microservices based on functionality. The count of integration points could increase in dozens or even hundreds while integrating microservices, thus, in turn, leads to a performance challenge due to frequent inter-process communication of microservices. Therefore, performance is essential to be addressed while designing MSA.
(4) Scalability is not straightforward to achieve. In microservices systems, business functionalities are written in different programming languages, loaded on different hardware, executed on different servers, and deployed across multiple clouds. Therefore, architects should devise strategies that allow microservices to be coordinated to scale or eventually identify which microservices must be scaled to meet the increased demand \cite{wolff2016microservices}. The survey conducted by Dragoni et al. \cite{dragoni2017microservices} also highlights the importance of scalability in MSA design.

Overall, we received a significant number of responses in which practitioners consider all of the remaining QAs as “important” or “very important”, such as reliability, usability, maintainability, and compatibility. It is because, MSA has a significant influence (e.g., positive impact) on most of QAs \cite{waseemMSAdevops}. For example, MSA can help to improve reliability through scalability and redundancy with the flexibility of increasing the instances of services \cite{Singleton2016}.

\begin{leftrightbar}

\textbf{Research Implication 4}

We found that security is one of the important QAs during the design of microservices systems. We also identified the design patterns (e.g., Access token) that support security in microservices systems. However, there is still a lack of knowledge of state-of-the-practice security techniques and tools as well as their potential security vulnerability, which deserves an in-depth empirical investigation.
\end{leftrightbar}


\textbf{MSA design patterns}: A pattern is a reusable solution to problems recurring in a particular context. Our results show that the most often used MSA patterns are API gateway, Backend for frontend, and Access token (\textbf{Finding 10}). API gateway and Backend for frontend patterns are also known as external API patterns used to address how different clients (e.g., mobile and desktop version) of MSA-based systems access the individual services. In contrast, Access token pattern is employed to address the problem of microservices security. We identified several reasons for using these patterns in the Results section (see Section \ref{sec:QADP}). The result regarding the frequent use of security pattern (Access token) is partially aligned with the result that security is considered as an important QA in MSA design. The aggregated responses regarding QAs (e.g., security), patterns (e.g., Access token), and design challenges (e.g., addressing security concerns) suggest that security requires more attention during the design of microservices systems.

On the other hand, practitioners reported the frequent use of deployment patterns, such as Service instance per container, Service instance per host, and Service instance per VM, which suggests that practitioners prefer to deploy service instances in isolation. The reasons are that (i) it is easy to monitor, manage, and redeploy each service instance, (ii) deployment of service instances in isolation helps to achieve scalability, performance, and high cohesion by increasing the number of e.g., containers, hosts, VMs, and (iii) deployment of service instances in isolation facilitates the development of microservices by using multiple languages. Another challenging aspect of MSA-based systems is the communication between microservices \cite{chelliah2017architectural}, and several design patterns provide efficient communication mechanisms (e.g., synchronous and asynchronous communication) between microservices. According to the results, many practitioners use Messaging, Remote procedure invocation, and Domain-specific protocol patterns for microservices communication. It is because, these communication patterns (i) provide loose runtime coupling, (ii) increase microservices availability, and (iii) use popular communication technologies (e.g., RESTful APIs\footnote{\url{https://restfulapi.net/}}, gRPC\footnote{\url{https://grpc.io/}}, Apache Thrift\footnote{\url{http://thrift.apache.org/}}).

Moreover, our survey results cover the data management patterns, such as API composition, Database per service, Sagas, Shared database, and CQRS, which are also used frequently in the industry. Benefits like a simple method to query data, supporting polyglot database strategies, promoting data consistency, providing ACID transactions, and supporting multiple denormalization could be the possible reasons for frequent use of these data management patterns in the industry for microservices systems \cite{richardson2019microservices}. For instance, API composition pattern provides a simple method to query data in microservices systems, Database per service pattern allows the use of polyglot database strategies according to service needs, Sagas pattern ensures the data consistency across multiple microservices services, Shared database pattern provides straightforward ACID transactions to enforce data consistency. In contrast, the CQRS pattern improves the separation of concerns and supports multiple denormalization of databases for microservices systems. Furthermore, we found in the literature (e.g., \cite {richardson2015service}) that service discovery is a key but challenging issue in cloud-based microservices systems due to the containerized and virtualized environment and dynamically assigned network locations. Our survey results indicate that a fear number of practitioners frequently employ several patterns for service discovery. For instance, Service registry, Client-side discovery, Server-side discovery, and Self-registration. The reasons for using these patterns are (i) identification of the dynamically assigned locations to service instances for service clients (e.g., mobile or desktop browsers) and (ii) ability to manage the load balancing requests.

\begin{leftrightbar}
\textbf{Research Implication 5}

Our study results highlight the importance of quality attributes and the use of design patterns for microservices systems. Future studies can present decision models for selecting MSA design patterns, which can assist developers and architects in choosing appropriate patterns by considering the positive and negative effect of QAs for diverse design areas (e.g., communication, service discovery, monitoring and logging) of microservices systems.
\end{leftrightbar}

\textbf{Major MSA design challenges and their solutions}: The survey results show that a large number of participants responded that \textit{DC1: clearly defining the boundaries of microservices}, and \textit{DC2: addressing security concerns} as two major design challenges (\textbf{Finding 11}). Some other survey and interview-based studies (e.g., \cite {zhang2019microservice,di2018migrating, knoche2019drivers}) also reported the empirically identified challenges related to MSA design. Compared with the existing work of Zhang et al. \cite{zhang2019microservice}, Di Francesco et al. \cite{di2018migrating}, and Knoche et al. \cite{knoche2019drivers}, our study has independently identified several unique challenges, reasons, and solutions. Defining the boundaries of microservices is deciding about microservices functionality according to business capabilities or subdomains. This challenge mainly occurs due to a lack of business knowledge to establish the right boundaries, for example, in a new application domain. Our survey results also indicate that unclear boundaries have a negative impact on the monitoring and testing of microservices systems (see Table \ref{tab:designChallenges}). 

To address DC1, we have collected several suggestions from the interviews (see Section \ref{designCha&sol}). For instance, gaining a deeper understanding of business requirements, introducing and implementing miniservices \cite{GuptaMiniservice}, using DDD, and asking experts for their judgments.
Another frequently reported challenge is addressing security concerns (i.e., DC2). Microservices systems are vulnerable to various security attacks. Therefore, the security of microservices systems demands serious attention. In our study and several other studies, the security of microservices is recognized as one of the leading challenges faced by many practitioners. For instance, Ponce et al. \cite{ponce2021smells} developed a taxonomy of smells and refactorings for microservices security by reviewing the grey and peer review literature. In our recent work (i.e., \cite{ waseem2021nature}), security issues are also identified as one of the top issues (10.18\%) in microservices systems. The study result indicates that practitioners use patterns (e.g., Access token), cloud solutions (e.g., Role-Based Access Control), and development practices (e.g., DevSecOps) as solutions to address security challenges (see Section \ref{designCha&sol}). We also found that some review studies (e.g., \cite{waseemMSAdevops, Pereira2019SecMec, mao2020preliminary, pereira2021security}) collected a comprehensive set of standards, strategies, and mechanisms from grey and peer review literature that can be used as solutions to secure microservices systems.

\begin{leftrightbar}
\textbf{Research Implication 6}

We collected several strategies, design patterns, DSLs, and development methods to address MSA design challenges (see Table \ref{tab:DesignSol}). However, further research is needed to develop and empirically evaluate the solutions for addressing the challenges in MSA design as listed in Table \ref{tab:designChallenges}.
\end{leftrightbar}

\subsubsection{Monitoring of microservices systems}\label{subsec:discussion_monitoring}

\textbf{Monitoring metrics}: Infrastructure monitoring metrics involve the status of the infrastructure and the servers on which the microservices are running, while microservice metrics concern the metrics of a specific microservice.
As reported in Figure~\ref{fig:SQ29}, most of the respondents focus on resource usage, database connection, load balancing, and threads that serve to monitor the infrastructure. It is not surprising as the infrastructure monitoring metrics require less development, maintenance, and operation effort~\cite{Susan17}. Moreover, the infrastructure software (e.g., Docker, Kubernetes) generally provides tools to collect such metrics. The microservices metrics, such as availability, errors and exceptions, and endpoint success, are more complex to implement because they depend on the languages, libraries, and frameworks~\cite{Susan17}. Therefore, it is not unusual that the participants declared to use less of these application metrics (e.g., availability, errors and exceptions, and endpoint success). 
Furthermore, infrastructure and application metrics are used to monitor quality aspects, such as performance (e.g., resource usage, threads, database connections, and latency), availability (e.g., status of service), scalability (e.g., load balancing), and reliability (e.g., errors and exceptions). These QAs assume particular importance when building microservices systems. For example, more than 85\% of the respondents considered performance a very important or essential QA when designing microservices systems (see Table \ref{tab:MSAQAs}). These quality aspects are in line with companies' motivation when adopting the MSA style to develop or modernize their applications, i.e., to have scalable, high-performance, and resilient systems (e.g., \cite{AdamBertramResons}).

\begin{leftrightbar}
\textbf{Research Implication 7} 

Our study results highlight that practitioners prefer to use infrastructure monitoring metrics (e.g., resource usage, database connections) over application monitoring metrics (e.g., availability, errors and exceptions). The latter, although more challenging to implement and provides better information for monitoring microservices systems. Therefore, it would be interesting to propose empirically evaluated monitoring practices and tools for application metrics.
\end{leftrightbar}

\textbf{Monitoring practices}: The participants indicated log management, exception tracking, and health check API as the frequently used monitoring practices as shown in Figure~\ref{fig:monitoringpractices}. Microservices and their instances run on multiple machines (e.g., VMs, containers), and these instances are dynamically created and destroyed during the execution of the microservices. Thus, an adequately managed log data (e.g., centralized logging) helps to identify the service failure, their causes, and unexpected behaviour of microservices systems. In contrast, poorly managed log data can affect the performance of microservices systems adversely. Therefore, it is not surprising that 68.8\% (33.0\% very often +35.8\% often) of the participants use log management very often or often (see Figure \ref{fig:monitoringpractices}). Exceptions may occur when some services handle the requests of other services or instances. Typically, monitoring tools, e.g., Grafana, Zipkin (see Figure \ref{fig:monitoringtools}), report exceptions to a centralized exception tracking service that aggregates and tracks exceptions and notifies them to developers and debugging professionals for further actions (e.g., addressing the underlying issues)~\cite{richardson2018microservices}. Another monitoring practice, frequently mentioned by the participants, is health check API (see Figure \ref{fig:monitoringpractices}). From the survey results, we observed that practitioners use monitoring practices that allow developing production-ready microservices as also highlighted in~\cite{richardson2018microservices}.

\begin{leftrightbar}
\textbf{Research Implication 8}

Concerning the monitoring practices, we found that many practitioners frequently use log management, exception tracking, and health check API to better trace issues and identify causes of issues in microservices systems. The recognized issues and their causes, through these monitoring practices, can be used to propose empirically validated taxonomies for microservices issues. 
\end{leftrightbar}

\textbf{Monitoring tools}: The participants mentioned that the most commonly used monitoring tools are Jira, Datadog Kafka Dashboard, Spring Boot Actuator, Grafana, and Zipkin. Jira is developed as an issue tracking tool, but now it can be integrated with tools for monitoring microservices systems, such as Dynatrace \cite{AndreasJira}, Prometheus \cite{AleksJira}, and Grafana \cite{AleksJira}. To understand the opinions of the microservices practitioners about Jira, we also asked the interview question IQ3.3.1, and the results indicate that the practitioners have a different understanding of Jira as a monitoring tool for microservices systems, for instance, three interviewees suggested Jira integrated with IDE to monitor microservices systems, two interviewees considered issue tracking (e.g., Jira) as part of monitoring of microservices systems, and one interviewee was not sure about the use of Jira for monitoring purposes. On the other hand, Datadog Kafka Dashboard and Spring Boot Actuator are mainly used to gather monitoring data. One reason that the participants prefer to use such tools could be that most of them are application developers (65.1\%), therefore, they mainly focus on using monitoring tools for collecting application data instead of the ones more oriented to analyze or visualize data. The results show that practitioners do not prefer to use visualizing and alerting data tools for monitoring microservices systems concerning the dashboard and alerting components. For example, only 17.9\% (19 out of 106) of the participants use Grafana, and 16.9\% (18 out of 106) use Zipkin. One reason could be that 59.4\% of the participants declared that they have 0 to 2 years of experience in microservice development, therefore, they may not yet have well experienced the added value of the monitoring activities resulting from these tools' use.

\begin{leftrightbar}
\textbf{Research Implication 9}

Our survey reports that the practitioners frequently use tools to monitor network performance, resource usage, and errors and exceptions for microservices systems. Therefore, future studies can empirically investigate the need, understandability, and actual use of monitoring tools in the industry.
\end{leftrightbar}

\textbf{Monitoring challenges and solutions}: Most of the challenges faced by the participants during the monitoring activities focus on several aspects, such as collecting monitoring metrics data and logs from containers, and distributed tracing (see Table~\ref{tab:monitoringChallenges}). These results show that microservice monitoring challenges are mostly related to their highly distributed nature, autonomy, independence, and self-containment. Concerning the challenge of collecting monitoring data and container records, the practitioners treat it as moderate (48.1\%) or minor (23.6\%) severity (see Table~\ref{tab:monitoringSeverityImpact}). It is because there are solutions to deal with these challenges. For example, the practitioners recommend monitoring tools, such as Prometheus, Spring Boot Actuator, and Grafana, to search, analyze, and visualize monitoring metrics. On the other hand, the respondents pointed out that distributed tracing of microservices systems is challenging because of the complexity and dynamicity originating from traditional distributed systems. However, only a few participants declared to use monitoring tools, such as Zipkin and Jaeger, to support distributed tracing. It could be the reason why distributed tracing raises several challenges, such as distributed state, concurrency, and heterogeneity of microservices~\cite{ DanielSpoonhower}.

\begin{leftrightbar}
\textbf{Research Implication 10}

Our research indicates that many organizations use tools and frameworks to address monitoring challenges (see Table~\ref{tab:MonitoringSol}). However, further research can be explored to develop and empirically evaluate the solutions for addressing the challenges, such as having many components to monitor (complexity), performance monitoring, and distributed tracing.
\end{leftrightbar}

\subsubsection{Testing of microservices systems}\label{subsec:discussion_testing}

\textbf{Testing strategies}: Concerning testing microservices systems in the industry (RQ3), the results indicate that the most widely used testing strategies are unit, E2E, and integration testing (see Figure \ref{fig:SQ30}). Whether in microservices or monolithic systems, unit testing provides a type of analysis that allows examining the functionality of methods or code modules. However, unit testing evaluates components in isolation, consequently, it is also necessary to test the communication paths and interactions between microservices. This is why integration testing is also prevalent for testing of microservices systems. Unit and integration testing are robust testing strategies, but the complexity of microservices systems demands more systematic testing strategies. That is why E2E testing emerges as one of the preferred testing strategies for microservices systems in the industry. E2E testing tests whether a microservices system satisfies the business goals, i.e., verifies that the microservices system meets external requirements and achieves its goals from an end user perspective \cite{clemson2014}.
Figure \ref{fig:SQ30} also shows that A/B testing and consumer-driven contract testing are not frequently used to test microservices systems. In this regard, we have identified the following reasons for both testing strategies. 
About A/B testing, 33.0\% of the participants use this strategy very often or often, however, 53.8\% of the participants rarely or never use A/B testing, which shows a contradictory preference of practitioners. We identified the following reasons: (1) there are not enough libraries or tools to use A/B testing \cite{Renz2016}, (2) there are not enough users to perform A/B testing \cite{brillmark}, (3) A/B testing takes a long time (e.g., minimum seven days) to get results \cite{brillmark}, and (4) for some domains (e.g., healthcare, transportation) A/B testing does not produce significant results \cite{Lauren}. About consumer-driven contract testing, we identified two reasons: (1) development teams have communication obstacles when several people working on one microservice (e.g., over 8 people, see Figure \ref{fig:SQ8-SQ9} (Right)) and (2) microservices systems extensively use third party resources \cite{lehva2019}.

Additionally, concerning the testing strategies (see Figure \ref{fig:SQ30}), the survey results from practitioners partially coincide with the results of our recent mapping study from academia regarding testing of MSA-based applications \cite{waseemtestingMSA}, in which we identified 39 testing techniques used to test MSA-based applications. Among the testing strategies collected in our survey, nine strategies (i.e., unit testing, E2E testing, integration testing, user interface testing, component testing, scalability testing, A/B testing, service component testing, and consumer-driven contract testing) (see Figure \ref{fig:SQ30}) can be found in the results of our recent SMS \cite{waseemtestingMSA}. Nevertheless, we found no evidence about integration contract testing in the SMS, despite its popularity (47.2\% of the participants very often or often use this testing strategy) in the industry.

\begin{leftrightbar}
\textbf{Research Implication 11}

End-to-End testing is a practical solution for testing each microservice or the whole microservices system. Nevertheless, using this critical testing strategy can cause several challenges in the testing process, such as delivery delays, generating bugs that are difficult to isolate, and liability conflicts. Future work should address the challenges related to E2E testing to make it more beneficial for testing microservices systems.
\end{leftrightbar}

\textbf{Testing tools}: Our survey results indicate that the respondents use a variety of tools to test microservices systems. The most used testing tool is JUnit (51.9\% of preferences) that is used for unit testing. Unit testing is one of the most straightforward but useful testing strategies. It can be used for testing microservices systems and other types of applications or services. In monolithic systems, applications structure all their functionalities into a single set \cite{dragoni2017}. Nevertheless, in microservices systems, functionalities are isolated, which makes testing of each microservice possible. Although other tools (e.g., Cucumber, Sauce Labs, Travis) offer a more complete and automated suite of tests for microservices systems, the respondents still chose traditional testing strategies (e.g., unit testing, integration testing) as shown in Figure \ref{fig:SQ30} and eventually selected popular tools (e.g., JUnit, JMeter) as listed in Figure \ref{fig:SQ36}. The respondents also highlighted that unit testing is part of the core testing strategies for microservices systems (Clemson also agrees with this observation \cite{clemson2014}). Other testing strategies, such as integration, component, consumer-driven contract, and E2E testing are also part of these core testing strategies. Some of the tools (e.g., JUnit, Selenium, Mockito) presented in Figure \ref{fig:SQ36} partially support these core testing strategies (e.g., integration and E2E testing). However, most of the mentioned tools (e.g., JUnit, JMeter) are not dedicatedly developed to test microservices systems.

\begin{leftrightbar}
\textbf{Research Implication 12}

The survey and interview results indicate a need for more investments in testing tools specific to microservices systems. Such tools can systematically and automatically test all components and elements of microservices systems (e.g., microservices, microservice collaboration, communication, data processing, networking, databases, CI/CD pipelines).
\end{leftrightbar}

\textbf{Testing challenges and solutions}: Table \ref{tab:TestingChallenges} shows 10 testing challenges identified in this survey. These challenges range from manual testing to polyglot technologies for testing microservices systems. The results indicate that the microservices testing challenges are mostly related to creating and \textit{TC1: Creating and implementing manual tests}, \textit{TC2: Integration testing of microservices}, \textit{TC3: Debugging of microservices on container platforms}, and \textit{TC4: implementing automated testing}. Compared to monolithic systems, practitioners need more expertise about the system design and the flow between microservices to successfully test microservices systems. One reason is that it is necessary to control and verify all communication points, as well as the functionality of all microservices. Therefore, more than one testing strategies are required to test all the components (e.g., APIs, messaging formats, services, databases) that characterize microservices systems.

On the other hand, the responses from the participants show that performing manual testing of microservices systems is challenging because of the complexity of microservices systems. This challenge can be addressed by using automated testing tools, such as Cucumber, Sauce Labs, and Travis. These tools provide comprehensive control over the test cases that need to be executed on each microservice. At this point, a useful approach to manage the complexity of microservices systems is Behavior-Driven Development (BDD). As suggested by a respondent (see Section \ref{sec:ChallengeAndSolutions}), BDD provides a common vocabulary among stakeholders (such as domain experts and developers). The vocabulary of the BDD approach can be used to establish test cases for testing microservices systems, such as automated acceptance testing (e.g., \cite{rahman2015reusable}).

\begin{leftrightbar}
\textbf{Research Implication 13}

We collected several testing strategies, approaches, and guidelines as solutions to address the respective testing challenges (see Table \ref{tab:TestingSol}). However, these solutions are not enough to address all microservices testing challenges. Further research is needed to develop and empirically evaluate the solutions for addressing the testing challenges in microservices systems as listed in Table \ref{tab:TestingChallenges}.

\end{leftrightbar}


\subsection{Comparing the survey results with non-MSA-based systems}
\label{compMSAvsNonMSA}
In this section, we compare the key findings of our survey with the surveys that explore the design, monitoring, and testing of non-MSA-based systems.

\subsubsection{Design of microservices systems}

Although we explored studies related to application decomposition, approaches to high-level system design, architecting activities, architecture description, and design challenges reported in the industry, we did not find sufficient evidence to be compared with our results. Nevertheless, we realized that addressing quality attributes is a major concern in the design of both non-MSA-based and MSA-based systems. The results of the surveys conducted by Kassab et al. \cite{Kassab2018}, Ameller et al. \cite{Ameller2016} and our study show that some quality attributes (e.g., performance, availability and security) are the common interest to practitioners of both MSA-based and non-MSA-based systems. Kassab et al. \cite{Kassab2018} found that the most relevant quality attributes from an architectural design point of view in practice are usability, performance, modifiability, and availability. Ameller et al. \cite{Ameller2016} identified that the most important quality attributes for practitioners in service-based systems are dependability, performance, security, and usability.

Although these studies mention other quality attributes, performance appears to be the quality attribute of common interest between MSA-based and non-MSA-based systems. Aspects related to a system's responses to perform actions for a certain period of time are concerns that must be addressed in any system. Other quality attributes, such as availability and security are also of interest to practitioners. Both attributes address recovery concerns from failures and security incidents, which are relevant in MSA-based systems. Scalability, on the other hand, is also relevant because the design of MSA-based systems must consider aspects of concurrency and partitioning \cite{Fowler2016Book}. Concurrency allows each task to be broken down into smaller pieces, while partitioning is essential to allow these smaller pieces to be processed in parallel. Therefore, scalability becomes more relevant when designing MSA-based systems, as these systems often address requirements that involve constant traffic growth. In these cases, each microservice must be able to scale with the entire MSA-based system without suffering from performance degradation.

\subsubsection{Monitoring of microservices systems}
To compare our results concerning the monitoring of MSA-based systems with non-MSA-based systems, we explored studies related to monitoring metrics, practices, tools, challenges, and solutions reported in the industry. However, we did not find sufficient evidence to be compared with our results. We only identified one literature survey in which Cassar et al. reviewed runtime monitoring instrumentation techniques for non-MSA-based systems \cite{Cassar2017}. The results of this survey and our study show that only one monitoring practice (i.e., log management) is commonly used by practitioners of both MSA-based and non-MSA-based systems. Log management practice facilitates the scanning and monitoring of log files generated by servers, applications, and networks. The generated log files are used to detect and solve various types of problems (e.g., performance and security issues) of MSA-based and non-MSA-based systems.

\subsubsection{Testing of microservices systems}

We explored several survey studies on testing conducted in the context of non-MSA-based systems (e.g., \cite{Lee2012, RaulamoJurvanen2019, Kochhar2019, Garousi2013, QuesadaLopez2019, Jahan2019, Garousi2015, DiasNeto2017, Garousi2017, Garousi2020}), and we compared our key findings on testing strategies, tools, and challenges with these surveys. Regarding testing strategies, the results of our survey and the surveys related to non-MSA-based systems (e.g., \cite{Lee2012, RaulamoJurvanen2019, Kochhar2019, Garousi2013, QuesadaLopez2019}) show that unit testing is commonly used as a \textit{de facto} strategy for testing all types of systems. Meanwhile, integration testing is also a commonly used testing strategy for both MSA-based and non-MSA-based systems (e.g., \cite{QuesadaLopez2019, RaulamoJurvanen2019,Garousi2015}). To our surprise, we did not find evidence regarding the use of scalability, integration contract, and consumer-driven contract testing strategies in the context of non-MSA-based systems. About testing tools, JUnit, JMeter, and Selenium are commonly used by practitioners of both MSA-based and non-MSA-based systems (e.g., \cite{Kochhar2019, Garousi2013, QuesadaLopez2019, Jahan2019}), for example, Selenium is used as a tool to perform testing on MSA-based systems (see Figure \ref{fig:SQ36}), and is also employed for testing non-MSA-based systems due to its test automation capabilities. Regarding testing challenges, the results of our survey and other surveys (e.g., \cite{QuesadaLopez2019, Garousi2017, Garousi2020}) show that two testing challenges (i.e., integration testing, creating and implementing automated tests) are commonly faced by practitioners of both MSA-based and non-MSA-based systems.

\subsection{Implications for practitioners}\label{sub_sec:Implications}

\begin{enumerate}
\item As shown in Figure \ref{fig:SQ7-SQ10} (Right), the majority of the participants use Scrum and DevOps as the development methods for microservices systems. It could be interesting to further explore what Scrum and DevOps practices are used to design, develop, and maintain microservices systems, the challenges that arise when MSA is used with Scrum and DevOps practices, and how practitioners address these challenges.

\item The results of SQ11 indicate that 36.7\% (39 out of 106) of the participants' organizations do not have a dedicated architecture team or an architect who is responsible for the architecture of microservices systems. It is then meaningful to explore the impact of the absence of architecture teams or architects on the quality of microservices systems as well as the countermeasures employed in practices.

\item Our survey results show that many organizations (42.3\%) use the combination of DDD and business capability to decompose monolithic applications into microservices. We assert that software organizations and practitioners develop guidelines that detail how to break a monolith into microservices using the DDD and business capability strategies together. Such guidelines would be beneficial for software teams (particularly inexperienced ones) to smoothly migrate from a monolithic system to a microservices system.

\item Our survey results highlight that the complexity of microservices systems poses several challenges to the design, monitoring, and testing of microservices systems. However, we did not get any satisfactory solutions from our survey that address such challenges. In this regard, a comprehensive investigation with practitioners is required to understand the nature of complexity and its mitigating strategies in the context of microservices systems.

\item We asked two survey questions (i.e., SQ30, SQ35) regarding monitoring and testing tools for microservices systems. However, most of the tools are general purpose tools that are not dedicatedly developed for microservices systems. The complex nature of microservices systems further exposes the weakness of the existing tooling, specifically in the context of intelligent monitoring and testing of microservices systems. Therefore, practitioners can put effort into developing and using such tools for improved productivity.
\end{enumerate}

\section{Threats to validity}
\label{sec:threats}
This section outlines the validity threats to our survey by following the guidelines described by Easterbrook et al. \cite{Easterbrook2002Selecting}.

\textbf{Internal validity} refers to the factors that could negatively affect the data collection and analysis process. We considered the following threats to the internal validity of this survey:
\begin{itemize}
    \item \textit{Participants selection}: When the survey invitations were sent through emails, those who responded may lack the required expertise to participate in our survey. We tried to minimize this issue by searching microservices practitioners through personal contacts and relevant platforms (see detailed approaches in Section \ref{Samplepopulation}). Additionally, the characteristics of the prospective participants were made explicit in the survey preamble.
    \item \textit{Closed-ended survey questions}: Most of the survey questions are closed-ended. The primary threat to this type of questions is an insufficient number of answer options. In this regard, the answer options of multiple-choice questions were defined by taking into account the existing literature. Moreover, we allowed the participants to answer the survey questions according to his/her experience in the free text area. As a result, we received several other answers besides the already provided answer options.
    \item \textit{Length of the survey}: 
    Our survey is composed of 39 survey questions, which may have affected the response rate of our survey. While almost 9000 emails were sent over three months to the potential practitioners, we only got 135 responses (i.e., 1.5\%), out of which 106 were valid. We tried to mitigate this threat, to some extent, by reducing the number of open-ended questions as much as possible to avoid memory bias \cite{Murphy15}. Second, in the survey preamble, we mentioned that completing the survey would take 20-25 minutes, which might help practitioners find their best available time to fill out the survey.
\end{itemize}

\textbf{Construct validity} in a survey focuses on whether the survey constructs are defined and interpreted correctly \cite{Easterbrook2002Selecting}. Design, monitoring, and testing of microservices systems are the core constructs of this survey. Having said this, we identified the following threats:

\begin{itemize}
    \item \textit{Inadequate explanation of the constructs}: This threat refers to the fact that the constructs are not sufficiently described before they are translated into measurements and treatments. To deal with this threat, we designed the survey based on our recent SMSs on MSA in DevOps \cite{waseemMSAdevops} and testing of microservices \cite{waseemtestingMSA}) as well as the existing literature on microservices systems (e.g., \cite{rajput2018hands, di2019architecting, alshuqayran2016systematic, DaveSwersky2018monitoring, JakeLumetta2018MSATesting, clemson2014}). Additionally, the authors had several internal meetings to identify issues regarding question format, understandability, and consistency. Moreover, we invited two experts who had experience in survey-based research to check the validity and integrity of the survey questions. The feedback collected from the experts helped us improve several survey questions that were found confusing or unclear. Finally, we evaluated the question format, understandability, and consistency of the survey questions by conducting a pilot survey, which was completed by ten microservices practitioners. Based on the feedback received from the pilot survey participants, we also rephrased a few survey questions (i.e., SQ21, SQ26, SQ33, SQ39).
    
    \item \textit{Fear of being evaluated}: This threat is related to the fact that some people, due to work restrictions, anxiety, or other reasons, are afraid of being evaluated. To mitigate this threat, at the beginning of the survey, we clearly mentioned that all the responses gathered through this survey will be treated with confidentiality and privacy and that the data (responses) gathered will be used for academic purposes only.
    
    \item\textit{Survey dissemination platforms}: The survey was disseminated on several social media and professional networking groups (see Table \ref{tab:platforms}). The main threat to survey dissemination platforms is the identification of the relevant groups. To address this threat, we read the existing discussion in the groups. After ensuring that the group members frequently discuss different aspects (e.g., techniques, tools, technology issues) of microservices, we posted our survey invitation.
    
    \item \textit{Inclusion of valid responses:} To identify valid answers, we asked three survey questions about the responsibilities of practitioners (DQ2), usage of MSA for developing applications (DQ3), and experience in the development of microservices systems (DQ5), and applied inclusion and exclusion criteria on received answers (see Table \ref{tab:inlu_exclu}). We removed those answers that were “meaningless”, “inconsistent”, and “logically senseless” (see the details in Section \ref{obtainingsample}). We also excluded the entire response from a respondent if more than 5 survey questions have inconsistent, randomly filled, or meaningless answers.

\end{itemize}

\textbf{External validity} is related to the extent to which the study results can be generalized to other cases. We sent almost 9000 emails to microservices practitioners and made dozens of posts in relevant groups available at social and professional networking platforms to reach the practitioners (see Table \ref{tab:platforms}). We get 106 valid responses from 29 countries (see Figure \ref{fig:countries}). The respondents' demographics show that they had various responsibilities (see Figure \ref{fig:roles} and worked in a wide range of teams and organizations in terms of size (see Figure \ref{fig:SQ8-SQ9}) and development method (see Figure \ref{fig:SQ7-SQ10}). Our participants had varying experiences on the development of microservices systems, from less than one year to more than ten years, in different domains (see Section \ref{sub_sec:demographics}). We also interviewed 6 respondents from four countries to understand the reasons behind the key survey findings. Although we have a relatively large sample size from the survey and interviews and variations in participants' demographics in terms of professional experience, company size, and work domains. Still, our findings may not generalize or represent all microservices practitioners' perspectives. For example, most of the survey respondents (i.e., 56.6\%) are from Pakistan and China. It would be interesting to perform another study with diverse practitioners in order to increase the external validity of the study results. 

\textbf{Conclusion validity} is related to the issues that affect the right conclusion in empirical studies. To mitigate this threat, the first author analyzed the survey data, and the second and third authors reviewed the survey data analysis results. Conflicts on data analysis results were resolved through mutual discussions and brainstorming among all the authors. Furthermore, the conclusion can be influenced as different researchers may interpret the inclusion and exclusion criteria for selecting responses differently. To minimize the effect of this issue, we defined and applied the inclusion and exclusion criteria on all the received responses (see Section \ref{obtainingsample}). We also confirmed the interpretations of the results and conclusions by arranging several brainstorming sessions among all the authors.

\section{Related work}
\label{sec:relatedWork}
In this section, we report the survey studies that investigated microservices systems from different perspectives as well as the difference between our survey and these existing surveys.
Knoche and Hasselbring surveyed the drivers, barriers, and goals of using microservices to modernize legacy systems along with the impact of microservices on runtime performance and transactionality \cite{knoche2019drivers}. The number of participants who are from Germany is 71 in this survey. The key results regarding the drivers are high scalability and elasticity, high maintainability, and short time to market. The key findings regarding the barriers that prevent the adoption of microservices are insufficient operation skills, resistance from operational staff, and insufficient development skills. The identified goals in the context of microservices are improve maintainability, time to market, and scalability. Regarding the impact of microservices on performance and transactionality, they found that the respondents considered performance degradation as a minor issue, whereas transactionality (e.g., cross-service ACID transactionality) is regarded as a major issue. Ghofrani and Lübke reported the challenges related to the design and development of microservices systems through a survey study with 25 practitioners \cite{ghofrani2018challenges}. The reported challenges are distributed nature of MSA, management of multiple repositories, issue tracking in MSA, sharing data among multiple services, dockers networking, and debugging of dependent microservices. Their survey also reported the goals for adopting microservices (e.g., scalability, agility, expendability), approaches for deriving services boundary (e.g., DDD, formal methods), and notations used for describing MSA.

A survey with 122 practitioners was conducted to understand how microservices are practiced in industry by Viggiato et al \cite{viggiato2018microservices}. A major focus of their survey is on identification of the programming languages, technologies, challenges, and advantages associated with microservices. The results of their survey indicated that popular programming languages used in microservices systems are Java, JavaScript, Node.js, C\#, and PHP, and the common Database Management Systems (DBMS) are PostgreSQL, MySQL, MongoDB, SQL Server, and Oracle. Regarding the communication protocols, 62\% of the participants in their survey declared that they used REST over HTTP. The key identified challenges are distributed transactions, microservices testing, faults diagnosing, and slow Remote Procedures Calls (RPCs). The major advantages that can be gained by employing microservices are independent deployment, scalability, maintainability, and flexibility to use several technologies for implementing microservices.
 
Haselbock et al. conducted interviews with 10 experts to investigate the design areas and their importance in microservices systems \cite{haselbock2018expert}. The design areas that they identified are service design (e.g., service partitioning), organization structure design (e.g., communication structures), monitoring and logging infrastructure, system design (e.g., integration, fault tolerance, service discovery, versioning, scalability, security), and automated testing. The participants of this survey assessed that monitoring and logging is the most important microservices design area. Baskarada et al. explored the opportunities and challenges related to MSA through in-depth interviews \cite{bavskarada2018architecting}. The survey results show that MSA provides better development and deployment agility along with operational scalability over the monolithic architecture. Some of the identified challenges are skills required for working on a different set of technologies, organizational culture, application decomposition into microservices, deciding about when to use orchestration and choreography in MSA, microservices testing, and performance degradation. Zhang et al. interviewed 13 practitioners from China to investigate the gap between literature and practice about microservices \cite{zhang2019microservice}. They took the nine characteristics of MSA (e.g., componentization via services, organization around business capabilities, smart endpoints, and dumb pipes) proposed by Fowler and Lewis \cite{fowler2014microservices} and characterized the gaps between vision and the reality of microservices. They also identified several benefits (e.g., independent upgrade, independent scale (up/down), independent development), and challenges (e.g., chaotic independence, unguided organizational transformation, complexity of API management) related to microservices.

Zhou et al. conducted an industrial survey with 16 practitioners to investigate the effectiveness of existing debugging practices for microservices systems \cite{zhou2018fault}. They specifically investigated faults of microservices systems, debugging practices, and the related challenges faced by practitioners. They reported 22 faults with their root causes, and also developed a benchmark system to replicate the identified debugging faults of microservices systems. The results of their study show that proper tracing and visualization analysis can help to locate various kinds of faults, and there is a need for more intelligent trace analysis and visualization techniques. Wang et al. conducted interviews with 21 practitioners and an online follow-up survey completed by 37 practitioners to investigate the best practices, challenges, and solutions concerning successfully developing microservices systems \cite{Wang2020}. Specifically, they investigated three aspects of microservices systems: architecture (e.g., microservice granularity), infrastructure (e.g., logging and monitoring), and code management (e.g., managing API change). They found that a clear sense of microservices ownership, controlled API management, process automation, and investment in robust logging and monitoring infrastructure are the best practices for the successful development of microservices systems. In addition, the study \cite{Wang2020} indicates that it is always not beneficial to use a plurality of languages (multiple languages) and utilize business capabilities to decompose applications into microservices. Finally, they revealed that managing the shared code between microservices is challenging.

Taibi et al. conducted an empirical study by interviewing 21 practitioners who migrated their monoliths to microservices. In their work, they identified the motivations (e.g., maintainability, scalability), issues (e.g., monolith decoupling, database migration and data splitting), and benefits (e.g., maintainability improvement, scalability improvement) for the migration from monoliths to microservices. They also proposed a migration process framework after comparing three migration processes adopted by the interviewed practitioners \cite{taibi2017processes}. Regarding the migration from monoliths to microservices, Di Francesco et al. also conducted interviews with 18 practitioners to report the migration activities and challenges. Domain decomposition, identification of the new system's services, and application of DDD practices are recognized as leading migration activities. In contrast, high coupling, identification of the microservice boundaries, and decomposition of the pre-existing systems are identified as leading migration challenges. Besides that, we also identified an important body of knowledge that reports dozens of patterns for implementing different aspects (e.g., data management, deployment) of microservices systems \cite{alshuqayran2016systematic, richardson2018microservices, marquez2018actual, marquez2018review, taibi2018architectural}. However, none of these studies report how often practitioners use these patterns when designing microservices systems.

Table \ref{tab:Comp_with_Exi} shows the comparison between the results of our survey and the existing studies. It is demonstrated in Table \ref{tab:Comp_with_Exi} that the results of our survey and the existing studies are significantly different. For instance, the existing studies do not report MSA architecting activities from practice, and do not discuss MSA patterns and QAs of microservices systems as well. Also, we did not find any study that reports the evidence from practice about monitoring and testing of microservices systems. Note that the symbol “\checkmark” in Table \ref{tab:Comp_with_Exi} indicates that our survey results are similar to some extent with the results of related work. In contrast, the symbol “-” indicates that our survey results are different from the results of existing studies.


{\renewcommand{\arraystretch}{1}
\footnotesize
\begin{longtable}{|l|c|c|c|c|c|c|c|c|c|c|c|}
\caption{A comparison of the results between this survey and the existing studies}
\label{tab:Comp_with_Exi}
\\\hline
\rowcolor[HTML]{C0C0C0} 
\cellcolor[HTML]{C0C0C0}{\color[HTML]{000000} } &
  \multicolumn{10}{c|}{\cellcolor[HTML]{C0C0C0}{\color[HTML]{000000} Existing Studies}} \\\cline{2-11} 
\rowcolor[HTML]{C0C0C0} 
\multirow{-2}{*}{\cellcolor[HTML]{C0C0C0}{\color[HTML]{000000} \textbf{This Survey Results}}} &
{\cite{knoche2019drivers}} &
{\cite{ghofrani2018challenges}} &
{\cite{viggiato2018microservices}} &
{ \cite{haselbock2018expert}} &
{\cite{bavskarada2018architecting}} &
{\cite{zhang2019microservice}} & 
{\cite{zhou2018fault}}&
{\cite{Wang2020}}& 
{\cite{taibi2017processes}} &
{\cite{di2018migrating}} \\\hline
{Decomposition strategies} &
  {-} &
  {\checkmark} &
  {-} &
  {-} &
  {-} &
  {\checkmark}&
  {-} &
  {\checkmark} &
  {-} &
  {-}
  \\ \hline
{Implementation strategies} &
  {-} &
  {-} &
  {-} &
  {-} &
  {-} &
  {-} &
  {-} &
  {-} &
  {-} &
  {-}
  \\ \hline
{Architecting activities} &
  {-} &
  {-} &
  {-} &
  {-} &
  {-} &
  {-} &
  {-} &
  {-} &
  {-} &
  {-}
  \\ \hline
{Description methods} &
  {-} &
  {\checkmark} &
  {-} &
  {-} &
  {-} &
  {-} &
  {-} &
  {-} &
  {-} &
  {\checkmark}
  \\ \hline
{Diagrams} &
  {-} &
  {-} &
  {-} &
  {-} &
  {-} &
  {-} &
  {-} &
  {-} &
  {-} &
  {-}
  \\ \hline
  {MSA components} &
  {-} &
  {-} &
  {-} &
  {-} &
  {-} &
  {-} &
  {-} & 
  {-} &
  {-} &
  {-}
  \\\hline
 {QAs} &
  {-} &
  {-} &
  {-} &
  {-} &
  {-} &
  {-} &
  {-} &
  {-} &
  {\checkmark} & 
  {-}
  \\ \hline
{MSA patterns} &
  {-} &
  {-} &
  {-} &
  {-} &
  {-} &
  {-} &
  {-} &
  {-} &
  {-} &
  {-}
  \\ \hline
{Design skills} &
  {-} &
  {-} &
  {-} &
  {-} &
  {-} &
  {-} &
  {-} &
  {-} &
  {-} &
  {-}
  \\ \hline
{Monitoring metrics} &
  {-} &
  {-} &
  {-} &
  {-} &
  {-} &
  {-} &
  {-} &
  {\checkmark} &
  {-} & 
  {-}
  \\ \hline
{Monitoring practice} &
  {-} &
  {-} &
  {-} &
  {-} &
  {-} &
  {-} &
  {-} &
  {\checkmark} &
  {-} &
  {-}
  \\ \hline
{Monitoring tools} &
  {-} &
  {-} &
  {-} &
  {-} &
  {-} &
  {-} &
  {-} &
  {\checkmark} &
  {-} &
  {-}\\ \hline
{Testing methodologies} &
  {-} &
  {-} &
  {-} &
  {-} &
  {-} &
  {-} &
  {-} &
  {-} &
  {-} &
  {-}
  \\ \hline
{Testing tools} &
  {-} &
  {-} &
  {-} &
  {-} &
  {-} &
  {-} &
  {-} &
  {-} &
  {-} &
  {-}\\ \hline
{Testing skills} &
  {-} &
  {-} &
  {-} &
  {-} &
  {-} &
  {-} &
  {-} &
  {-} &
  {-}&
  {-}
  \\ \hline
{Design challenges} &
  {\checkmark} &
  {-} &
  {\checkmark} &
  {-} &
  {\checkmark} &
  {\checkmark} &
  {-} &
  {\checkmark} &
  {\checkmark} &
  {\checkmark}\\ \hline
{Monitoring challenges} &
  {-} &
  {-} &
  {-} &
  {-} &
  {-} &
  {-} &
  {\checkmark} &
  {\checkmark} &
  {-} &
  {-}\\ \hline
{Testing challenges} &
  {-} &
  {-} &
  {-} &
  {-} &
  {-} &
  {-} &
  {-} &
  {-} &
  {-} &
  {-}\\ \hline

\end{longtable}}

\section{Conclusions} 
\label{sec:conclusions}
Through this mixed-methods study, we gained insight into microservices systems regarding how practitioners design, monitor, and test microservices systems in the industry. The study results are meaningful for both researchers and practitioners who intend to understand: (1) team structure, application decomposition and implementation strategies, MSA architecting activities, MSA design description methods, QAs of microservices systems, MSA patterns, and skills required to design and implement MSA; (2) microservices monitoring metrics, practices, and tools, (3) microservices testing strategies, tools, and skills, (4) the challenges and solutions related to design, monitoring, and testing of microservices systems, as well as (5) reasons behind key findings of the study. The main findings of this survey are summarized as follows:

\begin{enumerate}
    \item From 29 countries, 106 practitioners participated in our survey. 65\% of the participants have a role of application developers, and 59.5\% of the participants indicated that their organizations use the MSA style for building some applications. In most of the organizations, 2 to 3 people are working on one microservice. Regarding application domains, almost one-third (33.1\%) of the participants' organizations develop E-commerce applications. Moreover, Scrum (59.4\%) and DevOps (50\%) are the leading development methods employed in the MSA projects in the industry.
    
    \item For the interviews, we recruited 6 practitioners from four countries. Mainly they have design and development responsibilities. Their average work experience in the IT industry is 11.7 years, and their average work experience in microservices systems is 3.5 years. 
    
    \item The majority of the organizations (63.2\%) have a team or person for designing the MSA. Applications are mainly decomposed by using the combination of DDD and business capability strategies, responded by 42.3\% of the participants. 46.7\% of the participants indicated that they only modeled the high-level design of microservices systems. Concerning MSA architecting activities, practitioners are more likely to focus on architecture evaluation, implementation, and maintenance and evolution. More than half of the participants (50.9\%) used informal approaches (i.e., Boxes and Lines) for describing MSA. Flowchart, use case diagram, data flow diagram, activity diagram, and sequence diagram are the most popular diagrams for representing the design and architecture of microservices systems. Security, availability, performance, and scalability are ranked as most important QAs by the participants for designing microservices systems. Regarding MSA design patterns used in practices, API gateway, Database per service, Backend for frontend, and Access token are reported as the most often used patterns.
    \item Resource usage, load balancing, availability, and database connections are most frequently used monitoring metrics. About monitoring practices, more than 50\% of the participants pointed out that they used log management, exception tracking, and health check API most often. Jira (39.6\%) and Datadog Kafka Dashboard (38.6\%) are identified as the popular tool for monitoring microservices systems.
    \item Unit testing, E2E, and integration testing are recognized as the top three strategies to test microservices systems. The most popular tools among the participants for testing microservices systems are Junit, Jmeter, and Mocha.
    \item The most frequently reported challenges regarding the design of microservices systems are \textit{Clearly defining the boundaries of microservices} (69.8\%), \textit{addressing the security concerns} (46.2\%), and  \textit{managing microservices complexity at the design level} (34.9\%). Regarding monitoring challenges, we found that \textit{collection of monitoring metrics and logs from containers} (54.7\%), \textit{distributed tracing} (45.3\%), and \textit{having many components to monitor (complexity)} (41.5\%) are the leading challenges. Concerning testing challenges, practitioners highlighted \textit{creating and implementing manual tests} (50\%), \textit{integration testing of microservices} (43.4\%), and \textit{debugging of the microservices on a container platform} (42.5\%) as critical challenges.
    \item We received several recommendations (e.g., design patterns, approaches and tools) from practitioners to address some of the design, monitoring, and testing challenges. However, there are still several challenges that do not have dedicated solutions. For instance, \textit{recovering MSA from existing code}, \textit{failure zone detection}, \textit{performance testing of microservices}.
    \item We also compared the key findings of our survey with the surveys conducted with non-MSA-based systems (see Section \ref{compMSAvsNonMSA}). 
\end{enumerate}

We have also identified several promising research directions to improve the practices of the design, monitoring, and testing of microservices systems (see the research implications in Section \ref{analysis}). We argue that researchers and practitioners should work together to bring more dedicated solutions to address the challenges reported in this study.
\section*{Acknowledgments} 
\label{sec:ack}
This work is partially sponsored by the National Key R\&D Program of China with Grant No. 2018YFB1402800 and the NSFC with Grant No. 62172311, the Italian Ministry of Economy and Finance, Cipe resolution n. 135/2012 (project INCIPICT - INnovating CIty Planning through Information and Communication Technologies), the Department of Electronics and Informatics of the Federico Santa María Technical University (Concepción, Chile) and SISMA national research project funded by the MIUR under the PRIN 2017 program (Contract 201752ENYB). We would also like to thank all the participants who contributed their knowledge and insights to this industrial survey.

\clearpage
\label{sec:Appendix}
\begin{center}
 \textbf{Appendix A. Abbreviations used in this study}
 \end{center}

  {\renewcommand{\arraystretch}{1}
\footnotesize
\begin{longtable}{|l|l|}
\caption{Abbreviations used in this study}
\label{tab:Abbreviations}
 \\ \hline
AA & Architectural Analysis\\ \hline
ACID & Atomicity, Consistency, Isolation, Durability\\\hline 
ADL &Architectural Description Language \\\hline
AS & Architectural Synthesis\\  \hline
AE & Architectural Evaluation\\  \hline
AI & Architectural Implementation\\ \hline
AME & Architectural Maintenance and Evolution\\ \hline
API & Application Program Interface\\\hline
BDD & Behavior-Driven Development \\\hline
CQRS & Command Query Responsibility Segregation\\\hline
DevOps & Development and Operations\\\hline
DDD & Domain-Driven Design\\\hline
DQ & Demographic Question\\\hline
DSDM & Dynamic Systems Development Model\\\hline
DSL & Domain-Specific Language \\\hline
E2E & End-to-End\\\hline
FDD & Feature Driven Development\\\hline
IDC & International Data Corporation\\\hline 
IDE & Integrated Development Environment\\\hline
JAD & Joint Application Development\\\hline
MDD & Model-Driven Development\\\hline 
MSA & Microservices Architecture\\\hline
QA & Quality Attribute\\\hline
RAD & Rapid Application Development\\\hline
RQ & Research Question\\\hline
RUP & Rational Unified Process\\\hline
SOA & Service Oriented Architecture\\\hline
SLA & Service-Level Agreement\\\hline 
SMS & Systematic Mapping Study\\\hline
SQ & Specific Question\\\hline
UML & Unified Modeling Language\\\hline 
VM& Virtual Machine\\\hline
XP & Extreme Programming\\\hline
\end{longtable}
}
\clearpage
\begin{landscape}
  
\begin{center}
 \textbf{ Appendix B. Survey and interview questionnaires}
 \end{center}

 \begin{tcolorbox}[colback=gray!5!white,colframe=gray!75!black,title=\centering The Welcome Page of the Survey]
\justify
\textbf{Microservices Architecture (MSA)} is a cloud-native architectural style that is inspired from Service-Oriented Architecture (SOA). In general, microservices are organized as a suite of small granular services that can be implemented through different technological stacks.

This survey aims at understanding how MSA-based systems are designed, tested, and monitored in the software industry. The results of this survey will be used to develop an evidence-based body of knowledge to better and further support of design, monitoring, and testing of MSA-based systems.

The survey takes approximately \textbf{20-25} minutes to complete. This survey is comprised of 4 sections with the questions about:
\begin{enumerate}
    \item Background of the survey participants (e.g., their role in software project)
    \item The design aspect of MSA-based systems
    \item Monitoring of MSA-based systems
    \item Testing of MSA-based systems
\end{enumerate}
We are keen to invite you to participate, if 
\begin{enumerate}
    \item you are professionally working in the software industry and your organization has adopted/is adopting or is planning to adopt MSA; or
    \item you are an independent practitioner and have experience and/or expertise with MSA-based systems.
\end{enumerate}
Please note, all of the responses gathered through this survey will be treated with confidentiality and privacy. The results of this survey will be used for academic purposes only. We would very much appreciate your participation in this research project\\
\end{tcolorbox}

 
{\renewcommand{\arraystretch}{1}
 \footnotesize
 \begin{longtable}{p{1cm}p{10.3cm}p{10.3cm}}
    \caption{Questions on demographics of the survey}
    \label{tab:Demographics}
    \\\toprule 
   \textbf{ID} & \textbf{Survey Questions} & \textbf{Type of Answers}\\
    \midrule 
    DQ1 & Which country are you working in?& Selection from countries list\\
    \midrule
    DQ2 & (Multiple Choice) What are your major responsibilities in your company?& Application Developer / Database Developer/ Database Administrator / Architect / System Analyst/Tester / DevOps Engineer / Operations Staff / Business Analyst / Consultant / C-level Executive (CIO, CTO, etc)\\
    \midrule 
    DQ3 & Regarding the usage of Microservices Architecture (MSA) for developing applications &  (1) We use MSA style for building all applications in our organization/team. (2) The MSA style is only used for building some specific applications in our organization/team. (3) Our organization/team does not use MSA style at all
\\    \midrule 
DQ4 & How many years have you been working in the IT industry? & 0-2 Years / 3-5 Years / 6-10 Years / \textgreater 10 Years\\
\midrule 
 DQ5 & How many years have you been working with MSA-based systems? & 0-2 Years / 3-5 Years / 6-10 Years / \textgreater 10 Years\\
\midrule
DQ6 & Have you ever received any professional training (i.e., excluding higher education) related to MSA? & Yes/No\\
\midrule  
DQ7 & (Multiple Choice) What is the work domain of your organization? & E-commerce / Education / Embedded systems / Financial / Healthcare / Insurance / Internet / Manufacturing / Professional services / Real estate / Telecommunication / Transportation and Warehousing / Other\\
\midrule 
DQ8 & How large is your organization? & 1–20 employees / 21–100 employees / 101-500 employees / 501-1000 employees /  
\textgreater 1000 employees\\
\midrule 
DQ9 &  On average, how many people work (i.e., design, develop, test, deploy, etc.) on one microservice in your organization? & 1 / 2-3 / 4-5 / 6-7 / 8-9 / \textgreater 10\\
\midrule 
DQ10 & (Multiple Choice) Which of the following software development methods is (commonly) used in your MSA projects? & DevOps / SCRUM / Crystal Methods / DSDM / XP / FDD / JAD / Lean Development / RAD / Rational Unified Process / Iterative Development / Spiral / Waterfall (a.k.a. Traditional) / We don’t use any defined method / Other \\ 
\bottomrule 
    \end{longtable}
    }

 {\renewcommand{\arraystretch}{1}
 \footnotesize
 \begin{longtable}{p{1cm}p{8.3cm}p{12.3cm}}
    \caption{Questions on design of MSA-based systems}
    \label{tab:DesignofMSA}
    \\\toprule 
    \textbf{ID} & \textbf{Survey Questions} & \textbf{Type of Answers}\\
    \midrule
    \justify SQ11 & Does your organization have a dedicated individual or team for creating and managing architecture for MSA-based systems? & Yes / No / Yes, but with some other responsibilities\\
    \midrule
    SQ12 & (Multiple Choice) How do you decompose an application into microservices? & We define microservices corresponding to business capabilities. / We define microservices corresponding to DDD sub-domains. / We define microservices by using the combination of business capability and Domain-Driven Design (DDD). /  Other (please specify)\\
    \midrule

    SQ13 & Which strategy do you follow from design to implementation of MSA-based systems? & We model the high-level design for MSA-based systems that represents only major components (e.g., main microservices, databases, messaging systems), and then write the code based on that design model. / We model the detailed design for MSA-based systems that represents all components (e.g., design for individual microservice), and then write the code based on that design model. / We directly write the code for MSA-based systems without any design model. / We use a Model Driven Development (MDD) approach to generate the code from the design model. / Other (please specify) \\
    \midrule
    SQ14 & We consider Architecturally Significant Requirements (ASRs) when designing MSA-based systems (Architectural Analysis). & Strongly disagree / Disagree / Neutral / Agree / Strongly agree \\
    \midrule
    
    SQ15 & We design the candidate architecture solutions to address ASRs when designing MSA-based systems (Architectural Synthesis). & Strongly disagree / Disagree / Neutral / Agree / Strongly agree \\
    \midrule
    
    SQ16 & We evaluate the candidate architecture solutions when designing MSA-based systems to ensure design decisions are made right (Architectural Evaluation). & Strongly disagree / Disagree / Neutral / Agree / Strongly agree \\
    \midrule
    
    SQ17 & We refine the coarse-grained architecture design into the fine-grained detailed design so that developers can write the code based on the MSA (Architectural Implementation).  & Strongly disagree / Disagree / Neutral / Agree / Strongly agree \\
    \midrule
    
   SQ18 & To ensures the consistency and integrity of the MSA, we regularly fix issues in or update the MSA according to changed operational environments or new requirements (Architectural Maintenance and Evolution). & Strongly disagree / Disagree / Neutral / Agree / Strongly agree \\
   \midrule
   
   SQ19 & (Multiple Choice) Which of the following methods do you use for describing the architecture of MSA-based systems? & UML / Architectural description language / Domain-specific language / Boxes and Lines\\
   \midrule
 
  SQ20 & (Multiple Choice) Which of the following diagrams do you use for representing the design and architecture of MSA-based systems? & Composite structure diagram / Interaction overview diagram / Object diagram / Timing diagram / State machine diagram / Package diagram / Communication diagram / Functional flow block diagram / BPMN / Component diagram / Deployment diagram / Class diagram / Sequence diagram / Activity diagram / Data flow diagram / Use case diagram / Flowchart diagram\\
  \midrule
  
  SQ21 & (Multiple Choice) Which of the following entities can be considered as architectural design components of MSA-based systems? & Service interface (e.g., Pages, Grids/Panels) / Process (e.g., Use cases) / Domain (e.g., Domain objects, Factories/Repositories) / Service (e.g., Service gateways, Service clients) / Storage (e.g., Relations, MongoDB) /All the above / Other (please specify)\\
  \midrule
 
  SQ22 & (Multiple Choice) Please rate the importance of the following quality attribute when designing MSA-based systems. Please use the Likert-scale points \textit{“Very Important, Important, Somewhat Important, Not Important, and Not Sure”} to indicate the importance of QAs. & Availability / Compatibility / Functional suitability / Maintainability / Portability / Performance / Reliability / Scalability / Usability / Security / Resilience / Testability / Monitorability / Reusability / Interoperability\\
  \midrule
  
  SQ23 & (Multiple Choice) Please indicate how often you use the following MSA design patterns when designing MSA-based systems. Please use the Likert-scale points \textit{“Very Often, Often, Sometimes, Rarely, and Never”} to indicate the use of MSA patterns.
  & Access token / API composition / API gateway / Application events / Backend for frontend / Circuit breaker/ Client-side discovery / Command query / CQRS / Database per service / Domain-specific protocol / Event sourcing / Messaging / Sidecar / 3rd party registration/ Transactional outbox / Language-specific packaging format / Polling publisher / Remote procedure invocation / Sagas / Self-registration / Serverless deployment / Server-side discovery / deployment platform / Deploy a service as container / Service mesh / Deploy a service as a VM / Service registry / Shared database / Transaction log tailing / Other (please specify)\\
  \midrule
  
  SQ24 & (Multiple Choice) What challenges do you face when you design the architecture for MSA-based systems? & Clearly defining the boundaries of microservices / Finding appropriate modelling abstractions for microservices / Recovering MSA from existing code / Managing microservices complexity at the design level / Reaching scalability / Separating functional and operational concerns / Addressing data management at the design level / Addressing microservices communication at the design level / Addressing security concerns / Aligning team structure with the architecture of MSA-based system / Other (please specify)\\
  \midrule
 
 SQ25 & (Multiple Choice) Which of the following architectural challenges may encounter when monitoring and testing MSA-based systems?
 \begin{itemize}
     \item Monitoring of MSA-based systems
     \item Testing of MSA-based systems
 \end{itemize} & Clearly defining the boundaries of microservices / Finding appropriate modelling abstractions for microservices / Recovering MSA from existing code / Managing microservices complexity at the design level / Reaching scalability / Separating functional and operational concerns / Addressing data management at the design level / Addressing microservices communication at the design level / Addressing security concerns / Aligning team structure with the architecture of MSA-based system / Other (please specify)\\
 \midrule
 
 SQ26 & What solutions (e.g., practices, tools, team structures) can be employed to address or mitigate the challenges related to architecting for MSA-based systems? & Free text\\
 \midrule

 SQ27 & (Multiple Choice) What skills (including hard and soft skills) are required to properly design and implement MSA? & A practitioner should have the skill to break down a complex coding task into smaller tasks that can be implemented separately / A practitioner should have the skill to implement the functionality of a microservice by using suitable MSA patterns (e.g., circuit breaker) / A practitioner should have architectural knowledge of technology decisions / A Practitioner should have knowledge about the basic concepts of DevOps / A practitioner should have knowledge about container technologies / A practitioner should have knowledge of how to secure microservices / A practitioner should be able to work in a dynamic team structure / Other (please specify)\\
 \bottomrule
 \end{longtable}
 }
 
  {\renewcommand{\arraystretch}{1}
\footnotesize
 \begin{longtable}{p{1cm}p{8.3cm}p{12.3cm}}
    \caption{Questions on monitoring of MSA-based systems}
    \label{tab:MonitoringofMSA}
    \\ \hline
   \textbf{ID} & \textbf{Survey Questions} & \textbf{Type of Answers}\\
   \toprule
    SQ28 & (Multiple Choice) Which metrics do you collect for monitoring MSA-based systems? & Resource usage (e.g., CPU, memory) / Threads / Microservice’s open File Descriptors (FD) / Database connections / Load balancing / Language-specific metrics (e.g., garbage collector behaviour, RPC, and database latency) / Status of service availability / Service Level Agreement (SLA) / Latency (of both the service as a whole and its API endpoints) / Endpoint success / Endpoint response time / Errors and exceptions / Status of dependencies / Other (please specify)\\
    \midrule
    SQ29 & (Multiple Choice) Which of the following practices or techniques do you use to monitor MSA-based systems? & Log management / Audit logging / Distributed tracking / Exception tracking / Health check API / Log deployment and changes / Other (please specify) \\
    \midrule 
 
    SQ30 & (Multiple Choice) What tools do you use to monitor MSA-based systems? & Datadog Kafka Dashboard / Apache Mesos / Cacti / Grafana / iPerf / Jira / Nagios / Omnia / Pact / Prometheus / Raygun APM / Spring Boot Actuator / Zipkin / Other (please specify) \\
    \midrule 
    
    SQ31 & (Multiple Choice) What challenges do you usually face when monitoring MSA-based systems? & Having many components to monitor (complexity) / Collection of monitoring metrics data and logs from containers/ Monitoring of application running inside the container / Distributed tracing / Failures zone detection / Performance monitoring / Availability of the monitoring tools / Maintaining monitoring infrastructures / Analysing the collected data / Other (please specify) \\
    \midrule
    SQ32 & (Multiple Choice) How do you rate the severity (impact) of each challenge related to the monitoring of MSA-based systems? 
    Please use the Likert-scale points “Catastrophic, Major, Moderate, Minor, and Insignificant” to rate the severity of monitoring challenges. & Having many components to monitor (complexity) / Collection of monitoring metrics data and logs from containers/ Monitoring of application running inside the container / Distributed tracing / Failures zone detection / Performance monitoring / Availability of the monitoring tools / Maintaining monitoring infrastructures / Analysing the collected data / Other (please specify) \\
    \midrule
    SQ33 & What practices, techniques or tools do you use to improve the monitorability of MSA-based systems (i.e., making MSA-based systems easy to monitor)? & Free text \\
    \bottomrule
 \end{longtable}
 }
  {\renewcommand{\arraystretch}{1}
 \footnotesize
 \begin{longtable}{p{1cm}p{8.3cm}p{12.3cm}}
    \caption{Questions on testing of MSA-based systems}
    \label{tab:TestingofMSA}
    \\ \toprule
    \textbf{ID} & \textbf{Survey Questions} & \textbf{Type of Answers}\\
   \toprule
    SQ34 & (Multiple Choice) Which of the followings testing methodologies do you use to test MSA-based systems? & Unit testing / Integration contract testing / Consumer-driven contract testing / Integration testing / Component testing / Scalability testing / Service component testing / User Interface Testing / End to End Testing / A-B Testing / Other (please specify) \\
    \midrule
    SQ35 & (Multiple Choice) What tools do you use to test MSA-based systems? & Cucumber / FitNesse / Gatling / Jasmine / JMeter / Junit / Karma / Locust. IO / Microfocus UFT / Mocha / MockLito / Pdiffy / Perfecto / Phantom / Sauce Labs / Selenium / SoapUI / Tap / TestNG / Travis / Other (please specify) \\
    \midrule
    SQ36 &  (Multiple Choice) What challenges do you usually face when testing MSA-based systems? & Creating and implementing manual tests/ Creating and implementing automated tests / Integration testing of microservices / Independent microservices / Debugging of microservices on container platforms / Understanding about each testable microservice / Many independent teams / Performance testing of microservices / Polyglot technologies in MSA-based systems / Complexity and size of MSA-based systems / Other (please specify) \\ 
    \midrule
    
    SQ37 & (Multiple Choice) How do you rate the severity (impact) of each challenge related to testing of MSA-based systems? Please use the Likert-scale points “Catastrophic, Major, Moderate, Minor, and Insignificant” to rate the severity of testing challenges. & Creating and implementing manual tests can pose a challenge / Creating and implementing automated tests can pose a challenge / Integration testing of microservices can pose a challenge / Independent microservices can pose a challenge for testing / Debugging of microservices on container platforms / Understanding about each testable microservice / Many independent teams can pose a challenge for testing / Performance testing of microservices is challenging / Implementation of polyglot technologies in MSA-based systems can pose a challenge for testing / Complexity / size of the MSA-based system can pose the challenge for testing / Other (please specify) \\ 
    \midrule

    SQ38 &  (Multiple Choice) What skills (both soft and hard skills) are required to test MSA-based systems properly? & A tester has the ability to write good unit test cases to detect potential bugs / A tester has the ability to write good integration test cases to detect potential bugs / A tester should have basic knowledge about multiple databases (e.g., SQL, NoSQL, MongoDB) / A tester should have basic knowledge of Linux commands / A tester should have knowledge about test management tools (e.g., a tool for tracking test cases written by your team) / A tester should have knowledge about the test automation tools / A tester should have analytically and logically thinking / Other (please specify) \\ 
    \midrule
    
    SQ39 & What practices, techniques, or tools do you use to make MSA-based systems more testable (i.e., making MSA-based systems easy to test)? & Free text\\
    \bottomrule
 \end{longtable}
 }


 {\renewcommand{\arraystretch}{1}
 \footnotesize

 \begin{longtable}{p{1cm}p{8.3cm}}
    \caption{Interview questions on demographics information}
    \label{tab:IQDMinfo}
    \\ \toprule
    \textbf{ID} & \textbf{Interview Questions}\\
   \toprule 
    IQ1.1 & Which country are you working in?  \\
    \midrule
    IQ1.2 & What are your major responsibilities in your company?  \\
    \midrule
    IQ1.3 & How many years have you been working in the IT industry?  \\
    \midrule
    IQ1.4 & What is the work domain of your organization?  \\
    \midrule
    IQ1.5 & How many years have you been working with microservices systems?\\
   \bottomrule
 \end{longtable}
 }
 {\renewcommand{\arraystretch}{1}
 \footnotesize
 \begin{longtable}{p{1cm}p{8.3cm}p{12.3cm}}
    \caption{Interview questions on design of microservices systems}
    \label{tab:IQMsaDesign}
    \\ \toprule
    \textbf{ID} & \textbf{Questions Context} & \textbf{Interview Questions}\\
   \toprule
    IQ2.1& Our survey results show that most of the respondents (42.3\%) used a combination of DDD and business capability to decompose applications into microservices. Regarding this finding, we would like to ask the following three questions: & \textbf{IQ2.1.1}: Do you think that this finding reflects the actual use of application decomposition strategies (i.e., a combination of DDD and business capability) for microservices (Yes OR No)? Please give a reason(s) to support your answer. 
    
    \textbf{IQ2.1.2}: What could be the reasons for the frequent use of a combination of DDD and business capability strategy to decompose applications into microservices? 
    
    \textbf{IQ2.1.3}: Are you familiar with any other strategy that can be used to decompose applications into microservices? If yes, please provide the name of the strategy.
\\
    \midrule
    IQ2.2 & Our survey results show that 46.7\% of the respondents reported that they only created a high-level design for major microservices before implementing them. Regarding this finding, we would like to ask the following two questions: & \textbf{IQ2.2.1}: What could be the reasons for creating the only high-level design for major microservices?
    
    \textbf{IQ2.2.2}: What could be the possible risks if practitioners do not create a design or detailed design for microservices systems?
 \\
    \midrule
    IQ2.3 & Regarding MSA architecting activities when designing microservices systems investigated in this survey (Architectural Analysis (AA), Architectural Synthesis (AS), Architectural Evaluation (AE), Architectural Implementation (AI), and Architectural Maintenance and Evolution (AME)), we would like to ask the following two questions: &\textbf{IQ2.3.2}: What could be the reasons for the frequent use of Architecture Evaluation (AE) activity when designing microservices systems?
    
    \textbf{IQ2.3.3}: What are the reasons that practitioners pay more attention to architecting activities close to the solutions space (i.e., AE, AI, AME) instead of the problem space (i.e., AA, AS)? \\ 
    \midrule
   

    IQ2.4 &  Our survey results show that a large number of participants did not use architectural description languages (84.7\%) and domain-specific languages (72.7\%) for describing the architecture of microservices systems. Regarding this finding, we would like to ask the following question:& \textbf{IQ2.4.1}: What are the reasons that practitioners do not use architectural description languages and domain-specific languages for describing the architecture of microservices systems?
    
    \textbf{IQ2.4.2}: What are the reasons that practitioners prefer to use informal (e.g., Boxes and Lines) and semi-formal (e.g., UML) approaches for describing MSA?
 \\ 
    \midrule
    
    IQ2.5 & Our survey results show that security, availability, performance, and scalability are the most important quality attributes for microservices systems. Regarding this finding, we would like to ask the following question:& \textbf{IQ2.5.1}: What could be the reasons that practitioners consider security, availability, performance, and scalability the most important quality attributes for microservices systems?\\
    \midrule
    
    IQ2.6 & Our survey results show that API gateway and Backend for frontend as communication patterns, Database per service as data management pattern, and Access token as security pattern are reported as the most often used MSA design patterns in practice. Regarding this finding, we would like to ask the following question:& \textbf{IQ2.6.1}: What could be the reasons that these patterns are frequently used for designing microservices systems?\\
    \midrule
    
    IQ2.7 & Our survey results show that the following are the most prominent challenges of designing microservices systems
\begin{itemize}
    \item Clearly defining the boundaries of microservices
    \item Addressing security concerns
    \item Managing microservices complexity at the design level
\end{itemize}
 & \textbf{IQ2.7.1}: What could be the reasons that these design challenges happen?
 
 \textbf{IQ2.7.2}: What could be the potential solutions to address these design challenges? \\
    \bottomrule
 \end{longtable}
 }
 
 {\renewcommand{\arraystretch}{1}
 \footnotesize
 \begin{longtable}{p{1cm}p{8.3cm}p{12.3cm}}
    \caption{Interview questions on monitoring of microservices systems}
    \label{tab:IQmonitoring}
    \\ \toprule
    \textbf{ID} & \textbf{Questions Context} & \textbf{Interview Questions}\\
   \toprule
    IQ3.1& Our survey results show that the resource usage (66.9\%), load balancing (53.7\%), and availability (52.8\%) are the most frequently used monitoring metrics for microservices systems. Regarding this finding, we would like to ask the following question: & \textbf{	IQ3.1.1}: What could be the reasons that these monitoring metrics are frequently used?.\\
   \midrule
   IQ3.2& Our survey results show that the log management (68.8\%), exception tracking (60.4\%), and health check API (57.6\%) are the most frequently used monitoring practices for microservices systems & \textbf{IQ3.2.1}: What could be the reasons that these monitoring practices are frequently used?\\\midrule
   
   IQ3.3& Our survey results show that the majority of practitioners used Jira, Datadog Kafka Dashboard, Spring Boot Actuator, and Grafana to monitor microservices systems. However, not all the tools are dedicated for monitoring microservices systems, for example, Jira is a bug tracking tool. Regarding this finding, we would like to ask the following question: & \textbf{IQ3.3.1}: Do you agree that Jira can be used to monitor microservices systems? If so, please explain how Jira supports the monitoring of microservices systems?\\
   \midrule
   
   IQ3.4&Our survey results show that the following are the most prominent challenges of monitoring microservices systems.
   \begin{itemize}
       \item Collection of monitoring metrics data and logs from containers
       \item Distributed tracing
       \item Having many components to monitor (complexity)
   \end{itemize}
 & \textbf{IQ3.4.1}: What could be the reasons that these monitoring challenges happen?
 
 \textbf{IQ3.4.2}: What could be the potential solutions to address these monitoring challenges? \\\bottomrule
 \end{longtable}
 }

 {\renewcommand{\arraystretch}{1}
 \footnotesize
 \begin{longtable}{p{1cm}p{8.3cm}p{12.3cm}}
    \caption{Interview questions on testing of microservices systems}
    \label{tab:IQMsatesting}
    \\ \toprule
    \textbf{ID} & \textbf{Questions Context} & \textbf{Interview Questions}\\
   \toprule
    IQ4.1&Our results show that the most commonly used testing strategies to test microservices systems are unit testing (63.3\%), integration testing (59.4\%), and E2E testing (55.6\%). Regarding this finding, we would like to ask the following question: & \textbf{IQ4.1.1}: What could be the reasons that these testing strategies are frequently used?\\\midrule
   
    IQ4.2& Our results show that most of the practitioners used Junit (51.8\%), JMeter (26.3\%), and Mocha (21.7\%) to test microservices systems. Regarding this finding, we would like to ask the following question:& \textbf{IQ4.2.1}: What could be the reasons that these testing tools are frequently used? \\\midrule
    IQ4.3& Our results show that the following are the most prominent challenges of testing microservices systems.
    \begin{itemize}
        \item Creating and implementing manual tests
        \item Integration testing of microservices
        \item Debugging of microservices on container platforms 
    \end{itemize}
& \textbf{IQ4.3.1}: What could be the reasons that these testing challenges happen?

\textbf{IQ4.3.2}: What could be the potential solutions to address these testing challenges?
 \\\bottomrule
    
 \end{longtable}
 }

\end{landscape}
\clearpage
\bibliography{Survey}
\end{document}